# Evaluation of uncertainties in the estimates of correlated variables as the elemental-isotopic-abundances: properties of corresponding-indirect-measurement-system-specific-relationships

B. P. Datta[1]
Radiochemistry Laboratories (BARC), Variable Energy Cyclotron Centre, Department of Atomic Energy, 1/AF Bidhan Nagar, Kolkata 700064, India

## ABSTRACT

Isotopic-*abundances* $\left(\{\boldsymbol{Y_d}\}_{d=1}^{N}\right)$ of any multi-($\boldsymbol{N}$)-isotopic-element ($\boldsymbol{E}$) are *interdependent* and therefore needed to be measured **indirectly** from their rather accurately-measurable "$(\boldsymbol{N-1})$ abundance-*ratios*: $\{\boldsymbol{X_m} = \frac{\boldsymbol{Y_m}}{\boldsymbol{Y_1}}\}_{m=2}^{N}$", i.e. as: $(\boldsymbol{y_d} \pm \boldsymbol{\epsilon_d^Y}) = \boldsymbol{f_d}\left(\{(\boldsymbol{x_m} \pm \boldsymbol{u_m})\}_{m=2}^{N}\right)$; ($\boldsymbol{d}$ = 1, 2, ... $\boldsymbol{N}$). Even, corresponding-atomic-weight $(\boldsymbol{A_E})$ is determined by translating the *estimated*-ratios "$\left(\{(\boldsymbol{x_m} \pm \boldsymbol{u_m})\}_{m=2}^{N}\right)$" into the desired-estimate "$(\boldsymbol{a_E} \pm \boldsymbol{\epsilon_E^A})$" via concerned-*input(s)-output*-system-specific-relationship (**SSR)**. However, how good the *output-estimates* "$\boldsymbol{y_d}$***(s)*** and $\boldsymbol{a_E}$" should represent their ***desired***-*true-values* "$\boldsymbol{Y_d}$***(s)*** and $\boldsymbol{A_E}$" be indicated by the respective *uncertainties* "$\boldsymbol{\epsilon_d^Y(s)}$" and $\boldsymbol{\epsilon_E^A}$" **only**; i.e. correct-knowledge of even the ***output***-*uncertainties* be indispensable for the desired-*indirect*-measurements.

Usually, by assuming the **lab**-established-(*input*)-uncertainty(s) as "$\boldsymbol{u_m}$ ***(s)***" to be **accidental**-in-nature, a measure of any-corresponding-***non-correlated***-variable-specific-*output*-uncertainty is obtained by the popular-error-propagation-law. And, any-**correlated**-variable as the isotopic-*abundance*-$\boldsymbol{Y_d}$-specific-*output*-uncertainty "$\boldsymbol{\epsilon_d^Y}$" is evaluated by incorporating **additional**-(*correlation*)-correction-factor(s) in the error-propagation-formula, i.e. "*correlation*" is being assumed as an ***output***-uncertainty-*governing-parameter*.

---

[1]: The author had carried out this work as an employee (Scientific Officer) of the "Bhabha Atomic Research Centre, Department of Atomic Energy, Government of India". However, the author has retired from the service. His present address is as follows: VILLAGE: Ramnagar, POST OFFICE: Hatgobindapur, DISTRICT: Purba Bardhaman, West Bengal, INDIA, Pin code: 713407.

However, it is here clarified that**: (i)** the *output*-uncertainty "$\boldsymbol{\epsilon_d^Y}$" (or even "$\boldsymbol{\epsilon_E^A}$"**)** should, **like** the corresponding-*output-estimate* "$\boldsymbol{y_d}$" (or "$\boldsymbol{a_E}$") itself, be **independent** of whether the isotopic-abundances (and/ or their-ratios) are correlated; and**: (ii)** any *output*-uncertainty "$\boldsymbol{\epsilon}$" should actually be the corresponding-**SSR-**governed-***systematic***-parameter, irrespective of whether the *input*-uncertainty(s) "$\boldsymbol{u_m(s)}$" be purely-*random*-in-nature. That-is, like the ***non****-correlated*-variables-specific-**SSRs** discussed elsewhere, any *isotopic*-variable-specific-**SSR** (e.g.**:** $\boldsymbol{Y_d = f_d(\{X_m\}_{m=2}^N)}$) should be **bracketed** with certain *input(s)*-**to-***output*-**variation**-parameter(s) "$\boldsymbol{[MF]_m^d(s)}$" which really **prefix(s)**, for given the *input*-uncertainty(s) "$\boldsymbol{u_m(s)}$", the corresponding-*output-uncertainty* ($\boldsymbol{\epsilon_d^Y}$). Thus, an 'a priori' study of ***only*** the **SSR-***characterizing*-"**[*MF*](*s*)**" should help to *predict* the corresponding-output-uncertainty "$\boldsymbol{\epsilon}$" and to *enable* proper-experimental-preplanning (or, where the **SSR** itself is a *proposed*-**model**, to *ascertain* whether the modelling is being worth). We here show *how* even 'a priori' **predicted**-output-uncertainties (*for* determining the usually ***unknown***-isotopic-abundances-cum-atomic-weight) should be *authenticated*.

However the *output*-uncertainty, *either* "$\boldsymbol{\epsilon_d^Y}$" *or* "$\boldsymbol{\epsilon_E^A}$", is shown to ***vary*** as a function of the *measurable*-(**input**-ratio)-variable(s) "$\boldsymbol{\{X_m\}_{m=2}^N}$". That is, among the **two** different possible **(**namely, *measurable*-variable-[as-"$\boldsymbol{X_m}$"]-***independent*****:** "**F.1**" and, -$\boldsymbol{X_m}$-***dependent*****:** "**F.2**"**) families** of **all** conceivable physicochemical-**SSRs**-cum-**models** pointed out elsewhere, the isotopic-**SSRs** belong to the "**F.2**" family. Yet the point of interest is that, whereas certain other-(*isotopic*)-"**F.2**"-family-member as the popular-**IRMS**-measurement-**model** "$\boldsymbol{Y_\delta = f_\delta(X_S, X_W) = \left(\frac{X_S}{X_W} - 1\right)}$" is clarified elsewhere to **behave** as an extraneous-*error-**source*** in any **IRMS**-measurement, *both* "$\boldsymbol{Y_d}$ and $\boldsymbol{A_E}$" specific-**SSRs** are even exemplified to **act** as the (*isotope*-"***d***" and the element-"***E***" *specific*) ***sinks***-of-corresponding-***experimental***-(input)-*error-sum* (respectively).

## 1. INTRODUCTION

The term "isotopic-abundances ($\{\boldsymbol{Y_d}\}_{\boldsymbol{d=1}}^{\boldsymbol{N}}$) of any conceivable multi-(***N***)-isotopic-element (***E***)" should refer to, as well-known, "***N***" different-**correlated**-*quantities* (mathematically, -$Y_d$-*variables*, ***d*** = 1 to ***N***). Therefore the *measurement* of isotopic-abundances ($Y_d$-values) and thus of atomic-weight ($A_E$) of any element "***E***" is generally carried out ***indirectly*** by involving, grossly, *a-two-stage-process*: **(i)** rather accurate possible *real-world*-(mass-spectrometric)-*measurements* of the corresponding "($\boldsymbol{N-1}$)" isotopic-abundance-ratios ($\{\boldsymbol{X_m} = \frac{\boldsymbol{Y_m}}{\boldsymbol{Y_1}}\}_{\boldsymbol{m=2}}^{\boldsymbol{N}}$); and: **(ii)** *computations* of desired "$Y_d$-*values* and/ or $A_E$-*value*" from *measured*-$\boldsymbol{X_m}$-*value(s)* using their *system-specific-relationships* (**SSRs**) as:

$$Y_d = f_d(\{X_m\}_{m=1}^{N}) = \frac{X_d}{\sum_{m=1}^{N} X_m} = \frac{X_d}{1+\sum_{m=2}^{N} X_m}; with: d = 1,2 \ldots N \qquad (1)$$

And:

$$A_E = f_E(\{M_d, Y_d\}_{d=1}^{N}) = \sum_{d=1}^{N}(M_d \times Y_d) = \sum_{d=1}^{N}\left(M_d \times \frac{X_d}{1+\sum_{m=2}^{N} X_m}\right) \qquad (2)$$

where "$Y_1$" is, for the convenience of representation, referred to as the *reference*-isotopic-abundance (and thus: $X_1$ = **1**); and "$M_d$" stands for the $\boldsymbol{d^{th}}$-isotopic-mass (*see* also the **Appendix 1**, which gives a clarification on the notations used in this paper).

However, any (directly)-*measured-estimate* "$\boldsymbol{x_m}$ (of the *true*-isotopic-abundance-ratio: $\boldsymbol{X_m}$)" should, even after correcting for the possible mass-fractionation and/ or other detected method-specific-biases, be subject to certain *uncertainty* [1], i.e.: $\boldsymbol{X_m}$ = ($\boldsymbol{x_m} \pm \boldsymbol{u_m}$) and/ or: $\boldsymbol{x_m}$ = ($\boldsymbol{X_m} \mp \boldsymbol{u_m}$); with "$\boldsymbol{u_m}$" as, generally, the **lab**-established-$\boldsymbol{X_m}$-measurement-*uncertainty*. Therefore the *estimates* "$\{\boldsymbol{y_d}\}_{\boldsymbol{d=1}}^{\boldsymbol{N}}$, and $\boldsymbol{a_E}$" **(**of the **desired** *true*-isotopic-abundances "$\{\boldsymbol{Y_d}\}_{\boldsymbol{d=1}}^{\boldsymbol{N}}$" and the *true*-atomic-weight "$A_E$"**)** should also be at certain *uncertainties* (say: $\{\boldsymbol{\epsilon_d^Y}\}_{\boldsymbol{d=1}}^{\boldsymbol{N}}$, and: $\boldsymbol{\epsilon_E^A}$, respectively): $\boldsymbol{Y_d}$ = ($\boldsymbol{y_d} \pm \boldsymbol{\epsilon_d^Y}$) or: $\boldsymbol{y_d}$ = ($\boldsymbol{Y_d} \mp \boldsymbol{\epsilon_d^Y}$), and: $\boldsymbol{A_E}$= ($\boldsymbol{a_E} \pm \boldsymbol{\epsilon_E^A}$)or: $\boldsymbol{a_E}$= ($\boldsymbol{A_E} \mp \boldsymbol{\epsilon_E^A}$). That is, for a practical purpose, the Eqs. (1) and (2) should be re-represented as:

$$(y_d \pm \epsilon_d^Y) = f_d(\{x_m \pm u_m\}_{m=1}^{N}) = \frac{(x_d \pm u_d)}{1+\sum_{m=2}^{N}(x_m \pm u_m)}; with d = 1,2 \ldots N \qquad (1a)$$

And:

$$(a_E \pm \epsilon_E^A) = f_E(\{[m_d \pm u_d^M], [y_d \pm \epsilon_d^Y]\}_{d=1}^N)$$

$$= \sum_{d=1}^{N}\left([m_d \pm u_d^M] \times \frac{(x_d \pm u_d)}{1+\sum_{m=2}^{N}(x_m \pm u_m)}\right) \qquad (2a)$$

where $\boldsymbol{u_d^M}$ stands for the *uncertainty* in the *estimate* ($\boldsymbol{m_d}$) of the ***d****th*-*true*-isotope-mass "***Md***". However, "$\boldsymbol{m_d}$" is, in practice, used as its *recommended*-Table-data. Therefore, corresponding to any **given-source** of ***AE***-sample, the different possible measuring-**lab**-dependent-results ($^{Lab1}\boldsymbol{a_E}$, $^{Lab2}\boldsymbol{a_E}$, **…**) could be varied from one another provided the **lab**-specific-ratio-*estimate(s)* **"**$^{Lab1}\boldsymbol{x_m(s)}$, $^{Lab2}\boldsymbol{x_m(s)}$, …**"** are also different from one another. Similarly, the **labs**-estimated-*resultant*-uncertainties ($^{Lab1}\boldsymbol{\epsilon_E^A}$, $^{Lab2}\boldsymbol{\epsilon_E^A}$, **…**) could be varying from one another provided the measuring-**labs**-established-*ratio*-uncertainties ($^{Lab1}\boldsymbol{u_m(s)}$, $^{Lab2}\boldsymbol{u_m(s)}$, **…**) have also got varied from one another.

Moreover, any *desired result* **(**$\boldsymbol{y_d}$, or $\boldsymbol{a_E}$**)** is computed by simply replacing the ratio-*variable(s)* "$\boldsymbol{X_m(s)}$" in the corresponding **SSR (**Eq.(1), or Eq.(2)**)** by ***alone*** the ratio-*estimate(s)* "$\boldsymbol{x_m(s)}$" and, therefore **without** involving any *additional* **(i.e.** *other* than the ratio(s)-measurement**)** *source-of-error* in the overall process of the desired (*indirect*) measurement. Yet, a question is raised here**:** should **alone** the *accurately*-**measured**-ratio(s) "$\boldsymbol{x_m(s)}$" ensure the **desired**-*results* "$\{\boldsymbol{y_d}\}_{d=1}^{N}$, and $\boldsymbol{a_E}$" to also be *so accurate*? Should, even for any possible **bi-**isotopic-(***single***-*ratio*)-*measurement*-system (***d*** = **1** and **2**; $\boldsymbol{m} = [\boldsymbol{N-1}] =$ **1**), the *resultant*-uncertainties be *necessarily* **equal** to the corresponding ***single***-*ratio*-*uncertainty* ($\boldsymbol{\epsilon_1^Y} = \boldsymbol{\epsilon_2^Y} = \boldsymbol{\epsilon_E^A} = \boldsymbol{u_{m=(N-1)=1}}$)? Or, should e.g. "$\boldsymbol{\epsilon_d^Y} < \boldsymbol{u_m}$" or say "$\boldsymbol{\epsilon_E^A} \ll \boldsymbol{u_m}$" (i.e. the *output-estimates* "$\boldsymbol{y_1}$ and $\boldsymbol{y_2}$" or "$\boldsymbol{a_E}$" being ***more*** accurate than their single-*input*-*estimate* "$\boldsymbol{x_{m=1}}$") be impossible? Even, it may be a natural query**:** should the **correlation** (*interdependence*) of the isotopic-abundances "$\{\boldsymbol{Y_d}\}_{d=1}^{N}$" (and hence of their ratios "$\{\boldsymbol{X_m}\}_{m=2}^{N}$") decide in any possible way the *output*-uncertainties ($\{\boldsymbol{\epsilon_d^Y}\}_{d=1}^{N}$ and, $\boldsymbol{\epsilon_E^A}$ *values*)? That is to say, how should we ensure whether or not the *resultant-estimates* "$\{\boldsymbol{y_d}\}_{d=1}^{N}$, and $\boldsymbol{a_E}$", which are obtained at the end of the computational-steps as the "Eq.(1), and Eq.(2)", acceptably *represent* their desired true-values "$\{\boldsymbol{Y_d}\}_{d=1}^{N}$, and $\boldsymbol{A_E}$", respectively? Above all, can the *resultant*-uncertainty(s) "$\boldsymbol{\epsilon(s)}$" be predicted 'a priori' for enabling the *required* possible development of the method-of-ratio-measurement?

Thus, as even indicated by Eq. (1a) or (2a), the correct possible *evaluation* (*prediction*) of the resultant-uncertainty "$\boldsymbol{\epsilon}$" should be an integral part of any *indirect*-measurement. However, it may also be pointed out that if the *ratio-measurement*-uncertainty(s) can somehow be ensured to be ***nil*** (i.e. if**:** $\boldsymbol{u_m(s) = 0}$), then the *resultant*-uncertainty(s) should be ***nil*** ($\boldsymbol{\epsilon(s) = 0}$). Therefore, the basic question to be resolved here is**:** how exactly the **uncertainty**-*translation* "($\{\boldsymbol{u_m}\}_{m=2}^{N}$) → $\boldsymbol{\epsilon_d^Y}$" or "($\{\boldsymbol{u_m}\}_{m=2}^{N}$) → $\boldsymbol{\epsilon_E^A}$" or even "($\{\boldsymbol{u_m}\}_{m=2}^{N}$, $\{\boldsymbol{u_d^M}\}_{d=1}^{N}$) → $\boldsymbol{\epsilon_E^A}$" should take place?

However, it may before proceeding further be spelled that, as because certain *computations* are indispensable, any ***indirect***-*measurement* is also referred to as an ***input-output***-system [1]. That is, any **directly**-*measured* (and therefore, if detected, the measurement-method-specific-*biases*-**corrected**) *estimate* "$\boldsymbol{x_m}$" may also be referred to as the ***m<sup>th</sup>***-**input**-estimate, and its uncertainty "$\boldsymbol{u_m}$" as the ***m<sup>th</sup>***-*input*-uncertainty [1]. Similarly, any *resultant*-(**derived**)-estimate "$\boldsymbol{y_d}$" and its *uncertainty* "$\boldsymbol{\epsilon_d^Y}$" **(**or, an *estimated*-atomic-weight "$\boldsymbol{a_E}$" and its *uncertainty* "$\boldsymbol{\epsilon_E^A}$"**)** may be referred to as a **desired**- and/ or simply **output**-estimate and **output**-uncertainty, respectively [1].

Even, it may be re-noted that the ***cause***-of-concern here is ***two***-*fold***:** **(i)** any *measurement* (**input**-*estimate*) is subject to certain *uncertainty* ($\boldsymbol{u_m}$); and**:** **(ii)** the desired result (**output**-*estimate*) is obtained **only** at the *end* of certain *computational-process*. Over and above, it should be emphasized that ***no***-*user* (off-line/ on-line *software*) can in any possible way *enable* the *computational-step* **(**Eq. (1) and/ or Eq. (2), here**)** to yield the *result(s)* as might be **desired**, viz. as accurate as an **input**-*ratio*.

However, what exactly the term "***uncertainty***" should mean?

## 2. UNCERTAINTY

As indicated by e.g. Eq. **(1)** and Eq. **(1a)**, any input-**estimate** "$x_m$" should be equal to its unknown-**true**-value "$X_m$" provided that (the *input-uncertainty*)**:** $u_m$ = **0**. Otherwise (i.e. for**:** $u_m \neq 0$), the estimate "$x_m$" may turn out to be so much **higher** than its true-value "$X_m$" that**:** $x_m = (X_m + u_m)$ or even so much **lower** that**:** $x_m = (X_m - u_m)$. That is to say, the *uncertainty* " $u_m$ " should stand for the **possible,** and therefore the ***highest*** and ***unaccountable***, **error** in the **(**directly-*measured*, and hence the mass-fractionation and all other detected biases *corrected*, input**) estimate** "$x_m$ (cf. the Section 0.2 in [1])". Thus, as a natural consequence, the *resultant*-(**output**)-*uncertainty* "$\epsilon_d^Y$ (cf. Eq. 1a)" should also stand for the ***highest***-*possible-error* in the *desired-result* "(**output**-*estimate*)**:** $y_d$". Similarly, "$\epsilon_E^A$ (cf. Eq. 2a)" should represent the *highest-possible-**error*** in the *estimate* "$a_E$" (of the true atomic weight "$A_E$").

Clearly, the ratio-measurement-(**input**)-uncertainty "$u_m$" **(**which may itself turn out to be the measurement-*method*-specific and/ or the isotopic-*abundance-ratio*-specific, and which should **account** for the ***possible*** errors due to**: (i)** *detected-biases*-**correction**-*processes*-employed, and even**: (ii)** *small*-**undetectable**-*biases***)** should get ascertained, with the aid of relevant standards, by the experimenting-**lab**. Thus, it is not the purpose here to discuss how the uncertainty(s) "$u_m(s)$" of the required ratio-estimate(s) "$x_m(s)$" should be established, but it is intended to unfold the fact as to ***how*** really the *input*-**to**-*output* uncertainty-*transformation* "$u_m(s) \rightarrow \epsilon(s)$" should be taking place.

### 2.1 Output-uncertainty: conventional concept of evaluation

Generally, any **method**-of-*direct*-measurement, here, that of the isotopic-abundance-ratio ($X_m$), is used to be so developed that the corresponding ***uncertainty*** "$u_m$" **(**of any *measured* and hence of the **input**-estimate "$x_m$"**)** should get restricted by its ***limiting-***

*random*-value, i.e.**:** $\boldsymbol{u_m}$ = $\boldsymbol{\sigma_m}$ (with $\boldsymbol{\sigma_m}$ as the *standard deviation* [1] of the ratio "$\boldsymbol{X_m}$" measurements). Possibly, that is why, the **derived**-*estimates* as "$\boldsymbol{y_d}$" and "$\boldsymbol{a_E}$" (of the true-isotopic-abundance "$\boldsymbol{Y_d}$" and the true-atomic-weight "$\boldsymbol{A_E}$") are usually validated in terms of their *predicted-scatters* **(**cf. e.g. [2-10], and also the references therein**)** "$\boldsymbol{\rho_d}$"and "$\boldsymbol{\rho_E}$", respectively. The *predicted-scatter* is usually referred to as the combined standard [1] or probable [11] uncertainty. The *predicted-scatter*, e.g., "$\boldsymbol{\rho_d}$" of any **output**-estimate "$\boldsymbol{y_d}$" **(**cf. **Eq. (1a))** is computed here as:

$$\rho_d = \frac{1}{y_d}\left[\sum_{m=1}^{N}\left(\frac{\partial Y_d}{\partial X_m}\right)^2 \sigma_m^2\right]^{1/2} \qquad (3)$$

Moreover, as the isotopic abundances ($\{\boldsymbol{Y_d}\}_{\boldsymbol{d=1}}^{\boldsymbol{N}}$) and/ or their ratios "$\{\boldsymbol{X_m}\}_{\boldsymbol{m=2}}^{\boldsymbol{N}}$" are ***correlated***, the combined standard uncertainty "$\boldsymbol{\rho_d}$" is generally computed (cf. e.g. Refs. [8,9] and so) by incorporating suggested [1] *correlation-correction*-factor(s) in Eq. (3).

However, we here enquire**: (i)** can **ever** the *ratio* "$\boldsymbol{X_m}$", rather *any* ***direct***, *measurement*-uncertainty "$\boldsymbol{u_m}$" be ensured to be purely *random* in nature? **(ii)** Should, corresponding to a **given** (set of) **input**-*estimates* "$\{\boldsymbol{x_m}\}_{\boldsymbol{m=2}}^{\boldsymbol{N}}$" **(**and thus, for the generally unknown but **prefixed** *input-errors*/ *-uncertainties***:** $\{\boldsymbol{u_m}\}_{\boldsymbol{m=2}}^{\boldsymbol{N}}$**)**, there be any **room** for an **output**-*estimate* as the isotopic-abundance "$\boldsymbol{y_d}$" and *therefore* the corresponding-**error/** *uncertainty* "$\boldsymbol{\epsilon_d^Y}$" to **get**, *depending* on whether the **output**-*variables* "$\{\boldsymbol{Y_d}\}_{\boldsymbol{d=1}}^{\boldsymbol{N}}$" and/ or their **input**-*ratio-variables* "$\{\boldsymbol{X_m}\}_{\boldsymbol{m=2}}^{\boldsymbol{N}}$" are *correlated*, **varied**? However, these points are returned to the below.

Yet, we should re-note here that, it is rather because the isotopic-abundances ($\{\boldsymbol{Y_d}\}_{\boldsymbol{d=1}}^{\boldsymbol{N}}$) are **correlated**, any "$\boldsymbol{Y_d}$"-*measurement* is (really required to be) carried out **indirectly** from the correspondingly-*estimated*-ratios "$\{\boldsymbol{x_m}\}_{\boldsymbol{m=2}}^{\boldsymbol{N}}$".

Moreover, we should now resolve the basic possible queries as**: (i)** what should be the ***aim*** of any *measurement*? And**: (ii)** should there be a **room** for implying the ***uncertainty****-of-an-estimate* as being **different** from its-***accuracy***?

**2.2 On the uncertainty (*inaccuracy*) and accuracy of an estimate**

For any possible case of direct-measurement, the *minimization* of corresponding-possible-error (specifically, the elimination of, if detected, sources-of-systematic-errors) is well-known to be the motto behind *developing* the required-**method**-of-measurement. Thus, by really **any** required (direct/ indirect) *measurement*, it should mean the ***bracketing*** of **measured**-*estimate* with its (probable) **true**-*value*. Moreover, it may be emphasized, the foresaid statement should ever stand to be the **fact**, *even* though**: (i)** by any "**true**-*value*", it should refer to, as also to be commonly-known, a **relative**-*truth* (i.e. with reference to the "*corresponding-something*" as an ***attributed****-standard*), and/ or**: (ii)** any "*true-value*" is, **otherwise**, *indeterminate*. Thus, the *precision* ( $\boldsymbol{\sigma_m}$ ) of a required set of *bias-**free**-measurements* of any quantity should be the ***measures*** of the corresponding **"**(***least*** *possible*) *measurement*-**uncertainty"** and also that of *its* **"**(***best*** *achievable*) **accuracy"**.

Here, we should also take into account the facts that**: (i)** *a small*-measurement-*bias* might even get **undetected** (i.e. the variation *due to* certain *bias* can really get overshadowed by the *random*-variation); **(ii)** the **process** of correcting a detected-bias might itself stand for an ***indirect****-measurement*, i.e. should cause the *direct-measurement-uncertainty* to be **systematically** altered and thus, the corresponding-(*input*)-*uncertainty* ($\boldsymbol{u_m}$) be essentially *systematic* in nature. That is to say that the *standard-deviation*-of-measurements "$\boldsymbol{\sigma_m}$" may ***not*** always turn out to be *purely-statistical* in nature. Furthermore**: (iii)** neither the (*magnitude* of an) unknown **true**-*error* nor even the **possible**-*error* (*uncertainty*) of any ***derived****-result* and therefore, ***nor*** the derived-*result* **itself**, can get varied **for** whether the **cause** of the (unknown) *true-error* and/ or *possible-error* in a corresponding-***input****-estimate* is purely statistical or not. Thus, **for** rather assuming the *possible-**cause*** of a (real world) measurement-variation to be **either** random ***or*** systematic **or** both in nature, the idea of accordingly referring to the ***possible**-variation* of any corresponding **desired** (input/ output/ reported) *estimate* as **either** the *"uncertainty"* **or** the "*accuracy*" or "***so***" should **not** in practice be useful.

Thus, simply for avoiding the possible confusion in terms of terminologies their-selves, the ***possible*** *variation* of any *directly* or *indirectly* measured (i.e. of any *kind* of reported) *estimate* is referred to [12,13], even here, as either *uncertainty* or *inaccuracy* or *accuracy*.

**2.3 On "$u_m(s)$" to "$\epsilon$" translation and "correlation"**

The **result** of any *indirect*-measurement is obtained only at the **end** of certain *computations* in terms of the exactly-**corresponding**-*input-output*-**SSR/-model**, e.g. the ***d^th^**-estimated*-isotopic-abundance "$\boldsymbol{y_d}$" as **(**cf. Eq.(1)**)**: $\boldsymbol{y_d = f_d(\{x_m\}_{m=2}^{N})}$. Moreover, the *evaluation-step* (Eq.(1)) can, in terms of the **true**-*errors* ($\{\boldsymbol{\Delta_m}\}_{\boldsymbol{m=2}}^{\boldsymbol{N}}$ and $\boldsymbol{\Delta_d^Y}$) of the corresponding ***input***- and ***output***-estimates ($\{\boldsymbol{x_m}\}_{\boldsymbol{m=2}}^{\boldsymbol{N}}$ and $\boldsymbol{y_d}$, respectively), be rewritten as:

$$(Y_d \pm \Delta_d^Y) = f_d(\{X_m \pm \Delta_m\}_{m=2}^{N}) \qquad (1b)$$

And in terms of input- and output-***uncertainties*** ($\{\boldsymbol{u_m}\}_{\boldsymbol{m=2}}^{\boldsymbol{N}}$ and $\boldsymbol{\epsilon_d^Y}$, respectively) as:

$$(Y_d \pm \epsilon_d^Y) = f_d(\{X_m \pm u_m\}_{m=2}^{N}) \qquad (1a')$$

**Eq. (1b)** and/ or **Eq. (1a')** should make it clear that, though the desired-**result** (here, "$\boldsymbol{y_d}$") of *any*-indirect-measurement needs to be computed by simply **replacing** the concerned-*input-**variables*** (here, the abundance-ratios**:** $\{\boldsymbol{X_m}\}_{\boldsymbol{m=2}}^{\boldsymbol{N}}$) by their-***estimates*** **(**$\{\boldsymbol{x_m}\}_{\boldsymbol{m=2}}^{\boldsymbol{N}}$, i.e. as shown by the "Eq. **(1)**" **here** above**)**; the **required**-*translation* of **"measured-**(*input*)*-estimate(s)***"** into **"desired-**(*output*)*-estimate***"** **(**here**:** "$\boldsymbol{x_m(s)}$"→ "$\boldsymbol{y_d}$") and the **hidden**-*transformation* of "corresponding (unknown) true-*input-**error(s)***" into "true-*output-**error***" **(**here**:** "$\boldsymbol{\Delta_m(s)}$" → "$\boldsymbol{\Delta_d^Y}$" and therefore, the ***transformation*** of **"**corresponding-

*input*-***uncertainty(s)*"** into **"*output*-*uncertainty*"**, i.e. here**:** "[$^{\mathbf{MAX.}}\Delta_{m}(s) \rightarrow {}^{\mathbf{MAX.}}\Delta_{d}^{Y}] \equiv [\boldsymbol{u_m(s)} \rightarrow \boldsymbol{\epsilon_d^Y}]$"**)**, are complementary processes and/ or bound to take place (like the well-known oxidation and reduction processes) simultaneously.

2.3.1 *Should the inter-correlation-of-isotopic-abundances affect the uncertainties* ($\{\boldsymbol{\epsilon_d^Y}\}_{\boldsymbol{d=1}}^{\boldsymbol{N}}$) of *determining the abundances* "$\{\boldsymbol{Y_d}\}_{\boldsymbol{d=1}}^{\boldsymbol{N}}$", *and/ or "$\boldsymbol{\epsilon}$" be predictable?*

Eq. (1/ 1a) clarifies that (the **magnitude** of) any *estimated*-isotopic-*abundance* **"output: *$y_d$*"** should depend on ***only*** (the magnitude(s) of) the correspondingly-*measured*-abundance-*ratio(s)* **"input(s): $\boldsymbol{x_m(s)}$"**. In other words, the "***inter-dependency*** of isotopic-abundances ( $\{\boldsymbol{Y_d}\}_{\boldsymbol{d=1}}^{\boldsymbol{N}}$ , and/ or of their-ratios**:** $\{\boldsymbol{X_m}\}_{\boldsymbol{m=2}}^{\boldsymbol{N}}$ )", and even the required-***experimental***- and (if at all involved) *mathematical*-***processes*** of deciding the **input**-*estimate(s)* "$\boldsymbol{x_m(s)}$", ***cannot*** even be caused (by the user) to *affect* the **output** of the *computational-step***:** Eq. (1/ 1a) and/ or to *decide* the (magnitude of the) desired-*result* "***$y_d$***".

That is, once an input-output-**SSR/** model (here, Eq. **(1)** or Eq. **(2)**) and its input-*estimate(s)* should be **fixed**, the corresponding-output-*estimate* and ***therefore*** its (unknown) *true-error* will get automatically-***fixed***. Thus, it has also to be a fact that, if the *input*-uncertainty(s) "$\boldsymbol{u_m(s)}$" should be *preset* (**pre**-established), then the corresponding-*output*-uncertainty "$\boldsymbol{\epsilon}$" will naturally be ***pre****-fixed* **by** the *computational-process* (**SSR**/ model) involved. In other words, ***any*** possible *output*-**uncertainty (**e.g. "$\boldsymbol{\epsilon_d^Y}$" of determining the ***correlated***-variable "***$d^{th}$***-isotopic-abundance: ***$Y_d$***"**)** should, ***like*** the corresponding-*output*-estimate (as "***$y_d$***") itself, be the *concerned*-**SSR (**here, the "***$d^{th}$***-abundance-**specific**-Eq.**(1),** rather the function**: *$f_d$*"**)** *governed*-**systematic**-parameter. *Therefore*, any "$\boldsymbol{\epsilon}$" is likely to be **predictable** a priori.

Moreover, even for a **given** set of input-estimates (as "$\{\boldsymbol{x_m}\}_{\boldsymbol{m=2}}^{\boldsymbol{N}}$"), any two corresponding similar computational-steps/ **SSRs**/ functions **(**viz.**: "$\boldsymbol{Y_d = f_d(\{X_m\}_{m=2}^{N})}$"** and **"$\boldsymbol{Y_d^{/} = f_d^{/}(\{X_m\}_{m=2}^{N})}$") cannot** *generally* yield the output-estimates ("$\boldsymbol{y_d}$" and "$\boldsymbol{y_d^{/}}$") **of** exactly the **same**-*magnitude* or -*true*-(output)-*error*. Thus, it is clearly the **"intrinsic** *inputs*-**to**-*output* variation-***rates*"** [12] **of** the *concerned*-**SSR**/ -model **(**as the "***d^th^***-isotopic-abundance-"$\boldsymbol{Y_d}$"-specific-**Eq. (1)** or -function "$\boldsymbol{f_d}$"**)**, which should ***preset*** whether the *computational-step* **(**$\boldsymbol{x_m(s) \rightarrow y_d}$**)** will **act**, effectively, as **either** "an *extraneous*-***error*-*source***" **or** "a kind of *sink* of net-experimental-(input)-error" **or** "a *non-interfering-medium*" in the ***overall-process*** of the required-indirect-measurement (here, of-"$\boldsymbol{Y_d}$") and hence, in ***defining*** the corresponding (true) output-error ($\Delta_d^Y$) and/ or output-uncertainty ($\boldsymbol{\epsilon_d^Y}$).

Therefore, the **task** of *evaluating/ predicting* an **output**-uncertainty "$\boldsymbol{\epsilon}$" should reduce to the **process** of *identifying* the ***characteristic*** input(s)-**to**-output ***variation***-*parameter(s)* of the concerned **SSR** as Eq. (1) or Eq. (2) here.

**2.4 Evaluation of output-uncertainty**

It has been shown elsewhere [12-14] that, *irrespective* of what exactly an *indirect-measurement-system* and hence the *corresponding*-**SSR (**or, *proposed* measurement- and/ or input-output-**model)** as **Eq.(1)** might represent, the ***uncertainty*** as "$\boldsymbol{\epsilon_d^Y}$" in a corresponding ***derived***-*result* "$\boldsymbol{y_d}$, **(**cf. Eq. (1/ 1a)**)**" should be best (in fact, in the case of any *linear* **SSR**/ model, *exactly*) accountable/ predictable as:

$$\epsilon_d^Y = \sum_{m=1}^{N} (|[MF]_m^d| \times u_m) = [\sum_{m=1}^{N} (|[MF]_m^d| \times F_m)] \times {}^G u = ([UF]_d \times {}^G u) \quad (4)$$

That is, the corresponding (true) error "$\boldsymbol{\Delta_d^Y}$" should be governed as:

$$\Delta_d^Y = \sum_{m=1}^{N} ([MF]_m^d \times \Delta_m) \quad (4a)$$

where **(as** only different relevant-*relative-errors* should be *inter-comparable*, the [generally unknown] **true**-*input*- and -*output-errors* "$\Delta_m$ and $\Delta_d^Y$" [of determining any isotopic-abundance "$\boldsymbol{Y_d}$", cf. Eq. (1)] as well as the *corresponding*-uncertainties "$\boldsymbol{u_m}$ and $\boldsymbol{\epsilon_d^Y}$", respectively, are defined [12], even here, to be ***relative***, i.e.**:**$\boldsymbol{u_m}$= $^{\textbf{Max.}}|\Delta_m|$ = $^{\textbf{Max.}}\left|\frac{\Delta X_m}{X_m}\right|$ = $^{\textbf{Max.}}\left|\frac{x_m - X_m}{X_m}\right|$; and**:**$\boldsymbol{\epsilon_d^Y}$ = $^{\textbf{Max.}}|\Delta_d^Y|$ = $^{\textbf{Max.}}\left|\frac{\Delta Y_d}{Y_d}\right|$ = $^{\textbf{Max.}}\left|\frac{y_d - Y_d}{Y_d}\right|$; and**)** the factor "$[\boldsymbol{MF}]_{\boldsymbol{m}}^{\boldsymbol{d}}$" is a

theoretical constant, representing the **(***exactly*-corresponding-**SSR**/ model, i.e. here the function "$f_d$ [cf. Eq. (1)]", **specific)** *relative-**rate**-of-variation* of "(the **output**-variable, here, ***d**$^{th}$-desired*-isotopic-abundance) $Y_d$" *as* a function of the corresponding "*measurable*-cum-**input**-variable, here, the ***m**$^{th}$*-abundance-ratio) $X_m$":

$$[MF]_m^d = \left(\frac{\partial Y_d}{\partial X_m}\right)\left(\frac{X_m}{Y_d}\right) = \left(\frac{\partial Y_d/Y_d}{\partial X_m/X_m}\right), \quad (m = 1, 2 \ldots N) \qquad (5)$$

(It should of course be reminded that, *for* our ***isotopic*-SSR** as the Eq. (1) or (2): $m = 2, 3 \ldots N$). However, the "***m**$^{th}$***-input**-specific-*rate-of-variation*: $[MF]_m^d$" **(**which, as Eq.(5) clarifies, numerically equals the **relative-error-magnification-factor** [15,16] of *determining* "$Y_d$" with respect to its ***m**$^{th}$-measurable*-variable "$X_m$"**)** may also be referred to as the **(**corresponding-**SSR**/ model, here the function "$f_d$ (cf. Eq.(1)", *governed***)** "***m**$^{th}$**-individual**-error-magnification-factor*".

The factor "$F_m$" is defined as "$F_m = (u_m/{}^{G}u)$"; so that:

$$[UF]_d = (\epsilon_d^Y/{}^{G}u) = \sum_{m=1}^{N} (|[MF]_m^d| \times F_m) \qquad (6)$$

with "${}^{G}u$" as ***any*** *given-value* of the measurement/ input **uncertainty:** $u_m$ (viz. the measurement-accuracy which could be *preset to be achieved* **before** either the required-*development* of experimental-methodology or *establishing,* if it really happens to be so, $X_m$-specific "$u_m$"). And the factor "$[UF]_d$", which should give (cf. Eq.(6)) a *measure* of the concerned-**SSR**/ -model (here, the Eq.(1)) **specific** *relative-**rate**-of variation of* the **desired**-variable "$Y_d$" as a function *of* **all** the corresponding-**measurable**-variables "$\{X_m\}_{m=1}^{N}$" ***collectively****,* is referred to here as either "***collective**-error-magnification-factor*" or "***uncertainty-factor*** [13]".

However, if ***all*** different $X_m$-measurements (i.e. *input-estimates* "$\{x_m\}_{m=1}^{N}$") should happen to be subject to ***equal*** uncertainty (i.e. if**:** $u_m = {}^{G}u$, and/ or**:** $F_m = 1$, with**:** $m = 1, 2 \ldots N$), then the **Eq.(6)** will clearly be reduced as:

$$[UF]_d = \sum_{m=1}^{N} (|[MF]_m^d| \times F_m) = \sum_{m=1}^{N} |[MF]_m^d| \qquad (6a)$$

Thus (cf. **Eq.(6a)**), it should be better clear**:** **(i)** why the *uncertainty-factor* "$[\boldsymbol{UF}]_{\boldsymbol{d}}$" should stand ***for*** **(**or why "$[\boldsymbol{UF}]_{\boldsymbol{d}}$ as Eq.**(6a)**" should, in any **isotopic** case *of* "$\boldsymbol{N} > \mathbf{2}$" and "**(**$\boldsymbol{F_m}$ = [$\boldsymbol{u_m}$/$^{G}\boldsymbol{u}$]**)** ≠ **1**", give at least a **tentative** *measure* of**)** the *concerned*-**SSR**/ model (as the Eq. (1), and hence the function "$\boldsymbol{f_d}$") *specific* **rate**-of-variation of the **desired**-variable ($\boldsymbol{Y_d}$) as a *function* of **all** the corresponding-**measurable**-variables ($\{\boldsymbol{X_m}\}_{\boldsymbol{m=1}}^{\boldsymbol{N}}$) ***collectively***; i.e.**:** **(ii)** why the **evaluation** of *any*-concerned-**SSR**-(model)-**specific**-*theoretical-parameter* as the "**[*UF*]**" *alone* should make the corresponding-**output**-uncertainty "$\boldsymbol{\epsilon}$" to be ***predictable*** (cf. Eq. (4)) as**:** $\boldsymbol{\epsilon}$ **= ([*UF*]** × $^{G}\boldsymbol{u}$**)**.

Clearly, the *smaller* should be the *uncertainty-factor* (viz.**:** $[\boldsymbol{UF}]_{\boldsymbol{d}} <$ **1**, rather**:** $[\boldsymbol{UF}]_{\boldsymbol{d}} \ll$ **1**), the *better* accurate (*lesser* uncertain), than the *measured estimates* ***as a whole***, the desired-**result** ($\boldsymbol{y_d}$) should turn-out to be. That is, if a function as the "$\boldsymbol{f_d}$ (cf. Eq. (1)" represents a *proposed* evaluation-**model**, then the model as the "$\boldsymbol{f_d}$" itself should be a *better* one than the some other possible one of relatively higher "$[\boldsymbol{UF}]_{\boldsymbol{d}}$".

Thus, essentially, the query raised here is**:** should the *uncertainty-factor* (**[*UF*]**), corresponding to the isotopic **SSR** as either **Eq.(1)** or **Eq.(2)** or both, turn out to be **<1**? However, for evaluating this point, a note on the possible classification of **SSRs**/ models is believed here to be useful.

### 2.4.1 *Family characteristics of SSRs/ Models*

It has already been clarified elsewhere [12,14] that any conceivable **SSR**/ model/ function as "$\boldsymbol{f_d}$, (cf. **Eq.(1)**)" should, depending upon how its *input(s)*-**to**-*output* **variation-***characteristic(s)* being decided, belong to *either* of the **two** possible, namely the ***measurable-****variable-independent* (**F.1**) and-*dependent* (**F.2**), **families** only: ––***an*** established/ proposed *input-output-relationship* and therefore ***a*** *function* as **"**$\boldsymbol{f_d}$, ***with*****:** $\{|[\boldsymbol{MF}]_{\boldsymbol{m}}^{\boldsymbol{d}}| = \mathbf{1}\}_{\boldsymbol{m=1}}^{\boldsymbol{N}}$ (i.e. with **all** the "***N***" different (*measurable*-variables-dependent) *relative-**rates**-of-variations* "$[\boldsymbol{MF}]_{\boldsymbol{m}}^{\boldsymbol{d}}(\boldsymbol{s})$" (of the desired-variable "$\boldsymbol{Y_d}$") being equalling "**1**" in *magnitude*)**";** and ***an*** "$\boldsymbol{f_d}$ ***with*** (at least, any *single* "$\boldsymbol{m}$" dependent-rate-of-variation"$[\boldsymbol{MF}]_{\boldsymbol{m}}^{\boldsymbol{d}}$" not-equalling "**1**" in *magnitude*)**:** $\left|[\boldsymbol{MF}]_{\boldsymbol{m}}^{\boldsymbol{d}}\right| \neq$ **1"; ** be the members of the "**F.1**" and "**F.2**" families, respectively.

Clearly, for any**F.1** family member, **Eq. (4)** should reduce as:

$$\epsilon_d^Y = \sum_{m=1}^{N} (|[MF]_m^d| \times u_m) = \sum_{m=1}^{N} u_m = ([\sum_{m=1}^{N} F_m] \times {}^G u) = ([UF]_d \times {}^G u) \quad (4.F.1)$$

And, thus:

$$[UF]_d = (\epsilon_d / {}^G u) = \sum_{m=1}^{N} F_m = \sum_{m=1}^{N} (u_m / {}^G u) \quad (6.F.1)$$

Further, if the *input*-uncertainty "$\boldsymbol{u_m}$" should be **independent** of the *measurable* (**input**) variable "$\boldsymbol{X_m}$" (viz. $\boldsymbol{u_m} = {}^G\boldsymbol{u}$, i.e. if**:** $\boldsymbol{F_m} = \mathbf{1}$, with**:** $\boldsymbol{m} = \mathbf{1, 2} \ldots \boldsymbol{N}$), then **(**cf. Eq. (4.F.1)**)**:

$$\epsilon_d^Y = \sum_{m=1}^{N} u_m = (N \times u_m) = (N \times {}^G u) = ([UF]_d \times {}^G u) \quad (4.F.1a)$$

And **(**cf. Eq. (6.F.1)**)**:

$$[UF]_d = (\epsilon_d / {}^G u) = \sum_{m=1}^{N} (u_m / {}^G u) = N \quad (6a.F.1a)$$

Now let, however *for simply illustrating* the family-implications, the isotopic-**SSR** as "**Eq. (1)**" to be a member of the "**F.1**" family. Then the uncertainty ($\boldsymbol{\epsilon_d^Y}$) of determining an atom-fraction ($\boldsymbol{Y_d}$) should be governed **(**cf. Eq. (**4.F.1**), by **solely** the "$(\boldsymbol{N} - \mathbf{1})$" ratio-measurement-uncertainties, i.e.**)** as**:** $\boldsymbol{\epsilon_d^Y} = \sum_{m=2}^{N} \boldsymbol{u_m}$. That is, **then**, the *desired*-result ($y_d$) should**:** **(i)** at best, however **only** in the case of any ***bi-isotopic*** (**$N$ = 2**) element, turn out as accurate as the corresponding **single**-possible-*measured*-ratio ($\boldsymbol{x_m}$, as**:** $\boldsymbol{\epsilon_d^Y} = \sum_{m=2}^{N} \boldsymbol{u_m} = \boldsymbol{u_m}$)"; but can ***never*** be better accurate than the estimated-ratio "$\boldsymbol{x_{m=1}}$"; and**:(ii)** for any case of "$\boldsymbol{N} \geq \mathbf{3}$", be subject to the uncertainty as high as "$(\boldsymbol{N} - \mathbf{1})\boldsymbol{u_m}$".

If, however, **Eq. (1)** happens to be belonging to the "**F.2**" family, then the ***rate(s)*** of "$\boldsymbol{Y_d}$ **vs.** *individual*-$\boldsymbol{X_m}$" ***variation(s)***, i.e. "$[\boldsymbol{MF}]_{\boldsymbol{m}}^{\boldsymbol{d}}$***(s)***, cf. Eq. (5)", should be **dependent** on the *measurable-*$\boldsymbol{X_m}$*-variable(s)* and/ or on the *nature* of the "function**:** $\boldsymbol{f_d}$". That is, **then**, there should **on the one** be the possibility that "$[\boldsymbol{MF}]_{\boldsymbol{m}}^{\boldsymbol{d}}$***(s)*** $<$ **1**" or even "$[\boldsymbol{MF}]_{\boldsymbol{m}}^{\boldsymbol{d}}$***(s)*** $\ll$ **1**", thereby enabling to achieve**:** $[\boldsymbol{UF}]_{\boldsymbol{d}} < \mathbf{1}$, and hence "**(**output-uncertainty**:** $\boldsymbol{\epsilon_d^Y}$) $<$ (input-uncertainty**:** $\boldsymbol{u_m}$; cf. Eq.(4))". Moreover, if the *input-output-relationship* (the function "$\boldsymbol{f_d}$") should stand for a proposed-**model**, then a ***careful***-**F.2**-modelling of the function "$f_d$" may also offer the opportunity of achieving "$[MF]_m^d(s) \ll$ **1**" and/ or "$\epsilon_d^Y < u_m$". It may o**n the other** turn out to be that Eq. (1) is characterized by "$[MF]_m^d(s) > 1$, or so", thereby causing "$[UF]_d >$ **1**" (or even "$[UF]_d \gg$ **1**"), i.e.: "$\epsilon_d^Y > u_m$".

Thus e.g. while the **SSR** "$Y_R = f_R(X_S, X_W) = \frac{X_S}{X_W}$" should belong to the **F.1**-family [12], the apparently similar **SSR** "$Y_\delta = f_\delta(X_S, X_W) = \left(\frac{X_S}{X_W} - 1\right)$" is a member of the **F.2**-family [14]. And, one may verify [12,14,17,18] that the uncertainty ($\epsilon_R^Y$) of determining the ***simple-ratio*** "$Y_R$" should solely be governed as: $\epsilon_R^Y = [u_S + u_W]$; but the uncertainty ($\epsilon_\delta^Y$) in a corresponding-***differential-estimate*** ($y_\delta$) is decided as: $\epsilon_\delta^Y = \left(\left|\frac{X_S}{X_S - X_W}\right| \times [u_S + u_W]\right) = \left(\left|\frac{X_S}{X_S - X_W}\right| \times \epsilon_R^Y\right)$. Clearly, for "$|X_S - X_W| \rightarrow 0$", the *measurement*-accuracy "*u*" might improve, but the *differential*-estimate ($y_\delta$) **itself** should be increasingly ***erroneous***. These should explain why in any possible case of the isotope-ratio-mass-spectrometry (**IRMS)**, where "$|X_S - X_W| \rightarrow 0$" is rather a requirement, the **modelling** of "**IRMS**-*measurement* and/ or -*evaluation* as "$f_\delta$" rather than as "$f_R$" is illogical [14,18], and/ or why there should be confusion [7] in dealing with the **IRMS** data. Truly, such a fact has led us to study here the **intrinsic** *input(s)*-**to**-*output* **variation**-*behaviour* of the isotopic-**SSRs** "Eqs. (1) and (2)".

However, we are yet to examine whether the isotopic-**SSRs** (Eqs. (1) and (2)) should be belonging to the **F.1** or the **F.2** family; i.e. whether the uncertainties "$\{\epsilon_d^Y\}_{d=1}^N$ and $\epsilon_E^A$" of determining the *isotopic abundances* and *atomic weight* "$\{Y_d\}_{d=1}^N$ and $A_E$, respectively, of any possible element (***E***)" should be decided **(***like* the ***simple-ratio***-uncertainty "$\epsilon_R^Y$" here above**)** by the isotopic-ratio "$X_m$" specific *measurement*- and thus the -*input-uncertainties* "$\{u_m\}_{m=2}^N$" **only;** or (*like* the ***differential-ratio***-uncertainty "$\epsilon_\delta^Y$") by the "$X_m$-*variable(s)*" and the corresponding-measurement-uncertainty(s) "$u_m(s)$", ***collectively***.

**3. RESULTS AND DISCUSSION: prediction and verification of "$\epsilon(s)$"**

The **SSRs** "Eq.(1) and Eq.(2)" are isotopic-*multiplicity*-(***N***)-**specific**. Thus let us first consider the **simplest** possible isotopic-*measurement*-system "***bi-isotopic*** element (***N*** = **2**)" for our investigation.

**3.1 *Bi-isotopic cases***

Let's suppose that, for developing the required method of determining the isotopic composition and the atomic weight of **e.g.** natural chlorine **(**and hence for establishing the isotopic-ratio-*measurement*-(**input**)-uncertainty "$\boldsymbol{u_m} \equiv \boldsymbol{u_{37/35}}$"**)**, we have chosen a chlorine material of the known "$^{37}$**Cl/**$^{35}$**Cl**"-abundance-ratio**:** ($\boldsymbol{X_m} \equiv \boldsymbol{X_{37/35}}$) = **0.3199578**, and interested to know ***beforehand*** whether an *acceptable* "$\boldsymbol{u_m}$" should ensure the ***output***-uncertainties (i.e. "$\{\boldsymbol{\epsilon_d^Y}\}_{\boldsymbol{d=1}}^{\boldsymbol{2}}$ and $\boldsymbol{\epsilon_{Cl}^A}$" of determining the chlorine-isotopic-abundances "$\{\boldsymbol{Y_d}\}_{\boldsymbol{d=1}}^{\boldsymbol{2}}$" and the -atomic-weight "$\boldsymbol{A}_{\mathbf{Cl}}$", respectively) to also be *so restricted* **(viz.** whether "$[\boldsymbol{\epsilon_{(d=1)}^Y} \equiv \epsilon_1^Y \equiv \boldsymbol{\epsilon_{35}^Y}] = [\epsilon_2^Y \equiv \boldsymbol{\epsilon_{37}^Y}] = \boldsymbol{\epsilon_{\mathrm{Cl}}^A} = \boldsymbol{u_m}$" be *true*?**)**.

**3.1.1 Prediction of output-uncertainty(s)**

One can easily verify that the abundance-ratio "($\boldsymbol{X_m} \equiv \boldsymbol{X_{37/35}}$) = **0.3199578**" should imply that**: (i)** the ***true*** $^{35}$Cl and $^{37}$Cl *abundances* ($\{\boldsymbol{Y_d}\}_{\boldsymbol{d=1}}^{\boldsymbol{2}}$, cf. **Eq. (1)**) to be "**(**$\boldsymbol{Y_{(d\,=\,1)}} \equiv \boldsymbol{Y_{35}}$**)** = $\left(\frac{X_d}{\sum_{m=1}^{2} X_m} = \frac{X_1}{X_1 + X_2} = \frac{\mathbf{1}}{\mathbf{1} + \boldsymbol{X_m}}\right)$ = **0.7576**" and "**(**$\boldsymbol{Y_{(d=\,2)}} \equiv \boldsymbol{Y_{37}}$**)** = $\left(\frac{X_2}{X_1 + X_2} = \frac{X_m}{\mathbf{1}+\boldsymbol{X_m}}\right)$ = **0.2424**", respectively; and**: (ii)** the corresponding ***true*** *atomic-weight* "**(**$\boldsymbol{A}_{\mathbf{Cl}}$ **[**cf. **Eq.(2)]**, however for considering the **true**-isotopic-masses as**:** **[**$\boldsymbol{M_1} \equiv \boldsymbol{M_{35}}$**]** = **34.968852682**, and**:** **[**$\boldsymbol{M_2} \equiv \boldsymbol{M_{37}}$**]** = **36.965902602)**" to be**:** $\boldsymbol{A}_{\mathbf{Cl}} = \sum_{d=1}^{N}(\boldsymbol{M_d} \times \boldsymbol{Y_d}) = \sum_{d=1}^{2}\left(\boldsymbol{M_d} \times \frac{X_d}{\mathbf{1}+\boldsymbol{X_m}}\right)$ = **35.45293763**.

Moreover, even though we are dealing with an **input-output**-system involving a **single**-*input-variable* (and/ or the ***measurement*** of single possible isotopic-ratio) "$\boldsymbol{X_m} \equiv \boldsymbol{X_{37/35}}$", it can also be crosscheck that the ***output*** **(**governing-parameters as the *relative-**rates**-of-variations* "**[*MF*]*(s)***, cf. Eq.**(5)**" of the different *desired*-(**output**)-variables "$\boldsymbol{Y_{35}}$, $\boldsymbol{Y_{37}}$, and $\boldsymbol{A_{\mathrm{Cl}}}$" as a function of their **common** *input*-variable "$\boldsymbol{X_{37/35}}$", and/ or the output-variables-specific *uncertainty-**factors*** "**[*UF*]*(s)***, cf. Eq.**(6)**"; and thus [cf. Eq.**(4)**], the**)** ***uncertainties*** "$\boldsymbol{\epsilon_{35}^{Y}}$, $\boldsymbol{\epsilon_{37}^{Y}}$, and $\boldsymbol{\epsilon_{\mathrm{Cl}}^{A}}$ (namely, for determining the $^{35}$**Cl** and $^{37}$**Cl natural-***abundances* and the corresponding *atomic-weight* of chlorine, respectively,) should however turn out to be ***different*** from one another and, most importantly, from the "***input**-ratio-uncertainty***: $\boldsymbol{u_m} \equiv \boldsymbol{u_{37/35}} \equiv {}^{G}\boldsymbol{u}$" (i.e. as predicted in **Appendix 2:** $\boldsymbol{\epsilon_{35}^{Y}} \neq \boldsymbol{\epsilon_{37}^{Y}} \neq \boldsymbol{\epsilon_{\mathrm{Cl}}^{A}} \neq \boldsymbol{u_{37/35}}$). However, the interesting prediction (cf. **Appendix 2**) is that either "$\boldsymbol{\epsilon_{35}^{Y}}$" or "$\boldsymbol{\epsilon_{37}^{Y}}$" or "$\boldsymbol{\epsilon_{\mathrm{Cl}}^{A}}$" is "$<\boldsymbol{u_{37/35}}$", i.e. any desired-***output**-estimate* ($\boldsymbol{y_d}$ or, $\boldsymbol{a_{\mathrm{Cl}}}$) should even turn out ***less*** *erroneous* than the ***input***-(ratio)-*estimate* "$\boldsymbol{x_m} \equiv \boldsymbol{x_{37/35}}$" itself.

Actually, ***how*** might the "*input*-**to**-*output* **variation** *behaviour*" of any given *output*-variable-specific-**SSR** (-model) turn out to ***differ*** from **even** another similar one **(**e.g. the "$^{35}$**Cl** *abundance*-($\boldsymbol{Y_{35}}$)-specific-**Eq.(1)**" from **t**he "$\boldsymbol{Y_{37}}$-specific-**Eq.(1)**"**)** is, it may be noted, elaborated in Appendix **2**. In other words, *how* really any **output-**variable-specific-*uncertainty-**factor*** (**[*UF*]**, cf. Eq.**(6)**) and therefore, ***-uncertainty*** "$\boldsymbol{\epsilon}$, (cf. Eq.**(4)**)", should be *governed* by the **exactly** corresponding input-output-**SSR** is clarified in Appendix **2**.

Moreover, the predicted facts (cf. the Appendix **2**, and also Eq.**(4):** $\boldsymbol{\epsilon} =$ **(**$[\boldsymbol{UF}] \times \boldsymbol{u_m}$**)**) are further highlighted in **Fig. 1**, which depicts the different desired-*output*-variables ($\boldsymbol{Y_{35}}$, $\boldsymbol{Y_{37}}$ and, $\boldsymbol{A_{\mathrm{Cl}}}$) specific *plots* of **"output-**uncertainty ($\boldsymbol{\epsilon}$) **vs. input-**uncertainty ($\boldsymbol{u_m} \equiv \boldsymbol{u_{37/35}}$)**"** and/ or clarifies that **any** *output*-variable-**(SSR)-**specific "$\boldsymbol{\epsilon}$ vs. $\boldsymbol{u_m}$" *variation* stands for a "***line*** of *a* ***given*** *slope* ($[\boldsymbol{UF}]$)". That is, it may be noted that the ***slopes*** e.g. of even "$\boldsymbol{\epsilon_{35}^{Y}}$ vs. $\boldsymbol{u_m}$" and "$\boldsymbol{\epsilon_{37}^{Y}}$ vs. $\boldsymbol{u_m}$" variation-lines are distinctly ***different*** and, represent the ***predicted*** "$[\boldsymbol{UF}]_{35}$" and "$[\boldsymbol{UF}]_{37}$" *values*, respectively, shown in Appendix **2**. Above all, the rather *interesting feature* ($\boldsymbol{\epsilon} < \boldsymbol{u_m}$, cf. **Appendix 2**, i.e.) that any **variable**-*specific*-**SSR** as **"**Eq.**(1)"** or **"**Eq.**(2)"** should, in the overall process of the corresponding *indirect*-measurement, act as the (concerned-***variable**-specific*) ***sink*** of experimental-***error*** is better revealed in **Fig. 1**, which reflects the ***slopes*** of all the *three* variables-*specific* "$\boldsymbol{\epsilon}$ vs. $\boldsymbol{u_m}$" *variation-lines* as **"<1".**

It may in addition be noted (cf. **Appendix 2**) that the *uncertainty-factor* **"**$[\boldsymbol{UF}]_{\boldsymbol{d}}$**"**, and thus the *uncertainty* **"**$\boldsymbol{\epsilon_d^Y}$**"**, of determining the isotopic-abundance "$\boldsymbol{Y_d}$" are predicted to be ***inversely*** *proportional* to (the magnitude of) "$\boldsymbol{Y_d}$". And the **"**$[\boldsymbol{UF}]_{\boldsymbol{d}}$**"** should, for any ***bi-isotopic*** ($\boldsymbol{d}$ **= 1** and **2**) case as here, take the value as **"**$[\boldsymbol{UF}]_{\mathbf{1}} = \boldsymbol{Y_2}$**"** and/ or **"**$[\boldsymbol{UF}]_{\mathbf{2}} = \boldsymbol{Y_1}$**"**. Furthermore, because we are here dealing with the **single**-ratio "$\boldsymbol{X_m}$" *measurement* **(bi-**isotopic**)** system, the **SSR** (i.e. *output*-variable) **specific** "*input*-**to**-*output*-variation-***rate*** (**[*MF*]**)" and "*uncertainty*-***factor*** (**[*UF*]**)" are, by *magnitude,* identical (e.g.**:** $[\boldsymbol{UF}]_{\mathbf{35}}$ = $\left|[\boldsymbol{MF}]_{\boldsymbol{m}}^{\mathbf{35}}\right|$ = $\mathbf{0.2424}$, cf. Appendix 2).

Further, *any* **output-**variable (**SSR**) **specific** "*input*-**to**-*output*-variation-***rate*: [*MF*]**" and, in turn, the corresponding "*uncertainty*-***factor***: **[*UF*]**" and/ or "***uncertainty***: $\boldsymbol{\epsilon}$" is, it should be significant to note (cf. **Appendix 2**), shown to be ***decided*** by the *measurement*-(**input**)-*variable* "$\boldsymbol{X_m}$" itself (e.g.**:** $[\boldsymbol{MF}]_{\boldsymbol{m}}^{\mathbf{35}} = \boldsymbol{f_{35}(X_m)}$). That is, **firstly**, the *output*-uncertainty ($\boldsymbol{\epsilon}$) should be ***varying*** as a function of even (the *magnitude* of the isotopic-ratio) "$\boldsymbol{X_m}$". Thus, let's consider (in the place of the ***natural*** chlorine "($\boldsymbol{X_m} \equiv \boldsymbol{X_{37/35}}$) = **0.3199578**") a ***non-natural*** chlorine (**Cl**$^{/}$), e.g.:($\boldsymbol{X_m^{/}} \equiv \boldsymbol{X_{37/35}^{/}}$) = **0.6**. Then one can, like the natural case (Appendix 2), verify that the *uncertainties* ($\boldsymbol{\epsilon^{/}(s)}$) of determining the similar (kind of) *non-natural* chlorine isotopic abundances ( $\{\boldsymbol{Y_d^{/}}\}_{\boldsymbol{d=1}}^{\mathbf{2}}$) and the atomic weight ($A_{Cl}^{/}$) should be (*varied* from the natural case) *as***:** $\boldsymbol{\epsilon_{35}^{Y/}} = (\mathbf{0.375}\times\boldsymbol{u_m})$, $\boldsymbol{\epsilon_{37}^{Y/}} = (\mathbf{0.625}\times\boldsymbol{u_m})$, and**:** $\boldsymbol{\epsilon_{Cl}^{A/}} = (\mathbf{0.0131044}\times\boldsymbol{u_m})$. **— Secondly**, both of the kinds of isotopic-**SSRs** "**Eq.(1)** and **Eq.(2)**" should, *like* the above pointed out **"IRMS**-measurement-**model:** $\boldsymbol{f_\delta}$**"**, belong to the "**F.2**" *family* [12,14] of **SSRs-**cum-mathematical-**models**. However, what should be worth mentioning here is that the present-isotopic-**SSR** as either "function**:** $\boldsymbol{f_d}$, (cf. Eq. (1)" or "function: $\boldsymbol{f_E}$, (cf. Eq. (2)" and the "**IRMS-model:** $\boldsymbol{f_\delta}$", even though belong to ***one and the same*** *family* "**F.2**", are ***opposite*** by their intrinsic *input*-**to**-*output* variation features, namely, while any present-*isotopic*-**SSR** "function: $\boldsymbol{f_d}$, or $\boldsymbol{f_E}$" should act as an intrinsic ***sink*** of experimental-*error*, the IRMS-**SSR** "function: $\boldsymbol{f_\delta}$" has been shown [12,14,17,18] to act as an extraneous *error*-***source*** (in the respective cases of indirect measurements). These clarify why an '***a priori***' study of the *input*-**to**-*output* variation behaviour of the **SSR**/ *model*, representing a required *indirect*-measurement, should help to properly pre-plan and/ or to fulfil the desired goal of the ***proposed*** measurement.

Moreover, it should be important noting that **neither** of the *output-uncertainty* ($\boldsymbol{\epsilon}$) governing-parameters **(**cf. **Eqs**. (4-6)**) nor** their implications (cf. **Appendix 2**, and **Fig. 1**) did require to take into account the fact that the *isotopic abundances* "$(\boldsymbol{Y_1} \equiv \boldsymbol{Y_{35}})$ and $(\boldsymbol{Y_2} \equiv \boldsymbol{Y_{37}})$" are ***interdependent***. That is, the ***correlation*** of *isotopic-abundances* "$\{\boldsymbol{Y_d}\}_{\boldsymbol{d=1}}^{\boldsymbol{N}}$", which is signified above to be the cause for carrying out the *measurements* of isotopic-abundances *indirectly* from their *ratio(s)* "$\{\boldsymbol{X_m}\}_{\boldsymbol{m=2}}^{\boldsymbol{N}}$", should in no way be governing the corresponding *indirect*-measurement-***uncertainties*** " $\{\boldsymbol{\epsilon_d^Y}\}_{\boldsymbol{d=1}}^{\boldsymbol{N}}$" (and even "$\boldsymbol{\epsilon_E^A}$").

**3.1.1.1** *Multi-measurement-feature of bi-isotopic-systems*

So far we have only discussed how any ***measurement***-*variation* (i.e. input-ratio-uncertainty: $\boldsymbol{u_m}$) should get transformed as the ***resultant***-variation (*output-uncertainty*: $\boldsymbol{\epsilon}$). However the isotopic masses **(**$\{\boldsymbol{M_d}\}_{\boldsymbol{d=1}}^{\boldsymbol{N}}$), although used in practice as certain recommended Table-data (i.e. remain *invariant* between labs/ measurements**)**, are also subject to certain uncertainties: ($\{\boldsymbol{M_d} = [\boldsymbol{m_d} \pm \boldsymbol{u_d^M}]\}_{\boldsymbol{d=1}}^{\boldsymbol{N}}$). That is, though the *possible-mass-error* "$\boldsymbol{u_d^M}$" should ***never*** be the *cause* for *variation* between different possible **lab**-*estimates* "$(\boldsymbol{a_{Cl}} \pm \boldsymbol{\epsilon_{Cl}^A})^{\mathbf{Lab1}}$, $(\boldsymbol{a_{Cl}} \pm \boldsymbol{\epsilon_{Cl}^A})^{\mathbf{Lab2}}$, **...**" of any desired-**atomic-weight** (here: $\boldsymbol{A_{Cl}}$), it could be of interest to find out how any possible *isotopic-mass-variations* might affect the desired result "$\boldsymbol{a_{Cl}}$"

and/ or how much the *possible-mass-dependent*-errors " $\{u_d^M\}_{d=1}^N$ " should *cause* the ***resultant***-error "$\epsilon_{Cl}^A$" to be varied.

Nevertheless, like the **single**-*ratio*-$X_m$-*measurement* (-*input*) based input-**to**-output-***uncertainty***-*translation*: $u_m \rightarrow \epsilon_{Cl}^A$ (cf. **Appendix 2** and/ or the text, relating the **SSR** "**Eq.2**", above), one can verify that the corresponding **multiple**-inputs (*isotopic*-ratio and masses) based **uncertainty**-transformation "$(u_m,\ u_{35}^M,\ u_{37}^M) \rightarrow \epsilon_{Cl}^A$" is dictated by the *isotopic* (ratio "$X_m$" specific [cf. **Appendix 2**] as well as the) ***mass***-dependent *inputs*-**to**-*output* **variation**-*parameters* **(**of the concerned-atomic-weigh, $A_{Cl}$, defining **SSR** "**Eq.2**") shown in the **Appendix 3**.

In fact, it should be worth noting that the consideration as the **Appendix 3** is essentially a highlight of the *characteristics of evaluation* involving **more** than one ***measurable***-*input*. For example, it should be clear *why*: **(i)** the **rate** "**[*MF*]**, cf. Eq.(4) and/ or Eq.(**5**)" of any ***specified***-*input*-**to**-output-**variation** is also referred to as the ***individual*** (i.e. input-variable-specific) *error-multiplication-factor,* and**: (ii)** the "**[*UF*]**, cf. Eq.(4)/ (6)" as either the desired-***output***-*variable-* or the corresponding-**SSR**-***specific*** *error-magnification-factor,* or the ***collective***-*rate-of-variation* (of the desired-**output**-variable with respect to **all** the corresponding-***individual***-*input*-variables).

It may also be noted (cf. **Appendix 3**) that the ***d^th^***-isotopic-*mass* (input-variable**:** $M_d$, cf. Eq.(2)) dependent *rate-of-variation* ($[MF]_{M_d}^{A_{Cl}}$) of the desired-*atomic-weight* (output-variable**:** $A_{Cl}$) represents the ***d^th^***-isotopic-***mass-fraction***. And which is why, the **sum** of the individual-*mass*-($M_d$)-dependent ***rates***-*of-variations* (error-***multiplication***-*factors*) equals to unity **(**$\sum_{d=1}^{2}\left|[MF]_{M_d}^{A_{Cl}}\right|$ **= 1.0**; cf. the consideration for the *uncertainty-factor* "$[UF]_{A_{Cl}}$" in Appendix **3)**. However the important point here is that any "$[MF]_{M_d}^{A_{Cl}}$" is a ***fraction***, and therefore the **SSR** "Eq.(2)" should, **even** with reference to any possible isotopic-mass-dependent error ($\boldsymbol{u_d^M}$), act as the (desired-output, here**:** $\boldsymbol{A_{Cl}}$, specific) "***sink***-*of*-(*input*)-*error*" in the case of the corresponding-*indirect-measurement*. That is to say, the ***intrinsic*** *nature* of any **SSR** as "Eq.(2)" is such that the *output-error* (uncertainty***:*** $\boldsymbol{\epsilon_{Cl}^A}$) should turn out to be, it may be emphasized, ***less*** than the *mod*-***sum***-of-*input*-ratio-and-mass-*errors* (**sum**-of-input-uncertainties).

Yet (cf. **Appendix 3**), it should be pointed out that **(**though "$\boldsymbol{M_d}$" cannot be a *lab-measurable* variable as "$\boldsymbol{X_m}$", the *input*-to-*output* variation-*rate*)**:** $\left|[\boldsymbol{MF}]_{\boldsymbol{M_d}}^{\boldsymbol{A_{Cl}}}\right| \gg \left|[\boldsymbol{MF}]_{\boldsymbol{X_m}}^{\boldsymbol{A_{Cl}}}\right|$, i.e. the atomic-weight (here**:** $\boldsymbol{A_{Cl}}$) to be determined is predicted to be **more** vulnerable towards *a **given** variation* ($^{G}\boldsymbol{u}$) in any isotopic-mass "$\boldsymbol{M_d}$" than ***that*** in the isotopic-ratio "$\boldsymbol{X_m}$". This should in turn explain (cf. Appendix **3** for the *uncertainty-factor* "$[\boldsymbol{UF}]_{A_{Cl}}$") why the "*mass* ($\boldsymbol{M_d}$) **specific**-*uncertainty-factor/* -uncertainty **(**$^{(Mass)}\boldsymbol{\epsilon_{Cl}^A}$ = ($^{(Mass)}[\boldsymbol{UF}]_{A_{Cl}} \times {}^{G}\boldsymbol{u}$) = ($\mathbf{1.0} \times {}^{G}\boldsymbol{u}$))" is shown to be *so **higher*** as **≈100** times than the **"*ratio*** ($\boldsymbol{X_m}$) specific-*uncertainty-factor*/ -uncertainty **(**$^{(Ratio)}\boldsymbol{\epsilon_{Cl}^A}$ = ($^{(Ratio)}[\boldsymbol{UF}]_{A_{Cl}} \times {}^{G}\boldsymbol{u}$) = ($\mathbf{0.0103445} \times {}^{G}\boldsymbol{u}$))**"**.

### 3.1.2 Should the above predictions be true?

Let the preset/ pre-established **($^{37}$Cl/$^{35}$Cl ratio**-*measurement* cum detected-*biases-corrected***)** ***input***-uncertainty to be**:** ($^{\mathbf{Max.}}|\Delta_m| \equiv \boldsymbol{u_m} \equiv \boldsymbol{u_{37/35}}$) = $^{G}\boldsymbol{u}$ = **0.01**%. Then the correspondingly *estimated* **$^{35}$Cl** and **$^{37}$Cl** *abundances* **(**$\boldsymbol{y_{35}}$ and $\boldsymbol{y_{37}}$, cf. Eq.(1a)**)** and *atomic weight* **(**$\boldsymbol{a_{Cl}}$, cf. Eq.(2a)**)** of natural chlorine are predicted (cf. the **Appendix 2**) to be as uncertain (accurate) as **"(**[$^{\text{Pred.}}\epsilon_{35}^Y \equiv {}^{\text{Max.}}|\Delta_{35}^Y|$] = [$0.2424 \times \boldsymbol{u_m}$] **=)** $\mathbf{0.002424}$%**"**, "**(**[$^{\text{Pred.}}\epsilon_{37}^Y \equiv {}^{\text{Max.}}|\Delta_{37}^Y|$] = [$0.7576 \times \boldsymbol{u_m}$] **=)** $\mathbf{0.007576}$%" and **"(**[$^{\text{Pred.}}\epsilon_{Cl}^A \equiv {}^{\text{Max.}}|\Delta_{Cl}^A|$] = [$0.0103445 \times \boldsymbol{u_m}$] **=)** $\mathbf{0.000103445}$%**"**, respectively.

Now, let us consider that ***two*** different real world *measurements* of the chlorine standard "($\boldsymbol{X_m} \equiv \boldsymbol{X_{37/35}}$) = **0.3199578**" have caused the ***inputs*** (i.e. possible bias corrected

***ratio**-estimates*) to take values as**:** **(i)** $\boldsymbol{x_m}$= **[**($\boldsymbol{X_m}$+ $\boldsymbol{u_m}$) = ($\boldsymbol{X_m}$+ [0.0001 × $\boldsymbol{X_m}$])] = **0.3199898**; and**:** **(ii)** $\boldsymbol{x_m}$= **[**($\boldsymbol{X_m} - \boldsymbol{u_m}$) = ($\boldsymbol{X_m}$ − [0.0001 × $\boldsymbol{X_m}$])] = **0.31992580**.

Then one can easily verify that the above considered ***inputs*** cause the "$\boldsymbol{Y_{35}}$-specific-**Eq.(1)**", the "$\boldsymbol{Y_{37}}$-specific-**Eq.(1)**" and the "$\boldsymbol{A_{\mathrm{Cl}}}$-specific-**Eq.(2)**" to, in turn, yield their respective ***output**-estimates* "$\boldsymbol{y_{35}}$", "$\boldsymbol{y_{37}}$" and "$\boldsymbol{a_{\mathrm{Cl}}}$", really, *as accurate as* ***predicted*** (in the Appendix **2**). That is the "***input** no.* **(i)**" gives**:**

$\boldsymbol{y_{35}}$ = **0.75758164 (**= [$\boldsymbol{Y_{35}}$ − (0.00002424 × $\boldsymbol{Y_{35}}$)] = [**0.7576 −** $^{\mathbf{Pred.}}\boldsymbol{\epsilon^{Y}_{35}}$**])**;

$\boldsymbol{y_{37}}$ = **0.242418386 (**= [$\boldsymbol{Y_{37}}$ + (0.00007576 × $\boldsymbol{Y_{37}}$)] = [$\mathbf{0.2424}$ + $^{\mathbf{Pred.}}\boldsymbol{\epsilon^{Y}_{37}}$**])**; and**:**

$\boldsymbol{a_{\mathrm{Cl}}}$ = **35.45297430 (**= [$\boldsymbol{A_{\mathrm{Cl}}}$ + (0.00000103445 × $\boldsymbol{A_{\mathrm{Cl}}}$)] = (**35.45293763** + $^{\mathbf{Pred.}}\boldsymbol{\epsilon^{A}_{Cl}}$).

And, the "***input** no.* **(ii)**" yields**:**

$\boldsymbol{y_{35}}$ = 0.75761834 = **(**$\boldsymbol{Y_{35}}$ + $^{\mathbf{Pred.}}\boldsymbol{\epsilon^{Y}_{35}}$);

$\boldsymbol{y_{37}}$ = 0.24238166 = **(**$\boldsymbol{Y_{37}}$ − $^{\mathbf{Pred.}}\boldsymbol{\epsilon^{Y}_{37}}$); and**:**

$\boldsymbol{a_{\mathrm{Cl}}}$ = 35.45290095 = ($\boldsymbol{A_{\mathrm{Cl}}}$ − $^{\mathbf{Pred.}}\boldsymbol{\epsilon^{A}_{Cl}}$).

**3.1.2.1** *Consideration of even mass-uncertainty* ($\boldsymbol{u^{M}_{d}}$)*: Verification of "*$^{\mathbf{Pred.}}\boldsymbol{\epsilon^{A}_{Cl}}$*"*

Let the ***input***-ratio-uncertainty ($\boldsymbol{u_m}$) as well as the **-mass-uncertainty** ($\boldsymbol{u^{M}_{d}}$) to be 0.01%**:** $^{G}\boldsymbol{u}$ = **(**$\boldsymbol{u_m}$ = $\boldsymbol{u^{M}_{35}}$ = $\boldsymbol{u^{M}_{37}}$ **=)** $^{\mathrm{Max.}}|\Delta|$ = **0.01**%. Then the correspondingly estimated atomic weight ($\boldsymbol{a_{\mathrm{Cl}}}$) should turn out **(**cf. the **Appendix 3**: $^{\mathbf{Pred.}}\boldsymbol{\epsilon^{A}_{Cl}}$= [$\mathbf{1.0103445}$ × $^{G}\boldsymbol{u}$] **=)** $\mathbf{0.010103445}$% accurate. That is, even for the possible case where ***each*** *of the* ***three*** different input-estimates (isotopic-*ratio*: $\boldsymbol{x_m}$, and the -*masses*: $\boldsymbol{m_{35}}$, and $\boldsymbol{m_{37}}$) should be subject to ±**0.01%** ***error***, the true-***error*** ($|\Delta^{A}_{Cl}|$) of the correspondingly estimated atomic weight **"**$\boldsymbol{a_{\mathrm{Cl}}}$**"** is predicted to be **(**not only ***less*** than the **mod**-sum of ***three*** different *input*-***errors*** [i.e.**:** "**<((**3 × $^{G}\boldsymbol{u}$) = **0.03**%"] but also be**)** ***so restricted*** as **"**$|\Delta^{A}_{Cl}| \leq (\mathbf{Pred.}\,\boldsymbol{\epsilon^{A}_{Cl}} = \mathbf{0.010103}\%)$**"**.

Now, for examining the veracity of our above prediction, let us consider certain sets of "***three*** different ***input***-(*ratio* and *mass*)-***estimates*** having subject to *exactly* **±0.01%** errors", viz. those as the *example nos.* **(1)-(5)** in the **Appendix 4**. It may thus be noted (cf. Appendix 4) that the **true**-*output*-***error*** "$|\Delta^{A}_{Cl}|$" *either* has **equalled** the ***predicted*** *highest possible output*-***error*** "**Pred.**$\epsilon^{A}_{Cl}$ = $\mathbf{0.010103}$%" (cf. Example no. **(1)** and no. **(2)**), *or* is "<**Pred.**$\epsilon^{A}_{Cl}$ (cf. Example nos. **(3)-(5)**)"; but in ***no*** *case* has exceeded "**Pred.**$\epsilon^{A}_{Cl}$". Above all the prediction that the computational step (**SSR** as the "Eq.(2)") should act as a sink-of-input-errors is verified by the fact that "$|\Delta^{A}_{Cl}|$" has, really in every case, turned out *much* ***less*** than the total input-error of **0.03**%.

Thus, it should be clear that (cf. Appendix Nos. (2-4)), **irrespective** whether the *number* of real world ***measurable*** (input) *variable* to be ***one*** or *more than one,* the ***output*** (variable-*specific-uncertainty-factor* "**[*UF*]**" and hence the) *uncertainty* "$\epsilon$" should be decided by the *input(s)*-**to**-*output* **variation**-*characteristics* of the **exactly** corresponding **SSR**, which should in any of our **isotopic** cases act as the (***output***-variable as either the ***d^th^***-***isotopic***-*abundance* "$Y_d$" or the *atomic-weight* "$A_E$" *specific*) ***sink*** of experimental (input) error in the overall process of the desired-*indirect*-measurement.

Yet, it may be enquired: should the ***nature*** of the *above discussed facts* be **different** for cases of *isotopic-multiplicity* (***N***) as "> **2**"?

**3.2 *Tri-isotopic systems***

It should here be pointed out that, in terms of the required *evaluation* of desired result(s), the ***variation*** of the *isotopic-multiplicity* "***N***" should mean the *variation* of the *evaluation-process* "**SSR** as either Eq.(1) or Eq.(2)" by the ***number***-of-*input* (significantly, the *measurable*-"$X_m$") variables being involved. Therefore, even for "(***N*** > **2**)", the ***process*** of the inputs-**to**-output *uncertainty*-***transformation*** should be exactly analogous to the corresponding-*process* discussed above under the "*multi-measurement-feature* of **bi-**

**isotopic**-systems (cf. section **3.1.1.1**)". In other words, irrespective of exact isotopic-multiplicity ($\boldsymbol{N} \geq \mathbf{2}$), the evaluation/ prediction of any ***output***-*uncertainty* "$\boldsymbol{\epsilon}$" should be, as shown above, based on the "**Eq.(4)**".

However, the isotopic-mass "$\boldsymbol{M_d}$, cf. Eq.(2)" is never being a lab-measurable variable. Thus any possible *tri-isotopic* element, the determination of isotopic abundances "$\{\boldsymbol{Y_d}\}_{\boldsymbol{d=1}}^{\boldsymbol{3}}$ **(**cf. Eq.(1)**)**" and/ or atomic weight "$\boldsymbol{A_E}$ **(**Eq.(2)**)**" of which requires the *measurement* of ***two*** different isotopic-abundance-**ratios** "$\{\boldsymbol{X_m}\}_{\boldsymbol{m=2}}^{\boldsymbol{3}}$", should stand for the simplest possible *real-world*-***multi***-*measurement*-system. It is therefore thought worth re-noting the features of the *inputs*-**to**-*output* uncertainty-**translation**-process (cf. **Eq.(4)**) in terms of a tri-isotopic "($\boldsymbol{N} = \mathbf{3}$)" system, viz. Oxygen.

Actually, the point to be noted at this outset is that the ratio-($\boldsymbol{X_m}$)-dependent *individual* and *collective* ***rates***-of-*output* ***variations*** ("**[*MF*]**" and "**[*UF*]**", respectively) **cannot**, i.e. unlike a ***bi-isotopic***-case, be (by magnitude) the same for systems of "($\boldsymbol{N} > \mathbf{2}$)"; i.e. **e.g. (**even for a set of *equally uncertain input*-ratios, and hence for: $\boldsymbol{u_m} = {}^{G}\boldsymbol{u}$, and/ or: $\boldsymbol{F_m} = \mathbf{1}$ [with: $\boldsymbol{m} = \mathbf{2, 3 \ldots} \boldsymbol{N}$], cf. Eq.(6/ 6a)**)**: $[\boldsymbol{UF}]_{\boldsymbol{d}} = \sum_{\boldsymbol{m=2}}^{\boldsymbol{N}} \left(\left|[\boldsymbol{MF}]_{\boldsymbol{m}}^{\boldsymbol{d}}\right| \times \boldsymbol{F_m}\right) = \sum_{\boldsymbol{m=2}}^{\boldsymbol{N}} \left|[\boldsymbol{MF}]_{\boldsymbol{m}}^{\boldsymbol{d}}\right|$. Anyway, we will now exemplify the facts by choosing the **natural oxygen** [19] as the required isotopic-**standard**.

### 3.2.1 Predictive consideration

The oxygen isotopes, their masses (true $\boldsymbol{M_d}$-values), relative abundances (i.e. the true values of the measurable $\boldsymbol{X_m}$-variables), atom fractions (true $\boldsymbol{Y_d}$-values) and the *corresponding* "**Eq.(1)**, i.e. the *output-variable*: $\boldsymbol{Y_d}$" **specific** predicted input-**to**-output *variation*-governing-parameters **(**as: $[\boldsymbol{MF}]_{\boldsymbol{m}}^{\boldsymbol{d}}$, $[\boldsymbol{UF}]_{\boldsymbol{d}}$ and $\boldsymbol{\epsilon_d^Y}$; cf. Eqs. (4-6a)**)** are furnished in Table 1. Similarly, the true atomic weight of oxygen ($\boldsymbol{A_O}$) and the corresponding "Eq.**(2)**, i.e. the *variable*: $\boldsymbol{A_O}$" ***specific*** input-**to**-output *variation*-parameters are tabulated in Table 2. However, though already pointed out above (cf. ***bi-isotopic*** considerations), it may be re-noted here that the prediction of any *output-uncertainty* "$\boldsymbol{\epsilon}$" (cf. Table 1 and Table 2) did ***in no way*** require to consider that the isotopic-abundances "$\{\boldsymbol{Y_d}\}_{\boldsymbol{d=1}}^{\boldsymbol{3}}$" (and/ or their real-world-*measurable* ratios "$\{\boldsymbol{X_m}\}_{\boldsymbol{m=2}}^{\boldsymbol{3}}$") are correlated.

Moreover, it may be pointed out, the present predictions (cf. Table 1 and Table 2) are, by their basic-features, the *same as* those for the *bi-isotopic*-systems discussed above:

i) **(**Irrespective of whether the real-world-*measurement*-accuracy "$\boldsymbol{u_m}$" should be the ratio "$\boldsymbol{X_m}$" specific**)**, the achievable accuracy ($\boldsymbol{\epsilon_d^Y}$) of determining any isotopic-abundance "$\boldsymbol{Y_d}$" should be (*comparison between the* 2nd, 3rd and 4th *rows in* Table 1) the ***isotope*** "***d***" specific (i.e. like the bi-isotopic case, be decided by the *inputs*-**to**-*output* **variation**-*characteristics* of the exactly desired-*variable* "$\boldsymbol{Y_d}$" specific **SSR:** function "$\boldsymbol{f_d}$").

ii) The "$\boldsymbol{Y_d}$ vs. $\boldsymbol{X_m}$" *variation-rate* **(**$[\boldsymbol{MF}]_{\boldsymbol{m}}^{\boldsymbol{d}}$, cf. the Cols. 6 and 7 in Table 1 for specified "***d***" and "***m***"**)** and thus the achievable accuracy "$\boldsymbol{\epsilon_d^Y}$ (cf. Col. 8)" of determining any atom fraction "$\boldsymbol{Y_d}$" are, again like *bi-isotopic*-systems, shown to be ***dependent*** on the (magnitudes of the) measurable variables ($\boldsymbol{X_2}$ and $\boldsymbol{X_3}$). That is, "**Eq.(1)**" is re-ascertained to belong to the **F.2** family of **SSRs**.

iii) The uncertainty-factor "$[\boldsymbol{UF}]_{\boldsymbol{d}}$" and/ or the uncertainty "$\boldsymbol{\epsilon_d^Y}$" (cf. Col. 8 in Table 1**)** is, even for the tri-isotopic system here, predicted to be inversely proportional to the isotopic abundance "$\boldsymbol{Y_d}$" to be estimated.

iv) Table 1 and even Table 2 (*cf. Cols.* 2-6 *in* Table 2) show, **without** exception, that (the input-**to**-output **variation**-*rate*)**:** $|[\boldsymbol{MF}]| < 1$. That is, either of the *isotopic*-**SSRs** "Eq.(1) and Eq.(2)" is re-asserted to act, in the ***overall*** *process* of the corresponding ***indirect*** measurement, as a ***sink*** of net-input-***error***.

v) It could even here be verified that **(**the isotopic **mass** dependent *error-magnification-factors* [i.e. the *rates* of inputs-**to**-output *variations*] "$\left\{[\boldsymbol{MF}]_{\boldsymbol{M_d}}^{\boldsymbol{Ao}}\right\}_{\boldsymbol{d=1}}^{\boldsymbol{3}}$, cf. *Cols.* 2, 3 and 4 in Table 2" actually represent the corresponding mass fractions, and thus**):** $\sum_{d=1}^{3}[\boldsymbol{MF}]_{\boldsymbol{M_d}}^{\boldsymbol{A_O}} = \mathbf{1}$; and **(**the *mod-sum* of the **"**isotopic-**ratio**-specific *variation-rates* (cf. *Cols.* 5 and 6 in Table 2)**"):**$\sum_{m=2}^{3}\left|[\boldsymbol{MF}]_{\boldsymbol{X_m}}^{\boldsymbol{A_O}}\right| = \mathbf{0.00028}$. That is the atomic weight "$\boldsymbol{A_O}$" is, like the above discussed "$\boldsymbol{A_{Cl}}$", predicted to be more susceptible towards a (given) variation in any isotopic-*mass* "$\boldsymbol{M_d}$" than that in any isotopic-*ratio* "$\boldsymbol{X_m}$".

However, as the *isotopic masses* should remain ***invariable*** from one experiment/ lab to another, the expression of the collective error multiplication factor (i.e. the *uncertainty-factor* "$[\boldsymbol{UF}]_{\boldsymbol{A_O}}$") of determining the ***atomic weight*** involving, say for simplicity (cf. Eq.(6a)), equally accurate input-ratios "$\boldsymbol{x_2}$ and $\boldsymbol{x_3}$" should (as implied here above or in Table 2) reduce to:

$$[UF]_{A_O} = (\sum_{m=2}^{3}\left|[MF]_{X_m}^{A_O}\right|) = 0.00028$$

Therefore, even for a significantly higher ratio-measurement uncertainty as "$\boldsymbol{^{G}u} = \mathbf{1\%}$", the resultant-uncertainty ($\boldsymbol{\epsilon_O^A}$) should rather be insignificant:

$$\epsilon_O^A = ([UF]_{A_O} \times {}^{G}u) = \mathbf{0.00028\%}.$$

*However, should the predictions be true*?

### 3.2.2 Verification of the predictions

Let us **preset** the input-ratio-uncertainty as: $\boldsymbol{u_2} = \boldsymbol{u_3} = \boldsymbol{^{G}u} = \mathbf{0.01}\%$, i.e. consider the ***highest*** *possible unaccountable error* ($^{\mathbf{Max.}}\Delta_{\boldsymbol{m}}$) in either of the *estimates* ($\boldsymbol{x_2}$ and $\boldsymbol{x_3}$ [of the measurable ratios "$\boldsymbol{X_2}$ and $\boldsymbol{X_3}$", respectively]) to be ±0.01%.

Then, as predicted (see *Col.* 8 in Table 1), the ***correspondingly*** estimated *abundances* of $^{16}$**O**, $^{17}$**O**, and $^{18}$**O** (i.e.**:** $\boldsymbol{y_1}$, $\boldsymbol{y_2}$ and $\boldsymbol{y_3}$) should turn out 0.0000243%, 0.0100167% and 0.0099833% accurate, respectively. That is to say that, if should the predictions be correct, then the **true-error** ($\Delta_{\boldsymbol{d}}^{\boldsymbol{Y}}$) in the desired-(***d****th*)-*estimate* ($\boldsymbol{y_d}$) is expected to be ***restricted*** as follows: " $\left|\Delta_{16}^{Y}\right| \leq (\boldsymbol{\epsilon_{16}^{Y}} = \mathbf{0.0000243}\%)$ ", " $\left|\Delta_{17}^{Y}\right| \leq (\boldsymbol{\epsilon_{17}^{Y}} = \mathbf{0.0100167}\%)$ " and "$\left|\Delta_{18}^{Y}\right| \leq (\boldsymbol{\epsilon_{18}^{Y}} = \mathbf{0.0099833}\%)$".

And, if the isotopic *masses* are ***also*** (i.e. simply for exemplifying and/ or verifying the uncertainty "$\boldsymbol{\epsilon_{O}^{A}}$" of determining the atomic weight "$\boldsymbol{A_O}$") considered to be subject to 0.01% uncertainty **(**i.e. if: $^{\mathbf{Max.}}\Delta_{d}^{M} = \pm 0.01\%$, with:*d* = 1, 2 and 3**)**; then the *correspondingly* **estimated** *atomic weight* (*$a_O$*) is predicted (*cf. Col.* 7 in Table 2) to turn out **0.0100028**% accurate. That is, then, the corresponding (true) ***output-error*** should be so restricted as: "$\left|\Delta_{O}^{A}\right| \leq (\boldsymbol{\epsilon_{O}^{A}} = \mathbf{0.0100028}\%)$".

*However, do the* ***said****-output-errors follow the* ***said****-dictums*?

It may be reminded that the (present true) "$\boldsymbol{X_m}$ and, $\boldsymbol{Y_d}$**,** and $\boldsymbol{M_d}$" values are *known* (cf. Table 1). Therefore, as already shown for bi-isotopic case, the predictions here can also be verified in terms of the *known-****input****-errors* (i.e. by using the ***generated****-input-estimates*) as:

$$x_m = (X_m \pm u_m) = (X_m \pm {}^{\text{Max.}}|\Delta_m|) = (X_m \pm\ 0.01\%) \qquad (7)$$

And:

$$m_d = (M_d \pm {}^{\text{Max.}}|\Delta_d^M|) = (M_d \pm\ 0.01\%) \qquad (7a)$$

Such an endeavour is made in Table 3. However it may be noted that, in Table 3, we have considered the ***input****-estimates* **(**for both of the Eq.(1) and Eq.(2), *cf. Cols.* 2 and 6**)** having subject to the *uncertainty* ($^{\boldsymbol{G}}\boldsymbol{u} \equiv {}^{\mathbf{Max.}}|\boldsymbol{\Delta}|$) not only ***as*** exactly "$\pm\mathbf{0.01}\%$ (*cf. Example nos.* 1 and 2)" but also ***as*** high as "$\pm\mathbf{1.0}\%$ (*cf. Examples* 3 *and* 4); and presented the correspondingly *evaluated-****outputs*** **(**isotope "***d***" specific abundances: $\boldsymbol{y_d}$**-**values [*cf. Cols.* 3, 4 and 5] and, and the corresponding-atomic-weights: $\boldsymbol{a_O}$-values [*cf. Col.* 7]**)**, their ***observed*** *errors* ($\Delta_{d}^{Y}$ and $\Delta_{O}^{A}$-values) and also, for immediate comparison, their ***predicted****-true-errors* ($^{\mathbf{Pred.}}\Delta_{d}^{Y}$ and $^{\mathbf{Pred.}}\Delta_{O}^{A}$, cf. Eq. (4a)) and *-uncertainties* ($\boldsymbol{\epsilon_{d}^{Y}}$and $\boldsymbol{\epsilon_{O}^{A}}$, respectively).

However, it should be noted that the *example no.* 4 considers the *isotopic masses* as **constants**, thereby offering the opportunity to verify the "**predicted** *accuracies* of determining the atomic weight" for ***variable***, and also ***invariable***, isotopic-masses (*comparison between example no.* 3 *and no.*4).

At this stage, it might be kept in mind that Eq. (1) and Eq. (2) are *non-linear*. Yet, as Table 3 verifies, the predictions above (and/ or as of Tables 1 and 2) are correct for the ***highest*** *possible* (individual) *input-error* of not only $\pm 0.01\%$ (cf. example nos. 1 and 2) but also $\pm 1.0\%$ (cf. example nos. 3 and 4). Thus it may, for clarity, be pointed out that the **mod-sum** of *input-ratio*-**errors** is 0.02% (cf. either example no. 1 or no. 2) or as high as 2.0% (examples 3 and 4). However, the *output*-(***abundance***)-*errors* are, firstly, ***different*** for different isotopes (*comparison*, for a given example, *between column nos.* 3, 4 and 5); secondly, ***less*** (by the *magnitude*) than the corresponding-input-error-**sum**; and thirdly, in ***agreement*** with their predicted values **(**e.g.: "($\Delta^Y_{16} = {}^{\mathbf{Pred.}}\boldsymbol{\Delta^Y_{16}}$) $\leq {}^{\mathbf{Pred.}}\boldsymbol{\epsilon^Y_{16}}$" [*cf. any of the example nos.* 1-4; and/ or [*cf.* specifically, the *example no.* 2]: "($\Delta^Y_{16} = {}^{\mathbf{Pred.}}\boldsymbol{\Delta^Y_{16}}$) $< {}^{\mathbf{Pred.}}\boldsymbol{\epsilon^Y_{16}}$"**)**. Above all, the output error has in **no** case, numerically, exceeded the predicted uncertainty. Thus, simply for making the foresaid fact somewhat vivid, the *output error* in Table 3 has been *emboldened*, however, **only** in those cases where it has acquired its ***highest*** predicted value (cf. e.g. [example no. 1 and column 3]**:** $|\Delta^Y_{16}| = {}^{\text{Pred.}}|\Delta^Y_{16}| = {}^{\mathbf{Pred.}}\boldsymbol{\epsilon^Y_{16}}$; or [example 2 and column 4]**:** $|\Delta^Y_{17}| = {}^{\text{Pred.}}|\Delta^Y_{17}| = {}^{\mathbf{Pred.}}\boldsymbol{\epsilon^Y_{17}}$).

It may similarly be noted that, for the case of evaluating the *atomic weight* ($\boldsymbol{A_O}$, [*cf. Col.* 7 in Table 3), the mod-**sum** of *input*-**errors** (*cf. Cols.* 2 and 6 in Table 3) is as high as either **0.05%** (cf. example no. 1 or 2) or **5.0%** (cf. example no. 3) or **2.0%** (cf. example no. 4). However, the evaluated atomic weight ($\boldsymbol{a_O}$) is, as predicted, at a ***much less*** error than the corresponding-input-error-**sum**. Moreover, the observed-error "$\Delta^A_O$" is in *agreement* with its predicted value **(**i.e.: "($\Delta^A_O = {}^{\mathbf{Pred.}}\boldsymbol{\Delta^A_O}$) $< {}^{\mathbf{Pred.}}\boldsymbol{\epsilon^A_O}$" [*cf. Col.* 7 for *example no.* 2], and/ or: "($\Delta^A_O = {}^{\mathbf{Pred.}}\boldsymbol{\Delta^A_O}$) $\leq {}^{\mathbf{Pred.}}\boldsymbol{\epsilon^A_O}$" [*cf.* any of the *example nos.* 1-4]**)**.

Furthermore, the example **no.4 (**which considers the *isotopic masses* as certain ***constants***, i.e. which represents the *real-world-**invariability*** of isotopic-masses *between* experiments/ labs**)** confirms that the *computational step* "**Eq. (2)**" plays, in the overall process of the corresponding-indirect-measurement, the role of really a very good ***sink*** of *input*-(experimental)-***error*** (viz. the *net*-input-error-sum "**2%**" is caused to be reduced to "**0.00028%**" as the output error). This fact is, moreover, elaborated in **Fig.2**, which represents the "observed (true) ***output-error*** ($\Delta_O^A$)" as a function of the "*mod-**sum**-*of-*ratio-**errors***: $2\boldsymbol{u_m}$ (i.e. mod-sum of the possible *combinations of* the "errors: $+\boldsymbol{u_m}$ and $-\boldsymbol{u_m}$" of the known *equally* erroneous *two* different *input-**ratios***)". It may thus be noted that, irrespective of any input-ratio-error "$\boldsymbol{u_m}$-value"**:** $\Delta_O^A \ll \boldsymbol{(2u_m)}$**.**

Clearly, the present considerations assert that, ***irrespective*** of the isotopic-multiplicity (***N***), the findings as even those on a ***bi-isotopic**-system* should represent general facts, namely**: (i)** any *output-**uncertainty*** "$\boldsymbol{\epsilon}$" should get decided by the corresponding *output-* **variable-***specific* **SSR**, but *in **no** way* get varied for either the *measurable-**input***-(isotopic-ratio)-variables or the desired-***output***-(isotopic-abundance)-variables to be ***correlated***; **(ii)** both of the *isotopic*-**SSRs**, Eq.(1) and Eq.(2), belong to the "**F.2**" family of **SRRs/ models,** i.e. the *uncertainty* as either "$\boldsymbol{\epsilon_d^Y}$" **(**of determining an isotopic-abundance**:** $\boldsymbol{Y_d}$**)** or "$\boldsymbol{\epsilon_E^A}$" **(**of determining any atomic-weight**:** $\boldsymbol{A_E}$ **)** should vary as a function of even ***alone*** (the *magnitude(s)* of) any corresponding-isotopic-ratio(s) "$\boldsymbol{X_m(s)}$" to be measured; and**: (iii)** the isotopic-**SSR** either "Eq.(1)" or "Eq.(2)" should **(**unlike the other "**F.2**" family member as the **IRMS**-*measurement-**model***, which has been shown [12,14,17,18] to act a ***source of error*** for the correspondingly desired-output-estimate**)** *function* as however a ***sink*** of corresponding *net-**input***-(experimental)-***error***.

However how should the ***output**-uncertainty* "$\boldsymbol{\epsilon}$" be, in any **unknown** case, evaluated?

### 3.3 *An unknown case and the output-uncertainty "$\epsilon$"*

Let us consider *an abundance spectrum from the literature* for discussion.

The measured [4] estimates ($x_m$-values) of the isotopic-abundance-ratios of xenon and the corresponding reported [4] uncertainties, the correspondingly reported atom-fractions ($y_d$-values) and their probable errors ($\rho_d$-values), and the atomic weight ($a_{Xe}$) and its probable error ($\rho_{Xe}$), are reproduced in Table 4 (*cf. Cols.*4 and 5, and *the bottom leftmost box*, respectively). However it should be noted (cf. Col. 4 in Table 4) that "$^{Max.}|\Delta X_m|$" refers to the reported [4] *absolute* value of the $m^{th}$-ratio-*uncertainty* and, "$u_m$" to the (correspondingly computed) *relative* value. Similarly, "$^{Max.}|\Delta M_d|$" and "$u_d^M$" (cf. Col. 2) represent the reported [4] *absolute* and the (correspondingly computed) *relative* uncertainties in the isotopic-mass ($M_d$).

### 3.3.1 Prediction of the output-uncertainties "$\{\epsilon_d^Y\}_{d=1}^{9}$ *and,* $\epsilon_{Xe}^A$"

We have here also evaluated, corresponding to the reported [4] *inputs* (cols. 2 and 4 in Table 4), the desired *results* (outputs). This was so done for examining whether our computations cause the results to vary from the reported [4] ones. Our evaluated atom fractions ($y_d$-values) and atomic weight ($a_{Xe}$-value), along with their respective probable errors ($\rho_d$- and $\rho_{Xe}$-values, but which were computed using the Eq. 3 above) and *predicted* (cf. Eq.4) *uncertainties* ($^{pred.}\epsilon_d^Y$ and $^{Pred.}\epsilon_{Xe}^A$ values), are also furnished in Table 4 (*cf. Col.* 6 and *the bottom side middle box*, respectively). The discrepancy, if there is at all any between the reported [4] and our computed atom-fractions (*comparison* of $y_d$-**values** between *cols.* 5 *and* 6) and/ or between the estimates of atomic weight (*comparison* of $a_{Xe}$-**values** between the bottom side *left* and *middle* blocks), should mostly be due to the truncation of data.

However, it should be noted that the ***output***-uncertainties "$^{pred.}\epsilon_d^Y$ and $^{Pred.}\epsilon_{Xe}^A$" are here predicted (cf. Eq. 4) by considering the corresponding-***input***-uncertainties ($u_m$-values and also, for the case of atomic weight, $u_d^M$-values) to be exactly as those reported [4] and/ or tabulated in cols. 4 and 2, respectively, of Table 4.

Moreover, we should draw an attention to the "**process**" whatever being employed for *evaluating* the **uncertainty** on any possible *output-**estimate***. The method of evaluation whatsoever should demand to consider, at least in the back of the mind, any (unknown) ***estimate*** itself to be its ***true**-value*. Thus, irrespective of whether the foregoing statement should be a fact or not, we have here considered **(either** for *predicting* (cf. Eq. 4) the *uncertainties on* the reported [4] ***estimates*** of xenon-isotopic-abundances and atomic weight "$\{y_d\}_{d=1}^{9}$ and $a_{Xe}$, respectively"; **or** for *verifying* subsequently whether our predictions are correct or not**)** the ***measured*** cum reported [4] *isotopic-abundance-ratios* ($x_m$*-values*, cf. Col.4 in Table 4) as their **true**-*values*. That is, we have here evaluated the "$(Y_d$ **vs.** $X_m)$", "$(A_{Xe}$ **vs.** $X_m)$" and even "$(A_{Xe}$ **vs.** $[X_m, M_d])$" ***variation**-parameters* **(**as **[*MF*]**- and **[*UF*]**-values, cf. Eq.5 and Eq. 6a, respectively**)** by assuming that**:** $\{X_m = x_m\}_{m=2}^{9}$ and hence**:** $\{Y_d = y_d\}_{d=1}^{9}$, and**:** $(A_{Xe} = a_{Xe})$.

Here, it may also be re-noted that the determination of xenon isotopic abundances requires ***eight*** (***m*** = 2-9, cf. Eq.(1)) different ratio-$X_m$-measurements. Therefore, the **(***true-**error*** and, hence the**)** *uncertainty* of each of the **nine** *estimated*-isotopic-abundances ($\{y_d\}_{d=1}^{9}$) should be decided by the ***collective*** contribution of the **eight** different ***input**-isotopic-ratios* "$\{x_m\}_{m=2}^{9}$" specific **(***true-**errors*** and, thus the**)** *uncertainties* "$\{u_m\}_{m=2}^{9}$" **(**respectively**)**. And, the determination of atomic weight ($A_{Xe}$, cf. Eq.(2)) should **additionally** be subject to so many as the **nine** different isotopic-mass-uncertainties ($\{u_d^M\}_{d=1}^{9}$).

However it should be noted that, out of the ***eight*** (***m*** = 2-9) ***individual**-ratio-*$X_m$-**dependent** error-magnification-factors of determining any "isotope-***d***-specific-abundance**:** $Y_d$" **(**i.e. out of *eight* different "$Y_d$ vs. $X_m$" variation-***rates*:** $\{[MF]_m^d\}_{m=2}^{9}$**)**, **only** the corresponding-***highest**-*[***MF***]*-value* "Max.$|[MF]_m^d|$" and, of course, the *uncertainty factor* **(**i.e. the ***collective**-rate* "**[*UF*]**$_d$, cf. Eq.(6a)" of *eight* (***m*** = 2-9) different "$Y_d$ vs. $X_m$" variations**)** are, for simplicity, tabulated in Table 4 (cf. *Cols.* 7 and 8, respectively). Moreover, for the case of atomic weight ($\boldsymbol{A_{Xe}}$, cf. Eq.(2), *only* the corresponding-***collective*** "$\boldsymbol{A_{Xe}}$ **vs.** ($\boldsymbol{X_m}$, $\boldsymbol{M_d}$)" and thus the *collective* "$\boldsymbol{A_{Xe}}$ **vs.** $\boldsymbol{X_m}$" variation-rates (i.e. *both* "$\boldsymbol{X_m}$ and $\boldsymbol{M_d}$" as well as *only* "$\boldsymbol{X_m}$" **dependent** *uncertainty-factors*: $[\boldsymbol{UF}]_{A_{Xe}}$-values) are furnished in Table 4 (cf. the bottommost row/ box).

Yet, it could be of interest to note that even the **present** findings (cf. Table **4**) are, *by nature*, **no different** from the those for either ***bi***- or ***tri***-*isotopic cases* above, viz.: **(i)** both "$^{\text{Max.}}\left|[\boldsymbol{MF}]_m^d\right|$" and "$[\boldsymbol{UF}]_d$", and thus the *uncertainty* "$\boldsymbol{\epsilon_d^Y}$" (of determining the atom fraction: $\boldsymbol{Y_d}$), ***vary*** from one isotope (***d***) to another and: **(ii)** even *the* "$^{\text{Max.}}\left|[\boldsymbol{MF}]_m^d\right|$" is (irrespective of the isotope: ***d***) "**<1**", thereby reinforcing the above finding that the "**SSR:** Eq. (1)" should by its intrinsic nature act as the *isotope-**d**-specific* **error-sink** in the overall process of determining the isotopic abundance "$\boldsymbol{Y_d}$".

Similarly the "**Eq. (2)**" is also here predicted to *behave* as the isotopic-abundance-*pattern*- and/ or *element*-sensitive ***sink***-of-net-input-(*experimental*)-***error***, specifically in terms of the **required** *real-world*-(ratio-$\boldsymbol{X_m}$)-*measurements* **(**[cf. the bottommost row in Table 4 for the ratios-**specific**-*uncertainty-factor* of determining "$\boldsymbol{A_{Xe}}$"]**:** $^{\textit{Ratio}}[\boldsymbol{UF}]_{A_{Xe}} = \sum_{m=2}^{9}\left|[MF]_{X_m}^{A_{Xe}}\right|$ = **0.0116323)**. Thus say, however simply for understanding, that all the **eight** *estimated-ratios* ($\{\boldsymbol{x_m}\}_{\boldsymbol{m=2}}^{\boldsymbol{9}}$) are *equally* erroneous, viz.: $^{G}\boldsymbol{u}$ = $|\Delta_m|$ = **1.0%**. Then, though the net-**input**-error ($\sum_{m=2}^{9}|\Delta_m|$) becomes as high as "**8**%", the *highest*-possible-**output**-*error* "($^{\text{Max.}}\left|\Delta_{Xe}^A\right|$ = $^{\textit{Ratio}}\,\boldsymbol{\epsilon_{Xe}^A}$)" should turn out to be "**0.0116323%**" only **(cf.** Eq.(4)**:** $^{\textit{Ratio}}\,\boldsymbol{\epsilon_{Xe}^A}$ = **(**$^{\textit{Ratio}}[\boldsymbol{UF}]_{A_{Xe}} \times {}^{G}\boldsymbol{u}$**)** = (0.0116323 × $^{G}u$) = **0.0116323**%**)**.

Moreover, the present predictive implications are further clarified in **Fig. 3**, where it is assumed for simplicity that all the **required** *ratio-estimates* "$\{\boldsymbol{x_m}\}_{\boldsymbol{m=2}}^{\boldsymbol{9}}$" to be *equally* uncertain as "$\boldsymbol{u_m}$" and offered the *input*-**to**-*output uncertainty*-**variations** as "$\boldsymbol{\epsilon_d^Y}$ vs. $\boldsymbol{u_m}$"

**(**however, for only *three* typical isotopes [namely, "***d***" as the highest, middle and lowest abundant **$^{132}$Xe**, **$^{134}$Xe** and **$^{126}$Xe,** respectively]**),** and also "$\epsilon_{Xe}^{A}$vs. $u_m$". Thus, it may be noted that the ***slope*** ("**[*UF*]**", cf. **Eq.(4)**) of one isotope-***d***-specific "$\epsilon_{d}^{Y}$ vs. $u_m$" *line* is **different** from another and also from the atomic-weight-$A_{Xe}$-specific (i.e.**:** $\epsilon_{Xe}^{A}$vs. $u_m$) *line*. And, that the isotopic **SSR** "either Eq.(1) or Eq.(2)" should, in the case of the corresponding indirect-measurement, act as a kind of "***sink***-of-(experimental)-***error***" is reflected by the fact that any variable-specific ***output-uncertainty "ϵ"*** is "**<**$8u_m$ (rather "$\ll 8u_m$").

Moreover the isotopic mass (***M$_d$***) dependent *rate of variation* of atomic weight (i.e.**:** $\left|[MF]_{M_d}^{A_{Xe}}\right|$) could, like the chlorine or the oxygen case above, be shown to equal the corresponding (***d$^{th}$***) isotopic mass *fraction*(e.g.: $[MF]_{M_{124}}^{A_{Xe}}$ = 0.00089796; or: $[MF]_{M_{129}}^{A_{Xe}}$ = 0.25920437; and of course, it could be shown that**:** $\sum_{d=1}^{9}\left|[MF]_{M_d}^{A_{Xe}}\right| = 1.0$). That is, any *individual* isotopic-mass-***uncertainty***, which is though generally insignificant, should in the process of defining (cf. Eq.(2a) the *output estimate* ($a_{Xe}$) and hence, the *output uncertainty* ($\epsilon_{Xe}^{A}$), be further ***reduced*** by a *factor* equalling the corresponding isotopic mass *fraction*.

It should above all be pointed out that even the considerations on the ***unknown*** *measurement-system* [4] here did *in **no** way*, i.e. like those for the **known** either ***bi-*** or ***tri-*** *isotopic case* above, require to take into account the fact that the isotopic abundances ($\{Y_d\}_{d=1}^{9}$) are **interdependent**. That is to say, it is also in terms of the *unknown* case [4] clarified that**: (i)** the ***correlation*** "of the *measurable*-input (isotopic-ratio**:** $X_m$) and/ or the *desired*-output (isotopic-abundance**:** ***Y$_d$***) *variables*" **cannot** be an output-uncertainty ($\epsilon_{d}^{Y}$ or $\epsilon_{Xe}^{A}$) governing-parameter; **(ii)** the "*input(s) → output*" ***translation*** and the corresponding-**uncertainty**-*transformation* "$u_m(s) \rightarrow \epsilon$" should take place simultaneously, i.e. be governed by the input-**to**-output **variation**-*characteristics* of the exactly corresponding **SSR** (here**:** Eq.(1) or Eq.(2)); **(iii)** either of the **SSRs** "Eq.(1) and Eq.(2)" should, by its *intrinsic nature*, act as a sort of ***sink***-of-(*experimental*)-*error* in the overall process of the corresponding indirect measurement. Yet, it should be pointed out (cf. Table **4**) that the uncertainty ($\boldsymbol{\epsilon}$) of determining the atom fraction "$\boldsymbol{Y_d}$" (*Col.* 6) or even the atomic weight "$\boldsymbol{A_{Xe}}$" (*cf. the bottom side middle block*) is predicted to be higher than the corresponding reported [4] combined standard uncertainty **(**or than that evaluated using the Eq. (3) here**)**: $^{\text{Pred.}}\boldsymbol{\epsilon_d^Y} > \boldsymbol{\rho_d}$, and even: $^{\text{Pred.}}\boldsymbol{\epsilon_{Xe}^A} > \boldsymbol{\rho_{Xe}}$.

### 3.3.2 Cross-checking of predictions: experimental view-point

It may be recollected that we have, for our prediction, considered the reported [4] *estimates* (cf. col. 4 in Table 4) of isotopic-ratios as their *true-values*, i.e.**:** $\{\boldsymbol{X_m} = \boldsymbol{x_m}\}_{\boldsymbol{m=2}}^{\boldsymbol{9}}$ and hence: $\{\boldsymbol{Y_d} = \boldsymbol{y_d}\}_{\boldsymbol{d=1}}^{\boldsymbol{9}}$, and**:** $\boldsymbol{A_{Xe}} = \boldsymbol{a_{Xe}}$.

Thus, whether our predictions are correct or not **(**i.e. whether even the *reported* [4] *estimates* of the atom-fractions and the atomic-weight of xenon should be subject to the *uncertainty-values* as those predicted by the **Eq.(4)** here rather than by **Eq.(3)** or the so ***incorporating*** correction factors for inter-dependence of isotopic abundances**)** can, like the oxygen case (cf. Eqs. (7) and (7a)), be verified in terms of **generated** *input-estimates* of ***known*** *errors*, i.e. as: **(**$\boldsymbol{x_m} = (\boldsymbol{X_m} \pm \boldsymbol{u_m}) = (\boldsymbol{X_m} \pm^{\text{Max.}}|\Delta_{\boldsymbol{m}}|)$**)**, with: *m* = 2-9**)** and: **(**$\boldsymbol{m_d} = (\boldsymbol{M_d} \pm \boldsymbol{u_d^M}) = (\boldsymbol{M_d} \pm^{\text{Max.}}|\Delta_{\boldsymbol{d}}^{\boldsymbol{M}}|)$, with: *d* = 1-9**)**. Such an opportunity is offered in Table 5.

It may however be noted (cf. Table **5**) that**: (i)** Ex. Nos. 1 and 2 refer to the ***input*** ratio- and mass-*errors* "$\Delta_{\boldsymbol{m}}$ and $\Delta_{\boldsymbol{d}}^{\boldsymbol{M}}$" as **exactly** the reported [4] ratio- and mass-*uncertainties* "$\pm\boldsymbol{u_m}$ and $\pm\boldsymbol{u_d^M}$" (i.e. those tabulated in *Cols.*4 and 2 of Table **4**), respectively; **(ii)** Ex. Nos. 2a and 3 correspond to "$\{\Delta_{\boldsymbol{m}} = \pm\boldsymbol{u_m}\}_{m=2}^{9}$, but $\{\Delta_{\boldsymbol{d}}^{\boldsymbol{M}} = \boldsymbol{0.0}\}_{d=1}^{9}$"; **(iii)** Ex. No. 4 gives the results for "$\{\Delta_{\boldsymbol{m}} = \pm\boldsymbol{1.0}\%_m\}_{m=2}^{9}$, and $\{\Delta_{\boldsymbol{d}}^{\boldsymbol{M}} = \boldsymbol{0.0}\}_{d=1}^{9}$"; and**: (iv)** Ex. No. 5 refers to all the different ratio-measurement- and mass-errors as **±1.0**%. However, in **any** of the foresaid cases, the **results** of *abundance* (i.e. "$Y_d$ vs. $X_m$") *variations* are in Table 5 shown ***for***, instead of all the nine isotopes, only ***three*** typical isotopes, namely, the somewhat *lower-*, *highest-* and *middle*-abundant $^{124}$Xe, $^{132}$Xe and $^{136}$Xe(cf. Cols. 3, 4 and 5), respectively. Of course, this is so done for simplifying the presentation.

However it may be noted (cf. Table 5) that, irrespective of whether the input-*errors* are **small** as the reported [4] input-*uncertainties* (cf. Example Nos. 1-3) or so **high** as $\pm\mathbf{1.0}\%$ (cf. Example Nos. 4 and 5), the **results** of the said inputs-**to-**output variations (cf. cols. 3, 4, 5 and 7) supplement the *predictions*; e.g. the error of any correspondingly *estimated* isotopic abundance or atomic weight (i.e. any output-**error)** is**: (i)** by magnitude, ***less*** than the corresponding input-error-**sum** and, **(ii)** in *agreement* with its predicted (cf. Eq. 4a) value; **viz.:** $\Delta^{\boldsymbol{Y}}_{\mathbf{124}} = {}^{\textbf{Pred.}}\Delta^{\boldsymbol{Y}}_{\mathbf{124}}$, and even: $\Delta^{\boldsymbol{A}}_{\boldsymbol{Xe}} = {}^{\textbf{Pred.}}\Delta^{\boldsymbol{A}}_{\boldsymbol{Xe}}$ (cf. Cols. 3 and 7, respectively, for *any of the examples* 1-5). Moreover, the true-abundance-error ($\Delta^{\boldsymbol{Y}}_{\boldsymbol{d}}$) has, *in **no** case*, exceeded the corresponding predicted-uncertainty (${}^{\textbf{Pred.}}\boldsymbol{\epsilon}^{\boldsymbol{Y}}_{\boldsymbol{d}}$, cf. cols. 3, 4 and 5). That is, it may **e.g.** be noted that either: $\left(\left|\Delta^{\boldsymbol{Y}}_{\mathbf{124}}\right| = {}^{\text{Pred.}}\left|\Delta^{\boldsymbol{Y}}_{\mathbf{124}}\right|\right) = {}^{\textbf{Pred.}}\boldsymbol{\epsilon}^{\boldsymbol{Y}}_{\mathbf{124}}$ (*cf. Ex. No.* 1) or: $\left(\left|\Delta^{\boldsymbol{Y}}_{\mathbf{124}}\right| = {}^{\text{Pred.}}\left|\Delta^{\boldsymbol{Y}}_{\mathbf{124}}\right|\right) < {}^{\textbf{Pred.}}\boldsymbol{\epsilon}^{\boldsymbol{Y}}_{\mathbf{124}}$ (*cf. Ex. Nos.* 2-5). Similarly, the results of the *atomic-weight-variations* (**cf. col. 7**) also conform with the theory (Eq.(4/ 4a), i.e. the true-output-error "$\Delta^{A}_{Xe}$" has taken the value as ***either***: $(|\Delta^{A}_{Xe}| = {}^{\text{Pred.}}|\Delta^{A}_{Xe}|) = {}^{\textbf{Pred.}}\boldsymbol{\epsilon}^{\boldsymbol{Y}}_{\mathbf{124}}$ (*cf. Ex. Nos.* 2, 2a and 4) ***or*:** $(|\Delta^{A}_{Xe}| = {}^{\text{Pred.}}|\Delta^{A}_{Xe}|) < {}^{\textbf{Pred.}}\boldsymbol{\epsilon}^{\boldsymbol{A}}_{\boldsymbol{Xe}}$ (*cf. Ex. Nos.* 1, 3 and 5).

*Should the above shown agreement between the theory of indirect measurement uncertainty (Eq. 4) and the observations on correlated quantities be accidental*?

Actually, it had already been clarified elsewhere [12-14,17,18] that the method of evaluating the indirect measurement uncertainty "Eq.(4)" should *not* be dependent on the (*nature* of the *variables* of the *derived*) *system* to be applied. Yet, we may here examine our considerations in terms of some other isotopic data from the literature.

### 3.4 *Considerations for certain* [6] *isotopic estimates of* Nd

As indicated in Table 6, we have assumed the TIMS-***measured*** [6] **Nd**-isotopic abundance *ratios* ( $\{\boldsymbol{x_m}\}_{\boldsymbol{m=2}}^{\boldsymbol{7}}$ , cf. col. 4), the correspondingly *estimated* [6] isotopic abundances ($\{\boldsymbol{y_d}\}_{\boldsymbol{d=1}}^{\boldsymbol{7}}$, cf. col. 5) and Nd-atomic-weight ($\boldsymbol{a_{Nd}}$, cf. the bottom side 1st block) as their respective ***true*** *values*: $\{\boldsymbol{X_m} = \boldsymbol{x_m}\}_{\boldsymbol{m=2}}^{\boldsymbol{7}}$, $\{\boldsymbol{Y_d} = \boldsymbol{y_d}\}_{\boldsymbol{d=1}}^{\boldsymbol{7}}$, and "$\boldsymbol{A_{Nd}} = \boldsymbol{a_{Nd}}$". Thus we have evaluated, corresponding to the reported [6] ***input***-*ratio-uncertainties* ($\{\boldsymbol{u_m}\}_{\boldsymbol{m=2}}^{\boldsymbol{7}}$, cf. col.4), the *uncertainties* ($\boldsymbol{\epsilon(s)}$, cf. **Eq. (4)**) and the *probable errors* ($\boldsymbol{\rho(s)}$, cf. **Eq. (3)**) to be **expected** (i.e. as per our consideration) in the reported [6] ***results*** **"**($\{\boldsymbol{y_d}\}_{\boldsymbol{d=1}}^{\boldsymbol{7}}$, cf. col. 5) and ($\boldsymbol{a_{Nd}}$, cf. the bottom side 1st block)**"**; and furnished the same **(**however, along with **our** computed output-estimates ($\{\boldsymbol{y_d}\}_{\boldsymbol{d=1}}^{\boldsymbol{7}}$, cf. **col. 6**; and $\boldsymbol{a_{Nd}}$, cf. the bottom side **2nd block**)**)** in the Table 6 itself.

It may however be pointed out (cf. Table 6) that**: (i)** our computed output-estimates are **not** really different from the respective reported ones (*comparison between* the cols. 5 and 6; and *between* the bottom side 1st and 2nd boxes), and**: (ii)** the process (cf. **Eq.(4)**) of predicting any output-uncertainty "$\boldsymbol{\epsilon}$" did **in no way** (i.e. like the **bi-** or **tri-** or the **nona**-isotopic [**xenon]** case above) take into account the fact that the *isotopic abundances* "$\{\boldsymbol{Y_d}\}_{\boldsymbol{d=1}}^{\boldsymbol{7}}$" are ***correlated***.

*However, should the predictions be correct?*

For verifying the predicted uncertainties (**Pred.**$\boldsymbol{\epsilon_d^Y}$ and **Pred.**$\boldsymbol{\epsilon_{Nd}^A}$), we have here *generated* the required ***input***-*ratios* "$\{\boldsymbol{x_m}\}_{\boldsymbol{m=2}}^{\boldsymbol{7}}$" of ***known***-*errors* by considering three different "±" combinations of the reported [6] ratio-uncertainties **(**i.e. as**:**$\boldsymbol{x_m} = [\boldsymbol{X_m} \pm {}^{\mathbf{Max.}}|\Delta_{\boldsymbol{m}}|] = [\boldsymbol{X_m} \pm \boldsymbol{u_m}]$**,** with $\boldsymbol{u_m}$-*values* shown in **col. 4** of the **Table 6)**, computed the desired-***outputs*** (i.e. *atom fractions* "($\{\boldsymbol{y_d}\}_{\boldsymbol{d=1}}^{\boldsymbol{7}}$, cf. Eq.(1)" as well as the *atomic weight* "$\boldsymbol{a_{Nd}}$, cf. Eq.(2)"), and furnished also those, however, along with their *observed*-errors (**Δ*(s)***) and *predicted*-errors (**Pred.**Δ***(s)***, cf. Eq. (4a)) in the Table 6 itself (*cf. example nos.* 1, 2 and 3).

Thus, it may be noted that *even* the present results of ***input**-to-**output** variations* (cf. example nos. 1, 2 and 3in the Table 6) **corroborate** the theory (Eq. 4/ 4a). That is to note that, like the **bi-** or **tri-** or **nona**-isotopic case above, the presently *observed **output**-errors* are in agreement with their *predicted* values (see, for the *results* of "$\boldsymbol{Y_d}$ **vs.** $\boldsymbol{X_m}$" *variations*, the **Col.** No. 8 or 10 or 12)**: (i)**"$\Delta_d^Y$ = **Pred.**$\Delta_d^Y$" and/ or: ($|\Delta_d^Y|$ = **Pred.**$|\Delta_d^Y|$) ≤ "(**Pred.**$\epsilon_d^Y$ [cf. col. 6])"; and **(**see, for the *result* of "$\boldsymbol{A_{Nd}}$ vs. $\boldsymbol{X_m}$" *variation*, the **bottom** side block corresponding to the **Ex. No.** 1 or 2 or 3)**: (ii)** "$\Delta_{Nd}^A$ = **Pred.**$\Delta_{Nd}^A$" and/ or: ($|\Delta_{Nd}^A|$ = **Pred.**$|\Delta_{Nd}^A|$) ≤ "(**Pred.**$\epsilon_{Nd}^A$= **0.0005437%** [cf. the bottom **2nd** leftmost block])".

Here, it may also be pointed out (cf. the Table 6) that certain ***observed** output errors*, which have attained the corresponding *predicted **highest possible** values* "**Pred.**$\boldsymbol{\epsilon(s)}$" are emboldened; e.g. the example **no. 1** shows that**:** $|\Delta_{144}^Y|$= **Pred.**$|\Delta_{144}^Y|$ = (**Pred.**$\boldsymbol{\epsilon_{144}^Y}$ = **0.0333%)**. Similarly, the *input-error-combination* as example$\Delta_{124}^Y$ **no. 2** is shown to cause (cf. the corresponding **bottom side block**) the *error-in-atomic-weight* to attain its highest predicted value**:** $|\Delta_{Nd}^A|$ = **Pred.**$|\Delta_{Nd}^A|$ = (**Pred.**$\boldsymbol{\epsilon_{Nd}^A}$ = = **0.0005437%)**; and the combination **no. 3** has caused**:** $|\Delta_{150}^Y|$= **Pred.**$|\Delta_{150}^Y|$ = (**Pred.**$\boldsymbol{\epsilon_{150}^Y}$ = **0.1007%**).

Yet, it could be emphasized that**: (i)** the observed output-error "$|\boldsymbol{\Delta}|$" has ***in no case*** exceeded the predicted (cf. Eq. 4) uncertainty "**Pred.**$\boldsymbol{\epsilon}$", and**: (ii)** the *observed output-error* "$|\boldsymbol{\Delta}|$" **(**or even the *predicted uncertainty* "**Pred.**$\boldsymbol{\epsilon}$"**)** of any estimate as either an evaluated atom fraction "$\boldsymbol{y_d}$" or the atomic weight "$\boldsymbol{a_{Nd}}$" is, *irrespective of example*, ***less*** than the ***mod-sum*** of the *input errors* ($\sum_{m=2}^{7}|\boldsymbol{u_m}|$ = **0.311346**). Thus, the present considerations also underline the above pointed out facts that both of the isotopic **SSRs** "Eqs. **(1)** and **(2)**" should, in the process of translating the **input**-*estimates* into the **desired**-*outputs* (as an atom fraction**:** $\boldsymbol{y_d}$ and, the atomic weight**:**$\boldsymbol{a_{Nd}}$, respectively), cause the respective *isotope-**(d)*** and *isotopic-pattern* (**element**) specific ***reductions*** of the corresponding *collective-**input**-error*, thereby defining the (true) ***output***-*errors* and naturally (the *highest possible* output-errors called as the) *output*-***uncertainties***.

Moreover **(**assuming **all** the six input-ratios to be *equally* uncertain as "$\boldsymbol{u_m}$", cf. Eq.(4) and also Eq.(6a)**)**, the **SSRs**-dependent facts here are further elucidated in **Fig. (4a)**, which depicts the ***variations*** of the output-*uncertainties* "$\boldsymbol{\epsilon_d^Y}$ **(**however only corresponding to the *highest* abundant isotope (***d***)**:** **$^{142}$Nd**, the 2$^{nd}$-highest abundant and/ or the *reference* [6] isotope**:** **$^{144}$Nd**, and the *lowest* abundant isotope**:** **$^{150}$Nd)**" and "$\boldsymbol{\epsilon_{Nd}^A}$ **(**i.e. that for determining the atomic weight**:** $\boldsymbol{A_{Nd}}$**)**" *as a* ***function*** *of* the input-*uncertainty* "$\boldsymbol{u_m}$".

It should however be noted that the Fig.**(4a)** is, by implications, a ***replica*** of either the Fig.**1** (cf. the ***bi-isotopic*** system) or Fig.**3** (cf. the **Xe** system). That is, even the Fig.**(4a)**, which clarifies the ***slopes*** of the different "$\boldsymbol{\epsilon}$ vs. $\boldsymbol{u_m}$" ***lines*** as different, asserts that, for any non-zero *input*-uncertainty ($\boldsymbol{u_m} \neq 0$), the *output*-uncertainty ($\boldsymbol{\epsilon}$) should be decided by the exactly-corresponding-**SSR** (i.e. be the *output*-variable **specific**).

Here, it may also be pointed out that the *isotopic-masses* (cf. **Eq.(2)**) are in the present case considered as ***invariants***. That is, all the **four** different *plots* in **Fig. (4a)** should refer to the case(s) of employing the *one* and the *same set(s)* of measurable ***inputs***. However, as the figure makes it abundantly clear **(**i.e. as**:** **"slope:** $[\boldsymbol{UF}]_{\mathrm{A_{Nd}}}$**"** is **"<slope:** $[\boldsymbol{UF}]_{\boldsymbol{d}}$**"**, with the "***d***" as either **$^{150}$Nd** or **$^{142}$Nd** or even **$^{144}$Nd)**, a possible *error* in any of the **required** abundance-ratio-*measurements* (and hence in any of the six *input*-ratios) should **cause** the *estimated-atomic-weight* "$\boldsymbol{a_{Nd}}$" to be much ***less*** erroneous than the *estimate* "$\boldsymbol{y_d}$" (of any corresponding isotopic-***abundance*****:** $\boldsymbol{Y_d}$).

Furthermore, that the *true-error* ($\Delta_{Nd}^A$) in any possible estimate "$\boldsymbol{a_{Nd}}$" of the atomic weight ($\boldsymbol{A_{Nd}}$) should really never exceed the corresponding predicted *uncertainty* ($^{\mathbf{Pred.}}\boldsymbol{\epsilon_{Nd}^A}$) is clarified in **Fig. (4b)**, which depicts the *variation* of the "*observed* ***true*** *error* ($\Delta_{Nd}^A$)" as a function of the "***mod-sum*** of (certain typical "±" *combinations* of the known) *input*-**ratio**-*errors* ($\sum_{m=2}^{7}|\pm\Delta_m|$ = $\sum_{m=2}^{7}|\pm \boldsymbol{u_m}|$ = $\boldsymbol{6u_m}$)".

In fact, the **SSR** (as either Eq.(1) or Eq.(2)) specific *individual* error-magnification-factors "***MFs***" could here again be shown to be (*by magnitude*) "**<1**". Thus, for example, one can verify that the **Eq. (2)** specific ***rates*** of "$\boldsymbol{A_{Nd}}$ **vs.** $\boldsymbol{X_m}$" *variations* take the following values: $[\boldsymbol{MF}]_{\boldsymbol{X_2}}^{\boldsymbol{A_{Nd}}} = -0.004393$, $[\boldsymbol{MF}]_{\boldsymbol{X_3}}^{\boldsymbol{A_{Nd}}} = -0.001124$, $[\boldsymbol{MF}]_{\boldsymbol{X_4}}^{\boldsymbol{A_{Nd}}} = 0.0003858$, $[\boldsymbol{MF}]_{\boldsymbol{X_5}}^{\boldsymbol{A_{Nd}}} = 0.001992$, $[\boldsymbol{MF}]_{\boldsymbol{X_6}}^{\boldsymbol{A_{Nd}}} = 0.0014666$ and, $[\boldsymbol{MF}]_{\boldsymbol{X_7}}^{\boldsymbol{A_{Nd}}} = 0.002220$; which should in turn explain why any abundance-ratio ($\boldsymbol{X_m}$) specific (*true* or) *possible* **measurement** *error* should have (like the above discussed ***bi***- or ***tri***- or ***nona***-isotopic-**Xe**-case) a significantly ***reduced*** *effect* on the *estimated*-atomic-weight ($\boldsymbol{a_{Nd}}$) and hence, in defining the corresponding (*true output-error* or) output-*uncertainty*.

**4. Conclusions**

We have above chosen the usual mass spectrometric determination of *correlated quantities*, namely, the isotopic abundances "$\{\boldsymbol{Y_d}\}_{\boldsymbol{d=1}}^{\boldsymbol{N}}$" (and thus, also the atomic weight "$\boldsymbol{A_E}$") of any possible multi (***N***) isotopic element (***E***), as the *indirect measurement* (**input-output**) *systems* for discussion on the uncertainties ($\boldsymbol{\epsilon_d^Y}$ and $\boldsymbol{\epsilon_E^A}$) of the correspondingly derived *estimates* (any**:** $\boldsymbol{y_d}$ and, $\boldsymbol{a_E}$, respectively). However it is, in contrary to the usual conception, clarified above that the *inter-dependence* ***of*** the desired output-variables "isotopic-abundances**:** $\{\boldsymbol{Y_d}\}_{\boldsymbol{d=1}}^{\boldsymbol{N}}$" and hence, ***of*** the measurable input-variables "isotopic-abundance-ratios**:** $\{\boldsymbol{X_m}\}_{\boldsymbol{m=2}}^{\boldsymbol{N}}$", should ***in no way*** be governing the *output-uncertainty* ($\boldsymbol{\epsilon_d^Y}$ or $\boldsymbol{\epsilon_E^A}$). The output-uncertainty "$\boldsymbol{\epsilon_d^Y}$**"** or "$\boldsymbol{\epsilon_E^A}$" **(**however, *for* any **given** [preset/ pre-established and, if so, the abundance-ratio "$\boldsymbol{X_m}$" specific real-world-measurement and thus the] ***input***-uncertainty(s) "$\boldsymbol{u_m(s)}$"**)** is, like any **other** possible *kind* of *indirect*-measurement-(e.g. **non**-correlated-variable-specific)-uncertainty [12-14], shown above to really be ***prefixed*** by the (**characteristic** "*input(s)*-**to**-*output* **variation**-*parameter(s)*: [*MF*](*s*), and/ or [*UF*]" of the) corresponding ($d^{th}$ or $E^{th}$) input(s)-**to**-output-**SSR** (Eq.**(1)** or Eq.**(2)** above). Thus, the uncertainty ($\epsilon_d^Y$) of determining an isotopic-abundance ($Y_d$) is even exemplified above to **vary** as a function of the *isotope*, "***d***", ***alone***. Similarly, the output-uncertainty "$\epsilon_E^A$" is shown to be the *isotopic-abundance-pattern* and/ or *element* "***E***" (i.e. the ***exact***-function "$f_E$" (cf. Eq.2)")) specific.

That is to say that any desired **output**-*estimate* and hence the output-***error***, corresponding to any **given** (set of relevant) *input-estimate(s)* (and therefore, *for* the corresponding (set of usually) unknown but **fixed** input-*error(s)*), should, as elaborated above, get simultaneously ***prefixed*** by the concerned-**SSR**. Thus, as an immediate consequence, the *output-**uncertainty*** "$\epsilon$" (of determining even any correlated-variable) has been confirmed above to be the exactly corresponding *input-output*-**SSR** (as the Eq.(1) or Eq.(2)) governed ***systematic***-parameter (*irrespective* of whether the possible *cause* of real-world-measurement cum input-*uncertainty(s)* as "$u_m(s)$" should happen to be purely statistical or systematic or both); and therefore be ***predictable*** 'a priori'. It is in fact demonstrated above, considering even **unknown** *indirect measurement variables*, how exactly the *output-**uncertainty*** "$\epsilon$" should be *predictable* beforehand and/ or the evaluated (***predicted***) "$\epsilon$"-value be verified.

Moreover, the isotopic-**SSRs** "Eq. (1) and Eq. (2)" have been clarified above to belong (like the well-known **IRMS**-*measurement-model* "$Y_\delta = \left(\frac{X_S}{X_W} - 1\right)$") to the "**F.2** family" of the *two* (namely, the input-variable-*independent* "**F.1**" and -*dependent* "**F.2**") *families* [12] of **all** conceivable *input-output*-**SSRs** and -**models**. However, while really the **IRMS**-*measurement* (i.e. the specific "**F.2**") *model* "$Y_\delta = \left(\frac{X_S}{X_W} - 1\right)$" should act [14,17,18] as a ***source***-of-*error* in the corresponding process of the indirect measurement of "$Y_\delta$"; the present "**F.2**" family members as the *isotopic-abundance* specific **SSR** "Eq. (1): $\boldsymbol{Y_d} = \boldsymbol{f_d}(\{\boldsymbol{X_m}\}_{\boldsymbol{m=2}}^{\boldsymbol{N}})$" and also the *atomic weight* specific **SSR** "Eq. (2): $\boldsymbol{A_E} = \boldsymbol{f_E}(\{\boldsymbol{M_d}, \boldsymbol{Y_d}\}_{\boldsymbol{d=1}}^{\boldsymbol{N}})$" have *in contrast* been shown above to behave as the "$\boldsymbol{f_d}$" and "$\boldsymbol{f_E}$" specific ***sinks*** of (the mod-sums of) *experimental-errors* in the overall processes of determining "$\boldsymbol{Y_d}$" and "$\boldsymbol{A_E}$", respectively. These signify that, in the case of any *proposed* indirect measurement, an 'a priori' study of the corresponding **SSR** specific ***input***-to-***output*** *variation*-behaviour should help to ascertain ***beforehand*****: (i)** the *achievable-output-accuracy* (the *least possible output-uncertainty*), i.e. *whether* the corresponding real world measurement to be carried out would serve its *purpose* and/ or**: (ii)** in the case where the **SSR** might happen to be a proposed-mathematical-**model**, *whether* the "proposed-*model*" itself be *purpose-suiting*.

**REFERENCES**

1. ISO, *Guide to the Expression of Uncertainty in Measurement*, 1995**(**ISO/IEC Guide 98-3:2008).
2. Qi-Lu and Akimasa Masuda; The isotopic composition and atomic weight of molybdenum, Int. J. Mass Spectrom. and Ion Processes, 130(1994)65.
3. F. Schaefer, P. D. P. Taylor, S. Valkiers and P. De Bievre; Computational procedures for the treatment of measured or published isotopic abundance data, Int. J. Mass Spectrom. and Ion Processes, 133 (1994) 65.
4. S. Valkiers, Y. Aregbe, P. D. P. Taylor and P. De Bievre; A primary xenon isotopic gas standard with SI traceable values for isotopic composition and molar mass, Int. J. Mass Spectrom. and Ion Processes, 173 (1998) 55.
5. J. R. de Laeter, J. K. Bohlke, P. De Bievre, H. Hidaka, H. S. Pieser, K. J. R. Rosman and P. D. P. Taylor; Atomic weights of elements: review 2000, Pure Appl. Chem. 75 (2003) 683.

6. M. Zhao, T. Zhou, J. Wang. H. Lu and F. Xiang; Absolute measurements of neodymium isotopic abundances and atomic weight by MC-ICPMS, Int. J. Mass Spectrom., 245 (2005) 36.
7. W. Pritzkow, S. Wunderli, J. Vogl and G. Fortunato; The isotopic abundances and the atomic weight of cadmium by a metrological approach; Int. J. Mass Spectrom., 261 (2007) 74.
8. S. Burger, R. M. Essex, K. J. Mathew, S. Richter and R. B. Thomas; Implementation of Guide to the expression of Uncertainty in Measurement (GUM) to multi-collector TIMS uranium isotope ratio metrology, Int. J. Mass Spectrom. 294 (2010) 65.
9. Shuoyun Tong, Juris Meija, Lian Zhou, Brad Methven, Zoltán Mester, and Lu Yang; High-Precision Measurements of the Isotopic Composition of Common Lead Using MC-ICPMS: Comparison of Calibration Strategies Based on Full Gravimetric Isotope Mixture and Regression Models, Anal. Chem. 91 (2019) 4164(https://doi.org/10.1021/acs.analchem.9b00020).
10. Kristýna Kantnerová, Nils Kuhlbusch, Dieter Juchelka, Andreas Hilkert, Sebastian Kopf, Cajetan Neubauer; A guide to precise measurements of isotope abundance by ESI-Orbitrap MS. Nat Protoc 19 (2024) 2435(https://www.nature.com/articles/s41596-024-00981-5).
11. J. B. Scarborough, *Numerical Mathematical Analysis*, Oxford & IBH Publishing Co., Kolkata, 1966.
12. B. P. Datta, The theory of uncertainty for derived results: properties of equations representing physicochemical evaluation systems. *arXiv*: 0712:1732 [*physics.data-an*],2007.
13. B. P. Datta, Uncertainty factors for stage-specific and cumulative results of indirect measurements. *arXiv*: 0909:1651 [*physics.data-an*], 2009.

14. B. P. Datta, Prospects of ratio and differential (δ) ratio based measurement-models: a case study for IRMS evaluation. *arXiv*: 1511.01057[*physics.data-an*],2015**.**

15. B. P. Datta, P. S. Khodade, A. R. Parab, A. H. Goyal, S. A. Chitambar and H. C. Jain; Thermal Ionization Mass Spectrometry of $Li_2BO_2^+$ Ions: Determination of Isotopic Abundance Ratio of Lithium, *Int. J. Mass Spectrom. Ion. Processes* 116 (1992) 87 (and Erratum 121 (1992) 247).

16. B. P. Datta, P. S. Khodade, A. R. Parab, A. H. Goyal, S. A. Chitambar and H. C. Jain; Molecular Ion Beam Method of Isotopic Analysis: Effects of Error Propagation, a Case Study with $Li_2BO_2^+$, *Rapid Commun. Mass Spectrom.*7 (1993) 581.

17. B. P. Datta, Pros and cons of the technique of processing IRMS data as desired-δ values: a case study for determining carbon and oxygen isotopic abundance ratios as $CO_2^+$.*arXiv*:1101.0973 [*physics.data-an*], 2011.

18. B. P. Datta, Study of variations as desired-relative (δ), rather than absolute, differences: falsification of the purpose of achieving source-representative and closely comparable lab-results,*arXiv*:1401.1094 [*physics.data-an*], 2014.

19. Karlsruhe Nuclide Chart, 2006 **(**Zsolt Sóti, Joseph Magill, Raymond Dreher, Karlsruhe Nuclide Chart–New 10th edition 2018 , EPJ Nuclear Sci. Technol. 5, 6 (2019)**)**.

——  ——  ——

## Appendix 1: Notations and Clarifications

**1)** The index "***d***" refers to the term "***derived***/ *desired*" and thus the "***d^th^***"*output* (Variable/ estimate/ error or so).

**2)** The *index* "***m***" signifies "***measured/*** *measurable*" and therefore the "*m^th^*" ***input*** (Variable/ estimate/ uncertainty or so).

**3)** "$\boldsymbol{Y_d}$"**:** The **UPPER** CASE letter/ symbol as "$\boldsymbol{Y_d}$" is, by the usual convention [cf. section 4.1.1 in ref.**1**], referred to here as — the (**TRUE**-value of the *derived*/ desired [and therefore of the]) "***d^th^*** [*output* **VARIABLE** as the] *Isotopic* ***abundance***/ *Atom* ***fraction***"; and — the ***LOWER*** case symbol "$\boldsymbol{y_d}$" represents the ***corresponding*** (desired/ *derived* [***d^th^*** output]) "***estimate***".

**4)** "$\boldsymbol{X_m}$" represents the (**TRUE**-value of the ***measurement***/ *measurable* [and therefore of the]) "*m^th^*" [*input* VARIABLE as the] ***Isotopic****-abundance-****ratio***"; and "$\boldsymbol{x_m}$" stands for the ***corresponding*** (*m^th^ input*) "***estimate***".

**5)** "$\boldsymbol{M_d}$" stands for the **TRUE**-value of the "***Variable*: (*d^th^*)** *Isotopic* ***mass***"; and "$\boldsymbol{m_d}$" for the *corresponding* "***estimate***".

**6)** "$\boldsymbol{A_E}$"**:** the (**TRUE**-value of the *derived*/ desired) *output-****variable*** "***A****tomic weight* (of the element ***E***)" — [which is, however for uniquely distinguishing *from* the ***output*** variable as "$\boldsymbol{Y_d}$", represented as "$\boldsymbol{A_E}$"]; and the *corresponding* "***estimate***" as "$\boldsymbol{a_E}$".

7) "$u_m$" represents the "***uncertainty***" in (i.e. the ***accuracy*** of) the ***estimate*** "$x_m$" — (of the ***m*^th^** measurable [*input*] ***Variable*** as the *Isotopic-abundance-ratio* "$X_m$").

8) "$u_d^M$" is referred to as the "***uncertainty***" in (i.e. the ***accuracy*** of) the ***estimate*** "$m_d$" — (of the ***d^th^*** *Isotopic-mass* "$M_d$").

9) "$\epsilon$"**:** the **Output** *uncertainty/* ***accuracy*** — (which is, for uniquely distinguishing ***from*** the input-uncertainty "*u*", referred to here by the symbol "$\epsilon$").

10) "$\epsilon_d^Y$" stands for the "***uncertainty***" of the ***estimate*** "$y_d$" — (of the desired ***d^th^***- output-variable as the *Atom fraction* "$Y_d$").

11) "$\epsilon_E^A$" is referred to as the "***uncertainty***" in the output-***estimate*** "$a_E$" — (of the atomic weight variable "$A_E$").

12) "$\rho_d$" stands for combined standard [1] or probable [11] uncertainty in the output-estimate "$y_d$".

13) "$\rho_E$" refers to the combined standard/ probable uncertainty in the output-estimate "$a_E$".

14) $\Delta_m$**:** (True) error in the (***m^th^*** *measured/* **input**) *estimate* "$x_m$" — (of [the ***m^th^*** *Isotopic-abundance-ratio*] "$X_m$").

15) $\Delta_d^Y$**:** (True) error in the (***d^th^*** Output) ***estimate*** "$y_d$" — (of [the ***d^th^*** desired *Isotopic* ***abundance****/ Atom* ***fraction***] "$Y_d$").

16) $\Delta_d^M$**:** (True) error in the (input) ***estimate*** "$m_d$" — (of [the ***d^th^*** Isotopic mass] "$M_d$").

17) $\Delta_E^A$**:** (True) error in the [OUTPUT] ***estimate*** "$a_E$" — (of [the desired *Atomic weight* of the element ***E***] "$A_E$").

18) $[MF]_m^d$ **:** the ***d^th^***-desired **SSR/** *model/ function* "$f_d$" specific (relative) **rate of variation**, i.e. the *relative-rate of variation* of the ***d^th^*** output [desired-isotopic-abundance] variable "$Y_d$" as a function of the ***m^th^*** input [measurable] variable "$X_m$". The "$[MF]_m^d$" is also called as the ***m^th^*** *individual* (input-[measurable]-ratio "$X_m$" specific) ***error***-multiplication-[*magnification*]-***factor*** of determining the ***d^th^*** output [desired-isotopic-abundance] variable "$Y_d$".

19) "$[UF]_d$"**:** the desired-output [isotopic-abundance] variable "$Y_d$" (i.e. the exactly corresponding **SSR/** model/ function "$f_d$") specific ***uncertainty-factor***, representing the ***collective*** *rate of variation of* "$Y_d$" as a function *of* **all** the measurable variables "$\{X_m\}_{m=2}^N$" (together). The "$[UF]_d$" is also referred to as the ***collective****-error-magnification-**factor*** of determining [the isotopic-abundance] "$Y_d$".

20) $[MF]_{X_m}^{A_E}$ **:** *Individual*-isotopic-ratio ( $X_m$ ) specific ***error****-magnification-**factor*** of determining [the element ***E*** specific *Atomic weight*] "$A_E$".

21) $[MF]_{M_d}^{A_E}$ **:** *Individual*-isotopic-mass ( $M_d$ ) specific ***error****-magnification-**factor*** of determining [the *Atomic weight*] "$A_E$".

22) $[UF]_{A_E}$ **:** the **SSR/** model/ function "$f_E$ (cf. Eq. 2)" specific ***uncertainty-factor***, representing the ***collective*** *rate of variation of* "$A_E$" as a function *of* **all** the corresponding input-[ measurable]-variables (together), i.e. the ***collective****-error-magnification-**factor*** of determining "$A_E$".

—— —— ——

**Appendix 2:** *Input*-variable-(the *measurable*-**ratio:** $X_m \equiv X_{37/35}$)-based *variation*-parameters of the different *desired-output*-variables-(the $^{35}$**Cl** and $^{37}$**Cl abundances** "$Y_{35}$ and $Y_{37}$" and the Cl-**atomic-weight** "$A_{\text{Cl}}$")-*specific* **SSRs** as **Eq.(1)** and **Eq.(2)**

**(i)** $Y_1$ *-specific* [*variation-rate*: $([MF]_m^{(d=1)} \equiv [MF]_m^1 \equiv [MF]_{37/35}^{35}) = (-\frac{X_m}{1+X_m}) = -0.2424$, *uncertainty-factor*: $([UF]_{(d=1)} \equiv [UF]_1 \equiv [UF]_{35}) = \sum_{m=2}^{2} \left|[MF]_m^d\right| = \left|[MF]_m^1\right| = 0.2424$; and] ***uncertainty***: $(\epsilon_{(d=1)}^Y \equiv \epsilon_1^Y \equiv \epsilon_{35}^Y) = \sum_{m=2}^{2} (\left|[MF]_m^1\right| \times u_m) = (\left|[MF]_m^1\right| \times u_m) = ([UF]_1 \times u_m) = (\frac{X_m}{1+X_m} \times u_m) = (0.2424 \times u_m)$.

**(ii)** $Y_2$***-specific*** [*variation-rate*: $([MF]_m^2 \equiv [MF]_{37/35}^{37}) = (\frac{1}{1+X_m}) = 0.7576$; *uncertainty-factor*: $([UF]_2 \equiv [UF]_{37}) = \left|[MF]_m^2\right| = 0.7576$; and] ***uncertainty***: $(\epsilon_2^Y \equiv \epsilon_{37}^Y) = (\left|[MF]_m^2\right| \times u_m) = ([UF]_2 \times u_m) = (\frac{1}{1+X_m} \times u_m) = (0.7576 \times u_m)$.

**(iii)** $A_{\text{Cl}}$ ***-specific*** [*variation-rate*: $([MF]_{X_m}^{A_{Cl}} \equiv [MF]_{X_{37/35}}^{A_{Cl}}) = (\frac{X_m(M_2 - M_1)}{(1+X_m)(M_1 + [M_2 X_m])}) = 0.010344$; *uncertainty-factor*: $[UF]_{A_{Cl}} = \sum_{m=2}^{2} \left|[MF]_{X_m}^{A_{Cl}}\right| = \left|[MF]_{X_m}^{A_{Cl}}\right| = 0.0103445$; and] ***uncertainty***: $\epsilon_{Cl}^A = \sum_{m=2}^{2} (\left|[MF]_{X_m}^{A_{Cl}}\right| \times u_m) = (\left|[MF]_{X_m}^{A_{Cl}}\right| \times u_m) = ([UF]_{A_{Cl}} \times u_m) = (\frac{X_m(M_2 - M_1)}{(1+X_m)(M_1 + [M_2 X_m])} \times u_m) = (0.0103445 \times u_m)$.

——— ——— ———

**Appendix 3:** Isotopic-**mass**-based-*inputs* **to** *output* ***variation***-parameters of the **Cl**-*atomic-weight* ($A_{\text{Cl}}$) specific-**SSR "Eq.(2)"**

- Mass "$M_{35}$" dependent *rate of variation of* "$A_{\text{Cl}}$": $([MF]_{M_1}^{A_{Cl}} \equiv [MF]_{M_{35}}^{A_{Cl}}) = (\frac{M_1}{(M_1 + [M_2 X_m])} \equiv \frac{M_{35}}{(M_{37} + [M_{37} X_{37/35}])}) = 0.7472555$.
- Mass "$M_{37}$" dependent-*rate*: $([MF]_{M_2}^{A_{Cl}} \equiv [MF]_{M_{37}}^{A_{Cl}}) = (\frac{M_2 X_m}{(M_1 + [M_2 X_m])}) = 0.2527445$.

- *Uncertainty-factor* (for: $\boldsymbol{u_m} = \boldsymbol{u_1^M} = \boldsymbol{u_2^M}$ ; cf. Eq.6a) : $[\boldsymbol{UF}]_{A_{Cl}} = (\sum_{m=2}^{2} \left|[\boldsymbol{MF}]_{X_m}^{A_{Cl}}\right| + \sum_{d=1}^{2} \left|[\boldsymbol{MF}]_{M_d}^{A_{Cl}}\right|) = (\left|[MF]_{X_m}^{A_{Cl}}\right| + [\ \left|[\boldsymbol{MF}]_{M_1}^{A_{Cl}}\right| + \left|[\boldsymbol{MF}]_{M_2}^{A_{Cl}}\right|\ ]) = (\ \mathbf{0.0103445} + [\mathbf{0.7472555} + \mathbf{0.2527445}]) = (0.0103445 + \mathbf{1.0}) = \mathbf{1.0103445}$.
- ***Uncertainty***: $\boldsymbol{\epsilon_{Cl}^A} = (\sum_{m=2}^{2} [\left|[MF]_{X_m}^{A_{Cl}}\right| \times u_m] + \sum_{d=1}^{2} [\ \left|[\boldsymbol{MF}]_{M_d}^{A_{Cl}}\right| \times \boldsymbol{u_d^M}]) = ([\left|[MF]_{X_m}^{A_{Cl}}\right| \times u_m] + \sum_{d=1}^{2} [\ \left|[\boldsymbol{MF}]_{M_d}^{A_{Cl}}\right| \times \boldsymbol{u_d^M}]) = ([0.0103445 \times u_m] + [\mathbf{0.7472555} \times \boldsymbol{u_1^M}] + [\mathbf{0.2527445} \times \boldsymbol{u_2^M}])$.
- Or ( $\boldsymbol{u_m} = \boldsymbol{u_1^M} = \boldsymbol{u_2^M} = {}^G\boldsymbol{u}$)-based-u***ncertainty***: $\boldsymbol{\epsilon_{Cl}^A} = (\ [\boldsymbol{UF}]_{A_{Cl}} \times {}^G\boldsymbol{u}) = (\mathbf{1.0103445} \times {}^G\boldsymbol{u})$.

——— ——— ———

**Appendix 4:** Exemplification of ***output-error*** for *cases* where ***each*** of the *three* inputs ($\boldsymbol{x_m}$, $\boldsymbol{m_{35}}$, **and** $\boldsymbol{m_{37}}$) of the "$\boldsymbol{A}_{\text{Cl}}$" specific-**Eq.(2)** being subject to *exactly* **"±0.01%"** ***error***

1) **Input**-estimates as: **"**$\boldsymbol{x_m} = [(\boldsymbol{X_m} + \boldsymbol{u_m}) = (X_m + [0.0001 \times X_m])] = 0.3199898$, $\boldsymbol{m_{35}} = [(\boldsymbol{M_{35}} + \boldsymbol{u_{35}^M}) = (M_{35} + [0.0001 \times M_{35}])] = 34.972349567$, and: $\boldsymbol{m_{37}} = [(\boldsymbol{M_{37}} + \boldsymbol{u_{37}^M}) = (M_{37} + [0.0001 \times M_{37}])] = 36.969599192$" ***yield*** the **output** as: $\boldsymbol{a}_{\text{Cl}} = 35.45651960$ **(**= $[\boldsymbol{A}_{\text{Cl}} + (0.00010103 \times \boldsymbol{A}_{\text{Cl}})]$ = **[35.45293763 +** (***Error*** as "$\left|\Delta_{Cl}^A\right| = \mathbf{0.010103}\% = {}^{\text{Pred.}}\boldsymbol{\epsilon_{Cl}^A}$")**])**.

2) **Input**-estimates as: **"**$\boldsymbol{x_m} = [(\boldsymbol{X_m} - \boldsymbol{u_m}) = (X_m - [0.0001 \times X_m])] = 0.31992580$, $\boldsymbol{m_{35}} = [(\boldsymbol{M_{35}} - \boldsymbol{u_{35}^M}) = (M_{35} - [0.0001 \times M_{35}])] = 34.965355797$, and: $\boldsymbol{m_{37}} = [(\boldsymbol{M_{37}} - \boldsymbol{u_{37}^M}) = (M_{37} - [0.0001 \times M_{37}])] = 36.962206012$.**"** ***give*** the **output** as: $\boldsymbol{a}_{\text{Cl}} = 35.44935566$ **(**= $[\boldsymbol{A}_{\text{Cl}} - (0.00010103 \times \boldsymbol{A}_{\text{Cl}})]$ = **[35.45293763 –** (***Error*** as "$\left|\Delta_{Cl}^A\right| = \mathbf{0.010103}\% = {}^{\text{Pred.}}\boldsymbol{\epsilon_{Cl}^A}$")**])**.

3) **Input**-estimates as: **"(**$\boldsymbol{x_m} = [(\boldsymbol{X_m} + \boldsymbol{u_m}) = 0.3199898$, $\boldsymbol{m_{35}} = [(\boldsymbol{M_{35}} - \boldsymbol{u_{35}^M}) = 34.965355797$, and: $\boldsymbol{m_{37}} = [(\boldsymbol{M_{37}} + \boldsymbol{u_{37}^M}) = 36.962206012$)**"** ***yield*** the **output** as: $\boldsymbol{a}_{\text{Cl}} = \mathbf{35.45122125} = (\boldsymbol{A}_{\text{Cl}} - [0.00004841 \times \boldsymbol{A}_{\text{Cl}}])$ = **(35.45293763−** [***Error*** as "($\left|\Delta_{Cl}^A\right|$ = **0.004841**%) < ${}^{\text{Pred.}}\boldsymbol{\epsilon_{Cl}^A}$"]**)**.

4) **Input**-estimates as: **"(**$\boldsymbol{x_m} = [(\boldsymbol{X_m} - \boldsymbol{u_m}) = 0.31992580$, $\boldsymbol{m_{35}} = [(\boldsymbol{M_{35}} + \boldsymbol{u_{35}^M}) = 34.972349567$, and: $\boldsymbol{m_{37}} = [(\boldsymbol{M_{37}} + \boldsymbol{u_{37}^M}) = 36.969599192$)**"** ***give*** the **output** as: $\boldsymbol{a}_{\text{Cl}} = \mathbf{35.45644624} = (\boldsymbol{A}_{\text{Cl}} + [0.00009897 \times \boldsymbol{A}_{\text{Cl}}])$ = **(35.45293763+** [***Error*** as "($\left|\Delta_{Cl}^A\right|$ = **0.009897**%) < ${}^{\text{Pred.}}\boldsymbol{\epsilon_{Cl}^A}$"]**)**.

5) **Input**-estimates as**: "(**$x_m$ = **[**($X_m$ + $u_m$) = 0.3199898, $m_{35}$ = **[**($M_{35}$ + $u_{35}^{M}$) = 34.972349567, and**:** $m_{37}$ = **[**($M_{37}$ − $u_{37}^{M}$) = 36.962206012)**"** ***yield*** the **output** as**:** $a_{\text{Cl}}$ = **35.45472735** = **(**$A_{\text{Cl}}$ + [0.00005048 × $A_{\text{Cl}}$]**)** = **(35.45293763** + [***Error*** as "($|\Delta_{Cl}^{A}|$= **0.005048**%) < **Pred.**$\epsilon_{Cl}^{A}$]**)**.

——— ——— ———

**Table 1**

The natural oxygen isotopic details, and the correspondingly predicted *individual error magnification factor*s ($\{[\boldsymbol{MF}]_{\boldsymbol{m}}^{\boldsymbol{d}}\}_{\boldsymbol{m=2}}^{\boldsymbol{3}}$), *uncertainty-factor* ($[\boldsymbol{UF}]_{\boldsymbol{d}}$) and the *uncertainty* ($\boldsymbol{\epsilon_d^Y}$) of determining any of the three unknown but representative natural oxygen-isotope-fractions ($\{\boldsymbol{Y_d}\}_{\boldsymbol{d=1}}^{\boldsymbol{3}}$, cf. Eq.(1))

| Iso-tope ($d$) | Isotopic Mass ($M_d$) | Ra-tio no. ($m$) | Relative Abundance (i.e. the measurable ratio: $X_m$) | Atom/ Isotope Fraction ($Y_d$) | $[MF]_2^d$ (cf. Eq.(5)) | $[MF]_3^d$ (cf. Eq.(5)) | $[UF]_d$ [*1] ($\epsilon_d^Y = [[UF]_d \times {}^G u]$) |
|---|---|---|---|---|---|---|---|
| $^{16}O$ | 15.99491462 | 1 | 1.0 | 0.99757 | $\frac{-X_2}{1+X_2+X_3}$ $= -0.00038$ | $\frac{-X_3}{1+X_2+X_3}$ $= -0.00205$ | 0.00243 (0.00243 × $^G u$) |
| $^{17}O$ | 16.99913150 | 2 | $38.092565 \times 10^{-5}$ | 0.00038 | $\frac{1+X_3}{1+X_2+X_3}$ = 0.99962 | $\frac{-X_3}{1+X_2+X_3}$ $= -0.00205$ | 1.00167 (1.00167 × $^G u$) |
| $^{18}O$ | 17.99916040 | 3 | $20.549936 \times 10^{-4}$ | 0.00205 | $\frac{-X_2}{1+X_2+X_3}$ $= -0.00038$ | $\frac{1+X_2}{1+X_2+X_3}$ = 0.99795 | 0.99833 (0.99833 × $^G u$) |
| Atomic weight ($A_O$) = 15.99940493 | | | | | [*1]: (Assuming the estimates of "$X_2$ and $X_3$" to be equally accurate, i.e. for "$u_2 = u_3 = {}^G u$"; cf. Eq. (6a): $[UF]_d = (\lvert [MF]_2^d \rvert + \lvert [MF]_3^d \rvert)$ | | |

**Table 2**

The predicted *relative-rates* ($\left\{[\boldsymbol{MF}]_{\boldsymbol{M_d}}^{\boldsymbol{A_O}}\right\}_{\boldsymbol{d=1}}^{\boldsymbol{3}}$ and $\left\{[\boldsymbol{MF}]_{\boldsymbol{X_m}}^{\boldsymbol{A_O}}\right\}_{\boldsymbol{m=2}}^{\boldsymbol{3}}$) *of variations* of "$\boldsymbol{A_O}$ (i.e. the *atomic weight of oxygen*, cf. Eq.(2))" as a function of ***individual*** *isotopic mass* and *abundance-ratio* ($\boldsymbol{M_d}$ and $\boldsymbol{X_m}$, respectively); the corresponding ***collective*** rate of variation **(**i.e. the "$\boldsymbol{A_O}$ and hence, Eq.(2)" specific *uncertainty factor*: $[\boldsymbol{UF}]_{\boldsymbol{A_O}}$**)** and, the ***uncertainty*** ($\boldsymbol{\epsilon_O^A}$) of determining the atomic weight ($\boldsymbol{A_O}$)

| Atomic weight ($A_O$) | $[MF]_{M_1}^{A_O} = \frac{M_1}{D} = \frac{M_1}{M_1 + M_2X_2 + M_3X_3}$ | $[MF]_{M_2}^{A_O} = \frac{M_2X_2}{D}$ | $[MF]_{M_3}^{A_O} = \frac{M_3X_3}{D}$ | $[MF]_{X_2}^{A_O} = \frac{N_2}{D_X}$ where *1 | $[MF]_{X_3}^{A_O} = \frac{N_3}{D_X}$ where *2 | $[UF]_{A_O}$ where*3 |
|---|---|---|---|---|---|---|
| 15.999405 | 0.99729 | 0.0004037 | 0.0023062 | 0.0000237 | 0.0002562 | 1.00028 |

*1: $N_2 = X_2 \times (M_2 - M_1 + X_3 \times [M_2 - M_3])$ &: $D_X = (1 + X_2 + X_3) \times (M_1 + M_2X_2 + M_3X_3)$.

*2: $N_3 = X_3 \times (M_3 - M_1 + X_2 \times [M_3 - M_2])$.

*3: (Assuming the estimates of "$M_d$ and $X_m$" to be equally accurate, i.e. for "$u_1^M = u_2^M = u_3^M = u_2 = u_3 = {}^G u$", cf. Eq. (6a)):

$$[UF]_{A_O} = \left(\left|[MF]_{M_1}^{A_O}\right| + \left|[MF]_{M_2}^{A_O}\right| + \left|[MF]_{M_3}^{A_O}\right| + \left|[MF]_{X_2}^{A_O}\right| + \left|[MF]_{X_3}^{A_O}\right|\right) = 1.00028.$$

And (if "${}^G u$ = 1%", then the uncertainty (cf. Eq. (4)):

$$\epsilon_O^A = ([UF]_{A_O} \times {}^G u) = 1.00028\%.$$

**Table 3**

Output estimates **(**i.e. the atom fractions: $[\{\boldsymbol{y_d} \pm \Delta_{\boldsymbol{d}}^{\boldsymbol{Y}}\}_{\boldsymbol{d=1}}^{\boldsymbol{3}}$, cf. Eq.(1a)] and the atomic weight $[(\boldsymbol{a_O} \pm \Delta_{\boldsymbol{O}}^{\boldsymbol{A}})$, cf. Eq.(2a)] of oxygen**)** obtained against the relevant input estimates with **known** errors/ uncertainties ($\{\boldsymbol{x_m} = (\boldsymbol{X_m} \pm {}^{\text{Max.}}|\Delta_{\boldsymbol{m}}|)\}_{\boldsymbol{m=2}}^{\boldsymbol{3}}$, and $\{\boldsymbol{m_d} = (\boldsymbol{M_d} \pm {}^{\text{Max.}}|\Delta_{\boldsymbol{d}}^{\boldsymbol{M}}|)\}_{\boldsymbol{d=1}}^{\boldsymbol{3}}$; cf. Eq.7/ 7a)

| Ex. No. | Input ratios: $\{x_m\}_{m=2}^{3}$<br>(for **given:** $^{\boldsymbol{G}}\boldsymbol{u}$)<br>i) $x_2$<br>$(\Delta_2 \times 10^2)$<br>ii) $x_3$<br>$(\Delta_3 \times 10^2)$ | Output of Eq. (1a): $y_d$ and its Observed & Predicted **True-errors**, and also the Predicted-**uncertainty** | | | Input masses: $\{m_d\}_{m=1}^{3}$<br>i) $m_{16}$<br>$(\Delta_{16}^{M} \times 10^2)$<br>ii) $m_{17}$<br>$(\Delta_{17}^{M} \times 10^2)$<br>iii) $m_{18}$<br>$(\Delta_{18}^{M} \times 10^2)$ | Output of Eq. (2a): $a_O$<br>$(\Delta_O^A \times 10^2)$<br>$[^{\text{Pred.}}\Delta_O^A \times 10^2$;<br>Eq. (4a)]<br>$^{\textbf{Pred.}}\boldsymbol{\epsilon_O^A}$ |
|---|---|---|---|---|---|---|
| | | $y_{16} \times 10^2$<br>$(\Delta_{16}^{Y} \times 10^2)$<br>$[^{\text{Pred.}}\Delta_{16}^{Y} \times 10^2$,<br>cf. Eq. 4a]<br>$^{\textbf{Pred.}}\boldsymbol{\epsilon_{16}^Y}$ | $y_{17} \times 10^2$<br>$(\Delta_{17}^{Y} \times 10^2)$<br>$[^{\text{Pred.}}\Delta_{17}^{Y} \times 10^2$,<br>cf. Eq. 4a]<br>$^{\textbf{Pred.}}\boldsymbol{\epsilon_{17}^Y}$ | $y_{18}$<br>$(\Delta_{18}^{Y} \times 10^2)$<br>$[^{\text{Pred.}}\Delta_{18}^{Y} \times 10^2$,<br>cf. Eq. 4a]<br>$^{\textbf{Pred.}}\boldsymbol{\epsilon_{18}^Y}$ | | |
| 1 | (for: $^{\boldsymbol{G}}\boldsymbol{u}$ **= 0.01%)**<br>i) 0.00038096<br>(0.01)<br>ii) 0.00205520<br>(0.01) | 99.756976<br>$(\boldsymbol{-24.30 \times 10^{-6}})$<br>$[-24.30 \times 10^{-6}]$<br>**0.0000243%** | 0.038003791<br>(0.0099757)<br>[0.0099757]<br>**0.0100167%** | 0.205020447<br>(0.0099757)<br>[0.0099757]<br>**0.0099833%** | i) 15.99651411<br>(0.01)<br>ii) 17.00083141<br>(0.01)<br>iii) 18.00096032<br>(0.01) | 16.00100532<br>(**0.0100028**)<br>[0.0100028]<br>**0.0100028%** |
| 2 | (for: $^{\boldsymbol{G}}\boldsymbol{u}$ **= 0.01%)**<br>i) 0.00038089<br>(−0.01)<br>ii) 0.00205520<br>(0.01) | 99.756983<br>(−0.00001670)<br>$[-16.70 \times 10^{-6}]$<br>**0.0000243%** | 0.037996194<br>(**−0.0100167**)<br>[−0.0100167]<br>**0.0100167%** | 0.205020462<br>(0.**0099833**)<br>[0.0099833]<br>**0.0099833%** | i) 15.99331513<br>(−0.01)<br>ii) 16.99743159<br>(−0.01)<br>iii) 18.00096032<br>(0.01) | 15.9978127<br>(−0.0099515)<br>[−0.0099516]<br>**0.0100028%** |
| 3 | (for: $^{\boldsymbol{G}}\boldsymbol{u}$ **= 1%)**<br>i) 0.00038473<br>(1.0)<br>ii) 0.00207554<br>(1.0) | 99.754576<br>(**−0.002430**)<br>[−0.002430]<br>**0.00243%** | 0.03837907<br>(0.997546)<br>[0.997570]<br>**1.00167%** | 0.20704497<br>(0.997546)<br>[0.997570]<br>**0.99833%** | i) 16.15486377<br>(1.0)<br>ii) 17.169122815<br>(1.0)<br>iii) 18.179152004<br>(1.0) | 16.15944422<br>(**1.00028**3)<br>[1.000280]<br>**1.000280%** |
| 4 | i) 0.00037712<br>(−1.0)<br>ii) 0.00203444<br>(−1.0) | 99.759424<br>(**0.002430**)<br>[0.002430]<br>**0.00243%** | 0.03762091<br>(−0.997594)<br>[−0.997570]<br>**1.00167%** | 0.20295493<br>(−0.997594)<br>[−0.997570]<br>**0.99833%** | i) 15.99491462<br>(**0.0**)<br>ii) 16.99913150<br>(**0.0**)<br>iii) 17.99916040<br>(**0.0**) | 15.99936013<br>(**−0.000280**)<br>[−0.000280]<br>**0.000280%** |

## Table 4

Measured -cum-reported [4] isotopic-ratios ($\{\boldsymbol{x_m} \pm \boldsymbol{u_m}\}_{\boldsymbol{m=2}}^{\boldsymbol{9}}$) and the thereby estimated [4] atom fractions ($\{\boldsymbol{y_d} \pm \boldsymbol{\rho_d}\}_{\boldsymbol{d=1}}^{\boldsymbol{9}}$) and atomic weight ($\boldsymbol{a_{Xe}} \pm \boldsymbol{\rho_{Xe}}$) of Xenon; and also, for comparison, our evaluated estimates ($\boldsymbol{y_d}$ and $\boldsymbol{a_{Xe}}$) cum predicted uncertainties ($\boldsymbol{\epsilon_d^Y}$ and $\boldsymbol{\epsilon_{Xe}^A}$)

| Iso-tope ($d$) | Isotopic Mass: $M_d$ ($^{\text{Max.}}\lvert\Delta M_d\rvert$) [$u_d^M$ (%)] | $m$ | Inputs and Outputs of Eq. (1/ 1a) | | | Predicted parameters | |
|---|---|---|---|---|---|---|---|
| | | | Reported: $x_m$ ($^{\text{Max.}}\lvert\Delta X_m\rvert$) [$u_m$(%)] | Reported: $y_d \times 10^2$ ($^{\text{Max.}}\lvert\Delta Y_d\rvert$) [$\rho_d$(%)] | Our evaluated: $y_d \times 10^2$ [$\rho_d$(%)] $^{\text{Pred.}}\boldsymbol{\epsilon_d^Y(\%)}$*2 | $^{\text{Max.}}\lvert[MF]_m^d\rvert$ = (cf. Eq.(5)) | $[UF]_d$*1 ($\epsilon_d^Y$ = $[[UF]_d \times {}^G u)$ |
| $^{132}$Xe | 131.904154 (0.000001) [$75.8\times10^{-8}$] | 1 | 1.0 | 26.9086 (0.0033) [0.01226] | 26.908653 [0.01235] **0.02207** | $\lvert[MF]_5^{132}\rvert$ = 0.26401 | 0.73091 (0.73091 × $^G u$) |
| $^{124}$Xe | 123.905896 (0.000002) [$16.1\times10^{-7}$] | 2 | 0.003536 (0.000012) [0.339367] | 0.0952 (0.0003) [0.3152] | 0.095149 [0.3393] **0.3608** | $\lvert[MF]_2^{124}\rvert$ = 0.99905 | 1.72901 (1.72901 × $^G u$) |
| $^{126}$Xe | 125.904269 (0.000007) [$55.6\times10^{-7}$] | 3 | 0.0033077 (0.0000072) [0.217674] | 0.0890 (0.0002) [0.2247] | 0.089006 [0.2178] **0.2394** | $\lvert[MF]_3^{126}\rvert$ = 0.99911 | 1.72913 (1.72913 × $^G u$) |
| $^{128}$Xe | 127.903530 (0.000002) [$15.6\times10^{-7}$] | 4 | 0.070989 (0.000029) [0.040851] | 1.9102 (0.0008) [0.0419] | 1.910218 [0.0419] **0.0614** | $\lvert[MF]_4^{128}\rvert$ = 0.98090 | 1.69271 (1.69271 × $^G u$) |
| $^{129}$Xe | 128.9047794 (0.0000009) [$69.8\times10^{-8}$] | 5 | 0.98112 0.00041 [0.041789] | 26.4006 (0.0082) [0.0311] | 26.400618 [0.03126] **0.0418** | $\lvert[MF]_5^{129}\rvert$ = 0.73599 | 1.20290 (1.2029 × $^G u$) |
| $^{130}$Xe | 129.903508 (0.000001) [$77.0\times10^{-8}$] | 6 | 0.151290 (0.000047) [0.031066] | 4.0710 (0.0013) [0.0319] | 4.071010 [0.03224] **0.0506-** | $\lvert[MF]_6^{130}\rvert$ = 0.95929 | 1.64949 (1.64949× $^G u$) |
| $^{131}$Xe | 130.905082 (0.000001) [$76.4\times10^{-8}$] | 7 | 0.789055 (0.000076) [0.009632] | 21.2324 (0.0030) [0.0141] | 21.232407 [0.0144] **0.0276** | $\lvert[MF]_7^{131}\rvert$ = 0.78768 | 1.30627 (1.30627× $^G u$) |
| $^{134}$Xe | 133.9053945 (0.0000009) [$67.2\times10^{-8}$] | 8 | 0.387819 (0.000069) [0.017792] | 10.4357 (0.0021) [0.0201] | 10.435687 [0.02008] **0.03615** | $\lvert[MF]_8^{134}\rvert$ = 0.89564 | 1.52220 (1.52220× $^G u$) |
| $^{136}$Xe | 135.907220 (0.000008) [$58.9\times10^{-7}$] | 9 | 0.32916 (0.00017) [0.051647] | 8.8573 (0.0044) [0.0497] | 8.857252 [0.04845] **0.06457** | $\lvert[MF]_9^{136}\rvert$ = 0.91143 | 1.55377 (1.55377× $^G u$) |
| Reported [4] atomic weight ($a_{Xe}$) = 131.29275 $^{\text{Max.}}\lvert\Delta A\rvert$ = 0.00034 $\rho_{Xe}$(%) = 0.00026 | | | (Our evaluated) $a_{Xe}$ = 131.292761 $\rho_{Xe}$(%) = 0.0002617 $^{\text{Pred.}}\boldsymbol{\epsilon_{Xe}^A}$**(%) = 0.0004653** *2 *2: (cf. Eq. 4) Valid only for the set of data & errors [4] used here. | | | *1: (cf. Eq. 6a) Valid for "$\boldsymbol{u_m}$ = $^G\boldsymbol{u}$ (with: $m$ = 2, 3,... 9)"; i.e. here: $[UF]_d = \sum_{m=2}^{9}\lvert[MF]_m^d\rvert$ | |
| For the estimates of "$M_d$ and $X_m$" to be **equally** accurate, i.e. for "$\boldsymbol{u_m} = \boldsymbol{u_d^M} = {}^G\boldsymbol{u}$ (with: $m$ = 2, 3 ...9; and $d$ = 1, 2 ... 9)", Eq. (6a) predicts that: $[UF]_{Axe} = \sum_{m=2}^{9}\left\lvert[MF]_{X_m}^{Axe}\right\rvert + \sum_{d=1}^{9}\left\lvert[MF]_{M_d}^{Axe}\right\rvert$ = 1.0116323; and: $\boldsymbol{\epsilon_{Xe}^A}$= (1.0116323× $^G u$). But, if should the isotopic masses be treated as ***constants***, then: $^{\textit{Ratio}}[UF]_{Axe} = \sum_{m=2}^{9}\left\lvert[MF]_{X_m}^{Axe}\right\rvert$ = **0.0116323**; and: $^{\textit{Ratio}}\boldsymbol{\epsilon_{Xe}^A}$= (0.0116323× $^G u$). | | | | | | | |

**Table 5:** Typical ***outputs*** (atom fractions: $\{y_d \pm \Delta_d^Y\}$, with: "***d***" as **$^{124}$Xe**, **$^{132}$Xe** and **$^{136}$Xe**; and the atomic weight of Xenon: [$a_O \pm \Delta_{Xe}^A$]) of **Eq. (1a)** and **Eq. (2a)** corresponding to the relevant ***input*** *estimates* of **known** errors (as: $\{\pm\Delta_m\}_{m=2}^{9}$, and $\{\pm\Delta_d^M]\}_{d=1}^{9}$, **cf.** columns 4 and 2, respectively, **in Table 4**)

| Ex. No. | Input ratio-errors<br>i) $\Delta_2 \times 10^2$<br>ii) $\Delta_3 \times 10^2$<br>iii)$\Delta_4 \times 10^2$<br>iv)$\Delta_5 \times 10^2$<br>v)$\Delta_6 \times 10^2$<br>vi)$\Delta_7 \times 10^2$<br>vii)$\Delta_8 \times 10^2$<br>viii)$\Delta_9 \times 10^2$ | Output (cf. Eq, 1a): $y_d$ and its Observed & Predicted **True-errors** and also its Predicted **uncertainty** | | | Input mass-errors<br>i) $\Delta_{132}^M \times 10^2$<br>ii) $\Delta_{124}^M \times 10^2$<br>iii) $\Delta_{126}^M \times 10^2$<br>iv)$\Delta_{128}^M \times 10^2$<br>v)$\Delta_{129}^M \times 10^2$<br>vi)$\Delta_{130}^M \times 10^2$<br>vii)$\Delta_{131}^M \times 10^2$<br>viii)$\Delta_{134}^M \times 10^2$<br>ix)$\Delta_{136}^M \times 10^2$ | Output of Eq. 2a:<br>$a_{Xe}$<br>($\Delta_{Xe}^A \times 10^2$)<br>[$^{\text{Pred.}}\Delta_{Xe}^A \times 10^2$;<br>cf. Eq. (4a)]<br>$^{\text{Pred.}}\boldsymbol{\epsilon}_{Xe}^{A}$ |
|---|---|---|---|---|---|---|
| | | $y_{124} \times 10^2$<br>($\Delta_{124}^Y \times 10^2$)<br>[$^{\text{Pred.}}\Delta_{124}^Y \times 10^2$,<br>cf. Eq. 4a]<br>$^{\text{Pred.}}\boldsymbol{\epsilon}_{124}^{Y}$ | $y_{132} \times 10^2$<br>($\Delta_{132}^Y \times 10^2$)<br>[$^{\text{Pred.}}\Delta_{132}^Y \times 10^2$,<br>cf. Eq. 4a]<br>$^{\text{Pred.}}\boldsymbol{\epsilon}_{132}^{Y}$ | $y_{136} \times 10^2$<br>($\Delta_{136}^Y \times 10^2$)<br>[$^{\text{Pred.}}\Delta_{136}^Y \times 10^2$,<br>cf. Eq. 4a]<br>$^{\text{Pred.}}\boldsymbol{\epsilon}_{136}^{Y}$ | | |
| 1 | i) −0.339367<br>ii) 0.217674<br>iii) 0.040851<br>iv) 0.041789<br>v) 0.031066<br>vi) 0.009362<br>vii) 0.017792<br>viii) 0.051647<br>(see also Table 4) | 0.094806<br>(**−0.360**7)<br>[−0.3608]<br>**0.3608%** | 26.902889<br>(−0.021420)<br>[−0.021425]<br>**0.022070%** | 8.859928<br>(0.030215)<br>[0.030222]<br>**0.064568%** | i) $-75.81 \times 10^{-8}$<br>ii) $-16.14 \times 10^{-7}$<br>iii) $-55.6 \times 10^{-7}$<br>iv) $-15.6 \times 10^{-7}$<br>v) $-69.8 \times 10^{-8}$<br>vi) $-77.0 \times 10^{-8}$<br>vii) $-76.4 \times 10^{-8}$<br>Viii) $-67.2 \times 10^{-8}$<br>ix) $-58.9 \times 10^{-9}$ | 131.292717<br>($-33.51 \times 10^{-6}$)<br>[$-33.51 \times 10^{-6}$]<br>**0.0004653%** |
| 2 | i) −0.339367<br>ii) −0.217674<br>iii) −0.040851<br>iv) −0.041789<br>v) −0.031066<br>vi) −0.009362<br>vii) 0.017792<br>viii) 0.051647<br>(see also Table 4) | 0.094835<br>(−0.3302)<br>[−0.3302]<br>**0.3608%** | 26.911131<br>(0.009209)<br>[0.009208]<br>**0.022070%** | 8.862643<br>(0.060860)<br>[0.060855]<br>**0.064568%** | i) $75.81 \times 10^{-8}$<br>ii) $16.14 \times 10^{-7}$<br>iii) $55.6 \times 10^{-7}$<br>iv) $15.6 \times 10^{-7}$<br>v) $69.8 \times 10^{-8}$<br>vi) $77.0 \times 10^{-8}$<br>vii) $76.4 \times 10^{-8}$<br>Viii) $67.2 \times 10^{-8}$<br>ix) $58.9 \times 10^{-9}$ | 131.293372<br>(**0.0004653**)<br>[0.0004653]<br>**0.0004653%** |

| (**Table 5** continued) | | | | | | |
|---|---|---|---|---|---|---|
| Ex. No. | Input ratio-errors | Output (cf, Eq, 1a): $y_d$ and its Observed & Predicted Errors cum **uncertainty** | | | Input mass-errors | Output of Eq. 2a: $a_{Xe}$ |
| 2a | Errors are exactly the **same** as shown for the Example No. 2 | 0.094835<br>(−0.3302)<br>[−0.3302]<br>**0.3608%** | 26.911131<br>(0.009209)<br>[0.009208]<br>**0.022070%** | 8.862643<br>(0.060860)<br>[0.060855]<br>**0.064568%** | (i) to (ix): 0.0, i.e. $\{\Delta_d^M = \mathbf{0.0}\}_{d=1}^{9}$<br>(Isotopic masses are treated as **constants**) | 131.293370<br>(**−0.0004641**)<br>[−0.0004641]<br>**0.0004641%** |
| 3 | **i)**0.339367<br>**(ii)** to **(ix)**: Errors are exactly the **same** as shown for the Example No. 1 | 0.095451<br>(0.31723)<br>[0.31730]<br>**0.3608%** | 26.902715<br>(**−0.0220**66)<br>[−0.022070]<br>**0.022070%** | 8.859871<br>(0.029570)<br>[0.029576]<br>**0.064568%** | Isotopic masses are treated as **constants** | 131.29267065<br>$(-68.61 \times 10^{-6})$<br>$[-68.62 \times 10^{-6}]$<br>**0.0004641%** |
| 4 | (for: $^{G}\boldsymbol{u}$ **= 1%)**<br>(i) to (vi): +1.0, i.e.<br>$\{(\Delta_m = 1\%\}_{m=2}^{6}$<br>vii) −1.0<br>viii) −1.0 | 0.095770<br>(0.6527)<br>[0.6549]<br>**1.7290%** | 26.816123<br>(−0.3439)<br>[−0.3451]<br>**0.7309%** | 8.738527<br>(−1.3404)<br>[−1.3451]<br>**1.5538%** | See the Example No. 2a<br>(Isotopic masses are treated as **constants**) | 131.27754086<br>(**−0.0115**92)<br>[−0.011632]<br>**0.011632%** |
| 5 | (for: $^{G}\boldsymbol{u}$ **= 1%)**<br>(i) to (viii): +1.0, i.e.<br>$\{(\Delta_m = 1\%\}_{m=2}^{8}$<br>ix) −1.0 | 0.095571<br>(0.4438)<br>[0.4462]<br>**1.7290%** | 26.760462<br>(−0.5507)<br>[−0.5538]<br>**0.7309%** | 8.720389<br>(**−1.54**52)<br>[−1.5538]<br>**1.5538%** | (i) to (ix): −1.0, i.e. $\{\Delta_d^M = \mathbf{-1.0}\%\}_{d=1}^{9}$ | 129.97016539<br>(−1.007363)<br>[−1.007479]<br>**1.011632%** |

**Table 6**

"$\boldsymbol{Y_d}$ **vs.** $\boldsymbol{X_m}$" and "$\boldsymbol{A_{Nd}}$ **vs.** $\boldsymbol{X_m}$" variations for **known** variations in TIMS-measured [6] isotopic abundance ratios ($\{x_m\}_{m=2}^{7}$) of Neodymium

| Iso-tope ($d$) | Isotopic Mass: $M_d$ | $m$ | Let: $\{\boldsymbol{X_m} = x_m\}_{m=2}^{7}$ & $\{\boldsymbol{Y_d} = y_d\}_{d=1}^{7}$ | | | Example No. 1 | | Example No. 2 | | Example No. 3 | |
|---|---|---|---|---|---|---|---|---|---|---|---|
| | | | Reported: $x_m$ ($^{Max.}\lvert\Delta X_m\rvert$) [$u_m$(%)] | Reported: $y_d \times 10^2$ | Our evaluated: $y_d \times 10^2$ [$\rho_d$(%)] $^{\mathbf{Pred.}}\boldsymbol{\epsilon_d^Y}$**(%)** | $x_m = (\boldsymbol{X_m} + u_m)$ [$u_m$(%)] | $y_d \times 10^2$ ($\Delta_d^Y \times 10^2$) [$^{Pred.}\Delta_d^Y \times 10^2$, cf. Eq. 4a] | $x_m = (\boldsymbol{X_m} \pm u_m)$ [$u_m$(%)] | $y_d \times 10^2$ ($\Delta_d^Y \times 10^2$) [$^{Pred.}\Delta_d^Y \times 10^2$, cf. Eq. 4a] | $x_m = (\boldsymbol{X_m} \pm u_m)$ [$u_m$(%)] | $y_d \times 10^2$ ($\Delta_d^Y \times 10^2$) [$^{Pred.}\Delta_d^Y \times 10^2$, cf. Eq. 4a] |
| $^{144}$Nd | 143.910083 | 1 | 1.0 | 23.798(12) | 23.797738 [0.01458] **0.03332** | 1.0 | 23.789812 (−**0.0333**06) [−0.033317] | 1.0 | 23.798418 (0.002856) [0.002856] | 1.0 | 23.791850 (−0.024744) [−0.024750] |
| $^{142}$Nd | 141.907719 | 2 | 1.14101 (0.00027) [0.023663] | 27.153(19) | 27.153457 [0.02164] **0.4413** | 1.14128 [0.023663] | 27.150837 (−0.009650) [−0.009654] | 1.14128 [0.023663] | 27.160658 (0.026520) [0.026519] | 1.14128 [0.023663] | 27.153162 (−0.001086) [−0.001086] |
| $^{143}$Nd | 142.90981 | 3 | 0.51154 (0.00037) [0.072331] | 12.173(18) | 12.173495 [0.06458] **0.08804** | 0.51191 [0.072331] | 12.178243 (0.039001) [0.039014] | 0.51191 [0.072331] | 12.182648 (0.075188) [0.075186] | 0.51191 [0.072331] | 12.179286 (0.04757) [0.04758] |
| $^{145}$Nd | 144.912569 | 4 | 0.34848 (0.00012) [0.034435] | 8.293(7) | 8.293036 [0.03466] **0.06204** | 0.34860 [0.034435] | 8.293129 (0.001118) [0.001118] | 0.34836 [−0.034435] | 8.290417 (−0.031580) [−0.031580] | 0.34860 [0.034435] | 8.293839 (0.009683) [0.009686] |
| $^{146}$Nd | 145.913112 | 5 | 0.72228 0.00031 [0.042920] | 17.189(17) | 17.188630 [0.03770] **0.06148** | 0.72259 [0.042920] | 17.190280 (0.009600) [0.009603] | 0.72197 [−0.042920] | 17.181744 (−0.040065) [−0.040064] | 0.72259 [0.042920] | 17.191753 (0.018165) [0.018170] |
| $^{148}$Nd | 147.916889 | 6 | 0.24186 (0.00015) [0.062019] | 5.756(8) | 5.755721 [0.06013] **0.08820** | 0.24201 [0.062019] | 5.757372 (0.028693) [0.028703] | 0.24171 [−0.062019] | 5.752316 (−0.059165) [−0.059164] | 0.24201 [0.062019] | 5.757866 (0.03726) [0.03727] |
| $^{150}$Nd | 149.920887 | 7 | 0.23691 (0.00018) [0.075978] | 5.638(9) | 5.637922 [0.07304] **0.10073** | 0.23709 [0.075978] | 5.640327 (0.042647) [0.042661] | 0.23673 [−0.075978] | 5.633799 (−0.073125) [−0.073122] | 0.236730 [−0.075978] | 5.632245 (−**0.1007**0) [−0.10073] |
| Reported [6] atomic weight ($A_{Nd}$) = 144.2415 | | | (Our evaluated) $A_{Xe}$ = 144.241539 $\rho_{Xe}$(%) = 0.0002482 $^{\mathbf{Pred.}}\boldsymbol{\epsilon_{Nd}^A}$**(%) = 0.0005437** | | | $a_{Xe}$ = 144.241788 ($\Delta_{Nd}^A$= 0.00017308%) [$^{Pred.}\Delta_{Nd}^A$ = 0.00017314; Eq. 4a] | | $a_{Xe}$ = 144.240754 ($\Delta_{Nd}^A$= −**0.0005436**7%) [$^{Pred.}\Delta_{Nd}^A$ = −0.00054365; Eq. 4a] | | $a_{Xe}$ = 144.241302 ($\Delta_{Nd}^A$= −0.00016414%) [$^{Pred.}\Delta_{Nd}^A$ = −0.00016418;Eq. (4a)] | |

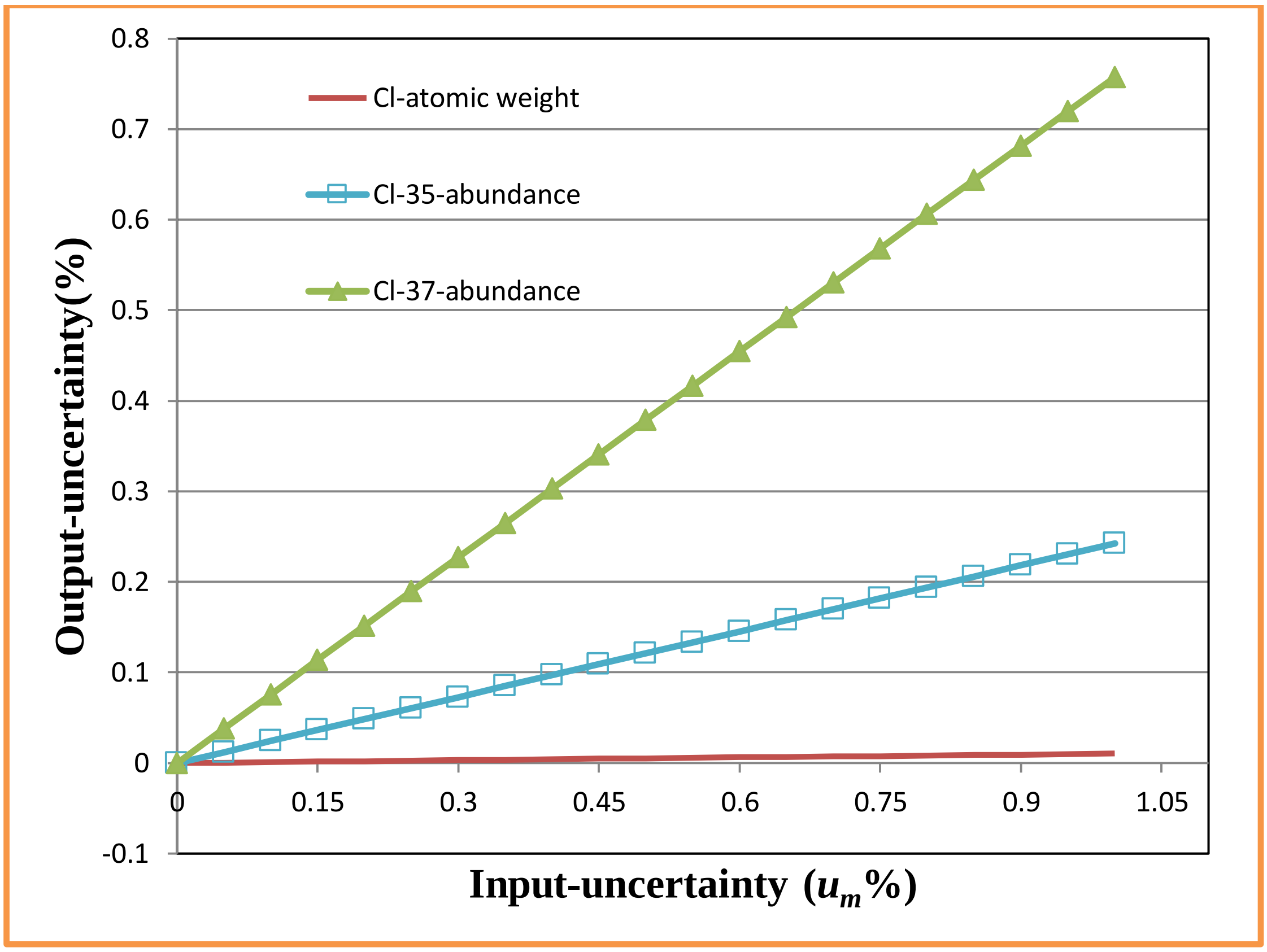

**Fig. 1:** Variation of ***uncertainties*** *of determining* the natural $^{35}$**Cl**-abundance, $^{37}$**Cl**-abundance and the atomic weight **(i.e. of the output-uncertainties** $\epsilon_{35}^{Y}$**,** $\epsilon_{37}^{Y}$ **and** $\epsilon_{Cl}^{A}$**,** respectively**)** as a function of the ***input-ratio-uncertainty*** **(**$u_m \equiv u_{37/35}$**)**

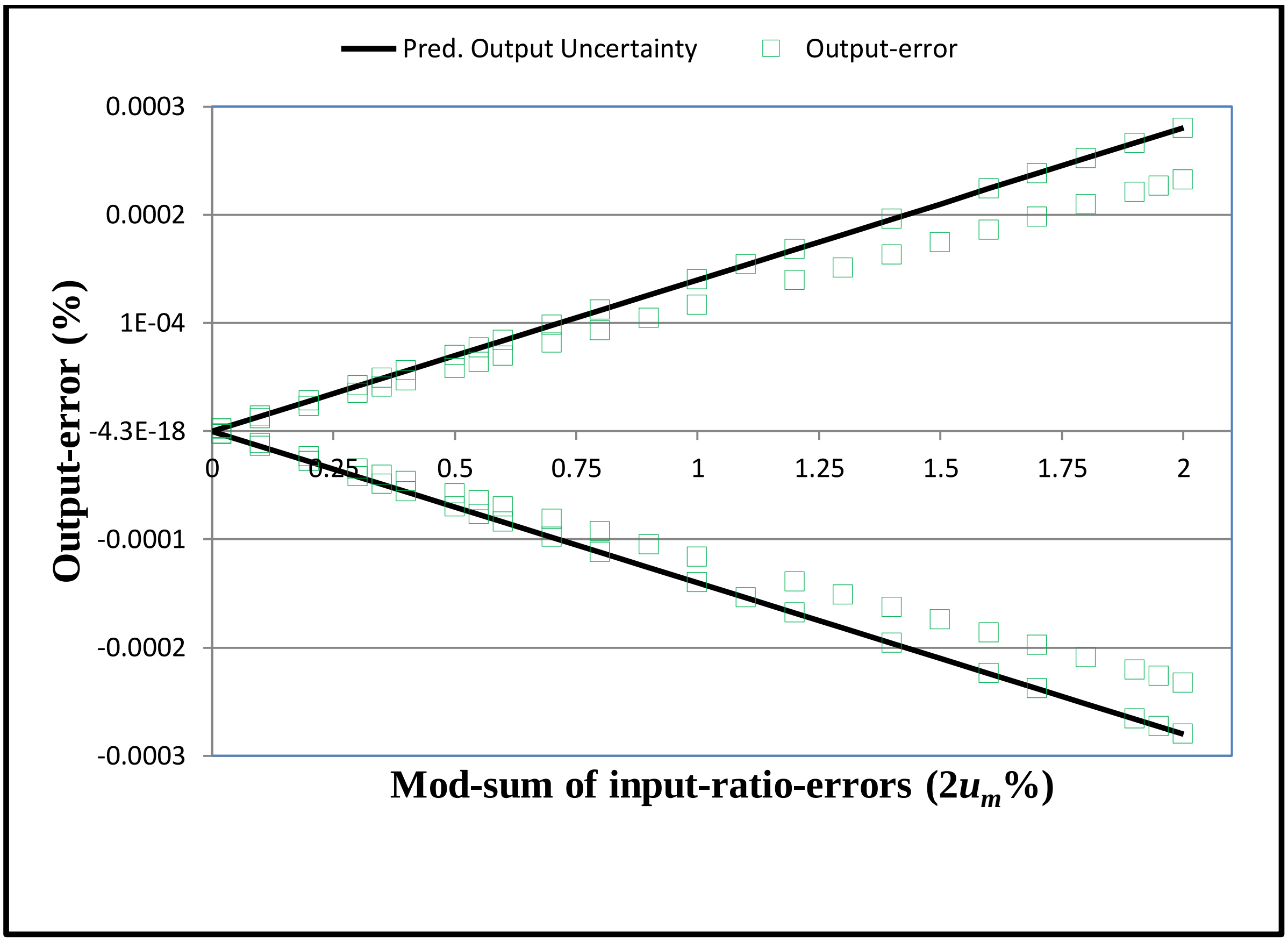

**Fig. 2:** *Variation* of the (observed **true**) *error* of determining the atomic weight of oxygen (i.e. of the *output error*"$\Delta_O^A$") as a function of the ***mod-sum*** of (the possible **"±"** ***combinations*** of the ***two***) *input-ratio-errors* ($\sum_{m=2}^{3}|\pm\Delta_m| = \sum_{m=2}^{3}|\pm u_m| = 2u_m$)

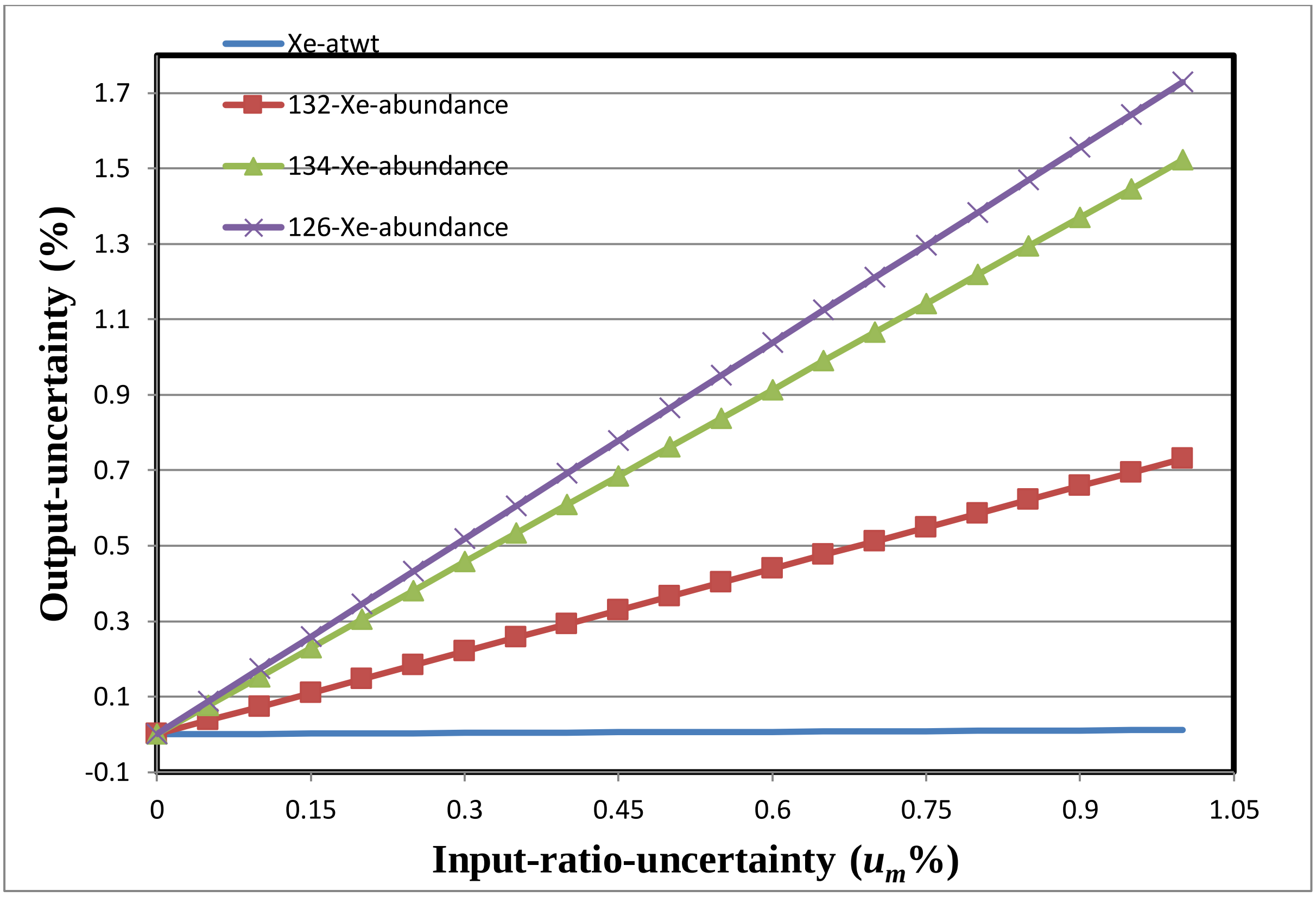

**Fig. 3: Variation of the *output*-uncertainties (of determining the abundances of "$^{126}Xe$, $^{132}Xe$ and $^{134}Xe$" and also the atomic weight of Xe i.e.) "$\epsilon_{126}^{Y}$, $\epsilon_{132}^{Y}$ and $\epsilon_{134}^{Y}$" and also $\epsilon_{Xe}^{A}$ (respectively) as a function of the *uncertainty* "$u_m$" of all the eight *equally* uncertain *input-ratios*.**

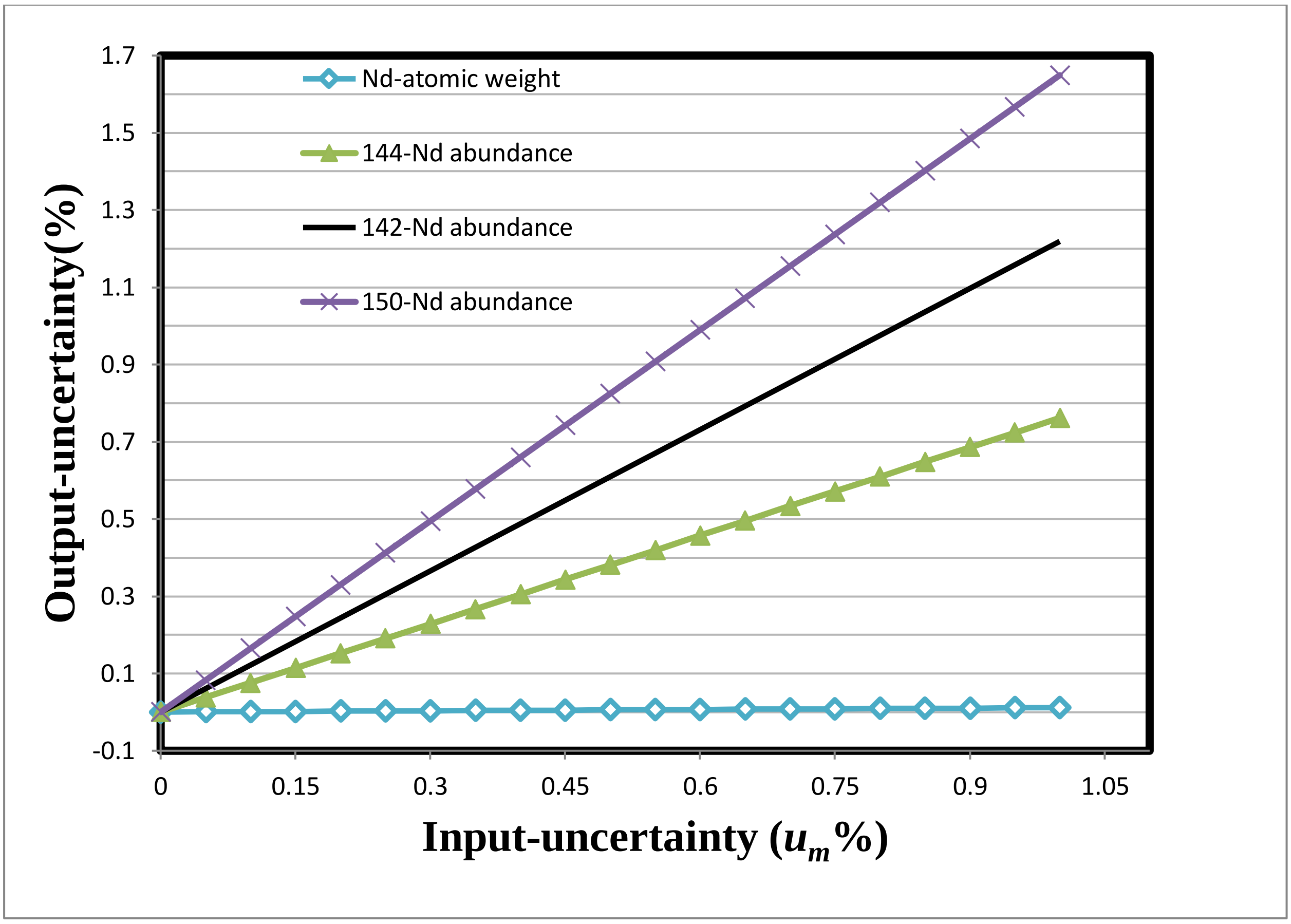

**Fig. (4a):** Variation of ***output**-uncertainty* (**ϵ**, i.e. for determining one or the other typical isotopic abundance or the atomic weight of **Nd**) as a function of (isotopic-ratio-measurement)-*uncertainty* **"$u_m$** (however for the specific case where all the *six* ***input**-ratios* to be ***equally*** uncertain as "$u_m$")**"**

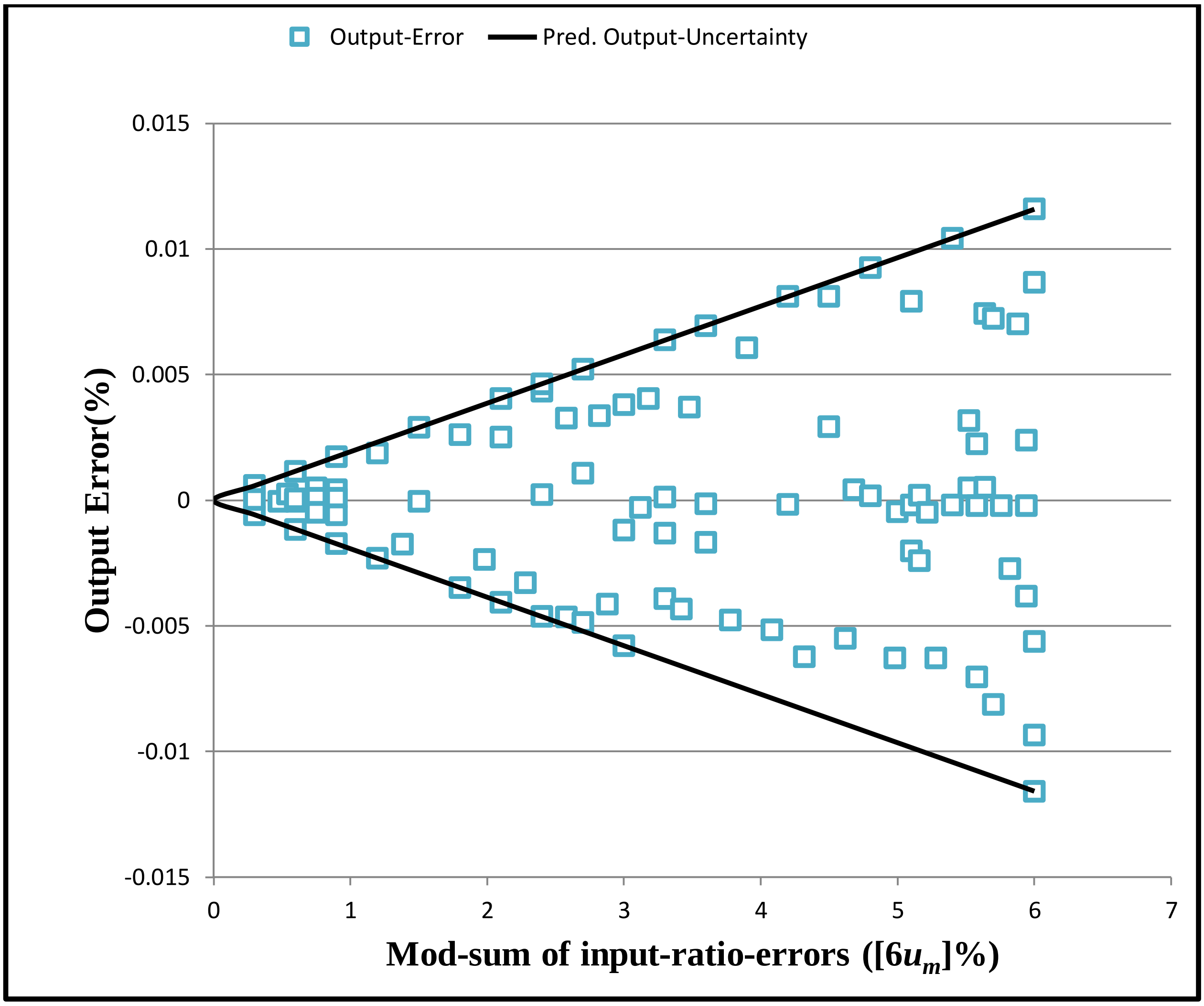

**Fig. (4b):** *Variation* of (observed **true** *error* of determining the atomic weight of Nd, i.e. of) the *output error* "$\Delta_{Nd}^{A}$" as a function of the ***mod-sum*** of (certain typical "±"***combinations*** of the known ***equally***-*erroneous* six) *input-ratio-errors* ($\sum_{m=2}^{7}|\pm\Delta_m| = \sum_{m=2}^{7}|\pm \boldsymbol{u}_m| = \boldsymbol{6u_m}$)

**Graphical Abstract**

$$Y_d = f_d(\{X_m\}_{m=2}^{N})\ ; with\ d = 1, 2 \ldots N$$

Thus: $(y_d \pm \epsilon_d^Y) = f_d(\{x_m \pm u_m\}_{m=2}^{N})\ ; with\ d = 1, 2 \ldots N$

How: $(u_m(s) \rightarrow \epsilon_d^Y)$? ➔ Answer: $(\epsilon_d^Y = f_d(\{u_m\}_{m=2}^{N}))$